\documentclass[11pt,openany]{book}
\usepackage[utf8]{inputenc}

\usepackage{url}
\usepackage{hyperref}
\usepackage{xcolor}
\hypersetup{
    colorlinks,
    linkcolor={black!50!black},
    citecolor={blue!50!black},
    urlcolor={blue!80!black}
}
\usepackage[natbib=true,backend=biber,sorting=nyt,style=apa]{biblatex} 
\usepackage{graphicx} 
\usepackage[margin=1in]{geometry} 
\usepackage[english]{babel}
\usepackage{ae}
\usepackage[Lenny]{fncychap} 
\usepackage{tikz,fp,ifthen,fullpage}
\usepackage{multicol}
\usepackage{pgfplots}
\usepackage{endnotes}
\usepackage{appendix}
\usepackage{setspace} 
\usepackage{amsmath}
\usepackage{amssymb}
\usepackage{times} 
\newtheorem{defi}{\textcolor{red}{Def}}  
\newtheorem{theorem}{Theorem}

\usepackage{array} 
\usepackage{multicol}
\usepackage{enumitem}
\usepackage{caption}
\usepackage{subcaption}
\usepackage{lipsum}
\usepackage{listings} 
\lstdefinestyle{mystyle}{
    backgroundcolor=\color{white},   
    commentstyle=\color{black},     
    keywordstyle=\color{blue},      
    stringstyle=\color{red},        
    basicstyle=\small\ttfamily,     
    breaklines=true,                
    captionpos=b,                   
    frame=single,                   
    framesep=3pt,                   
    rulecolor=\color{black},        
    numbers=left,                   
    numberstyle=\tiny\color{gray},  
    showstringspaces=false,         
}

\title{A computational model for the evolution of learning physical micro-contents in peer instruction methodology}
\author{Paco H. Talero L.}
\date{August $2023$}

\begin{document}

\begin{titlepage}
  \thispagestyle{empty}
  \begin{minipage}[c][0.17\textheight][c]{0.25\textwidth}
    \begin{center}
    \hspace*{-13mm}
      \includegraphics[ height=5cm]{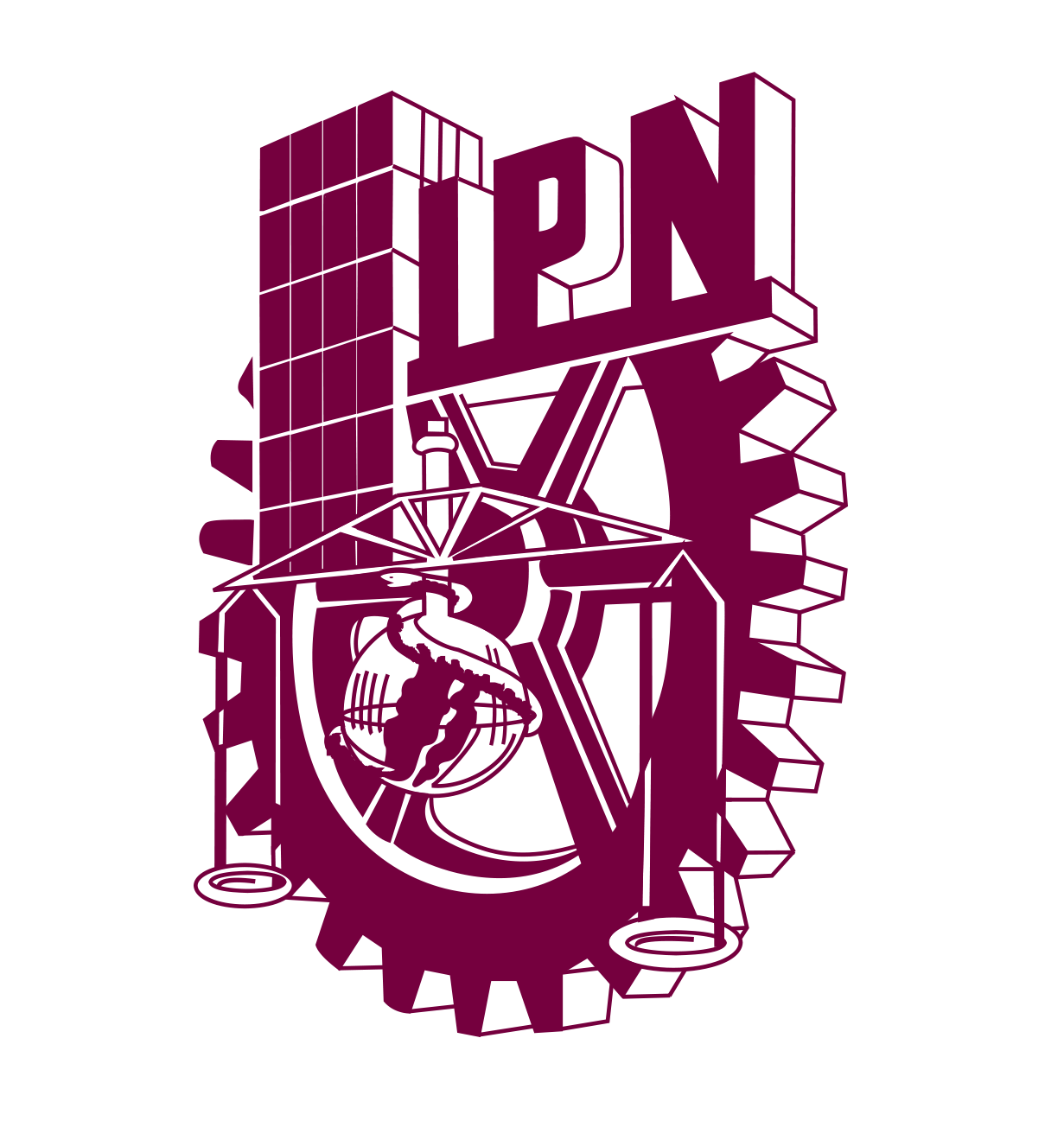}
    \end{center}
  \end{minipage}
  \begin{minipage}[c][0.195\textheight][t]{0.75\textwidth}
    \begin{center}
      \vspace{0.3cm}
             {\color{black}\textsc{\large Instituto Politécnico Nacional} }\\[0.5cm]
             \vspace{0.3cm}
                    {\color{black}\hrule height3pt}
                    \vspace{.2cm}
                           {\color{black}\hrule height2pt}
                           \vspace{.8cm}
                           \textsc{ \large Centro de Investigación en Ciencia Aplicada y Tecnología Avanzada, Unidad Legaria}\\[0cm] %
    \end{center}
  \end{minipage}
  \begin{minipage}[c][0.81\textheight][t]{0.25\textwidth}
    \vspace*{5mm}
    \begin{center}
      \hskip0.5mm
             \vspace{5mm}
             \hskip2pt
                 {\color{black}\vrule width3pt height13cm}
                 \hskip2mm
                     {\color{black}\vrule width1pt height13cm} \\
                     \vspace{5mm}
                     \hspace*{1mm}
                     \includegraphics[height=3.0cm]{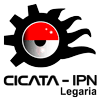}
    \end{center}
  \end{minipage}
  \begin{minipage}[c][0.81\textheight][t]{0.75\textwidth}
    \begin{center}
      \vspace{1cm}

      {\color{black}{\large\scshape A computational model for the evolution of learning physical micro-contents in peer instruction methodology}}\\[.2in]

      \vspace{1cm}            

      \textsc{\LARGE T\hspace{0.5cm}E\hspace{0.5cm}S\hspace{0.5cm}I\hspace{0.5cm}S}\\[1.5cm]
      \textsc{\large que para obtener el grado de:}\\[0.5cm]
      \textsc{\large DOCTOR EN CIENCIAS EN FÍSICA EDUCATIVA}\\[0.5cm]
      
      {\color{black}\textsc{\large presenta:}}\\[0.5cm]
      \textsc{\large {Paco Hernando Talero Lopez}}\\[1cm]          

      \vspace{0.5cm}

      {\large\scshape 
        {\color{black}Director de tesis}\\[0.3cm] {DR. César Eduardo Mora Ley }}\\[.2in]

      \vspace{0.5cm}
       
      \large{January $2024$}
    \end{center}
  \end{minipage}
\end{titlepage}
\renewcommand{\rmdefault}{ptm} 
\renewcommand{\figurename}{Fig}
\renewcommand{\tablename}{Tab}  
\chapter*{Abstract}
\thispagestyle{empty}
One of the most important active methodologies for physics learning developed in recent years is peer instruction. Its technique has allowed, among other things, to monitor students' conceptual learning. In this sense, \textcite{PhysRevSTPER.6.020105} has proposed a model that seeks to understand the dynamics of this methodology. However, her model is very phenomenological and overlooks fundamental aspects such as the cognitive process that students follow during interaction with their peers, the role of the instructor, the connection between content and student characteristics, and long-term learning. The objective of this thesis was to develop a computational model based on neurocognitive principles that aimed to address the shortcomings found in Nitta's work. The model was formulated, simulated, and validated based on several field studies on the learning of one-dimensional graphical interpretation and the falling of bodies among science and engineering students in Bogotá, Colombia.\\

Keywords: computational model, peer instruction, physics learning, neurocognitive principles,one-dimensional graphical interpretation, falling bodies. 
\thispagestyle{empty}
\vspace{10cm}
\begin{flushright} 
\large \emph{Y a mí se me ocurre que hemos caminado más de lo que\\ llevamos andado.
Se me ocurre eso. De haber llovido \\ quizá se me ocurrieran otras cosas.}\\
\large \textbf{JUAN RULFO} \\
\large \emph{\textbf{Nos han dado la tierra}}
\end{flushright}


\vspace{5cm}
\begin{flushleft}
   \large \textbf{Dedication}:\\
   \vspace{\baselineskip}
   
   \large \emph{To my sons, colleagues and friends, who have inspired, nurtured, enriched, and supported the ideas developed in this thesis.}
\end{flushleft}

\singlespacing
\thispagestyle{empty}
\vspace*{3cm}

\begin{flushleft}
\large \textbf{Acknowledgments}:\\
\end{flushleft}

I wish to express my sincere gratitude to all the individuals and institutions that have been pillar in the completion of this doctoral thesis.

First and foremost, I would like to extend my heartfelt thanks to Dr. César Mora Ley, my director, for his invaluable guidance, confidence, and accuracy in both academic and administrative aspects; his mentorship has been crucial to the success of this work. Likewise, I'm deeply appreciative of the constant support provided by the Faculty of Engineering and Basic Sciences of the Central University; their unwavering support has been indispensable throughout the entire research process. I'd like to offer my special thanks to the students who generously and voluntarily participated in the studies; their collaboration has been vital in obtaining significant results. Lastly, I want to give special recognition to my colleagues, whose rich discussions, valuable contributions, in-depth criticism, and support in field research have allowed this work to come to  fruition.
With profound gratitude,\\

Paco H. Talero L.

\pagenumbering{roman}
	\tableofcontents
	\listoftables
	\listoffigures

\pagenumbering{arabic}
\addcontentsline{toc}{chapter}{Introduction} 
\singlespacing
\chapter*{Introduction}

In recent years several physics departments, principally in the USA, have turned their attention to the field of physics education at a university level, forming the Physics Education Research (PER) group aimed at investigating this field with a scientific approach. Such an approach was proposed in the late $1990s$ by E.Redish and L.McDermott at the University of Maryland and the University of Washington, respectively \autocite{mcdermott1999resource}. These researchers argue that it is feasible to treat the teaching and learning of physics as a science rather than an art,  and that in fact with the existing research methodologies until then (mid-$2000$) and the results obtained in this field, the teaching-learning of physics could be considered an empirical science \autocite{mcdermott2001oersted,benegas2007tutoriales}. 

The results obtained by the PER have allowed us to understand, among other things, that: active methodologies are much more efficient than traditional expository methods when learning elementary physics \autocite{hake1998interactive};  the development of questionnaires or global instruments of multicultural inquiry in different areas of physics facilitate the homogenization of learning problems \autocite{hestenes1992force,beichner1994testing,ding2006evaluating}; is necessary to take cognitive science as a basis to understand the advantages and limitations of cognitive origin that students have when learning elementary physics \autocite{PhysRevSTPER.6.020105,grote1995effect};  quantitative instruments allow the measurement of the impact of instruction and the quantification of the teaching and learning process \autocite{hake1998interactive,bao2001concentration,talero2013diagrama};  and that the formulation of mathematical models allows to structure and generalize empirical results, as well as forecasting new research and application scenarios \autocite{mcdermott2001oersted,pritchard2008,PhysRevSTPER.6.020105}.

The PER has also identified some important issues related to instructional implementation and methodological research. Regarding the instruction, it has been observing: first, traditional instruction does not eliminate the misconceptions brought by students from their everyday life \autocite{redish1994implications}; second, students use similar and erroneous mental models when explain a physical fact \autocite{ding2006evaluating}; third, if instruction is based on active learning, then there are better results than traditional instruction \autocite{hake1998interactive}. About the methodology of investigation, it is observed that some lines of research use the pre and post-test methodology and ignore the process between the two moments, which introduces errors in the data they collect \autocite{heckler2010happens};  and second, the mathematical models that have been developed are described in totally global terms without taking into account students on an individual basis, which leaves an understanding of how individual dynamics influence global behavior \autocite{PhysRevSTPER.6.020105}.

Furthermore, among the most important active-learning methodologies is the peer instruction (PI), which was developed by Harvard physics professor Eric Mazur in the $1990s$. In the PI approach, students prepare the topic at home, and in class, they solve problems using multiple-choice questions; first, students work individually, and then they work with a partner, finally, the teacher answers questions and resolves any doubts. \autocite{mazur1999peer}.

Research indicates that the PI methodology can enhance student learning, but it must overcome several obstacles, including resistance from both students and teachers, as well as the need to improve the depth of knowledge that students acquire \autocite{crouch2001peer,PhysRevSTPER.6.020123}. 

In the context of the Peer Instruction (PI) methodology, several questions emerge. These inquiries encompass the instructor's significance, the learning process during interactions among peers and with the teacher, the time required to consolidate a small physics topic effectively, the impact of group size on the learning process, the role of cognitive confrontation quality in shaping learning outcomes, and the scope and depth of content to be taught. These aspects will be explored and addressed in the development of this thesis.

The general objective of this thesis is formulate, simulate and validate a model that explains the learning mechanism in PI, based on some principles of cognitive neuroscience and some results of the PER and validate it in engineering students in Bogotá (Colombia).

Nonetheless, there are some limitations to the model. Firstly, it's not feasible to validate all the model's implications in just one field study. Secondly, the model does not take into account the attitudinal behavior of students. Thirdly, the model doesn't consider the interferences that arise from studying similar topics in other courses simultaneously. Lastly, the model doesn't consider the learning or confusion that students may experience outside of class.

To shape the model, first an extensive and detailed observation of the target population was carried; then an epistemological framework was defined with respect to which the concept of model was defined; after, based on the works of 
\textcite{axelrod1997dissemination,grabowski2006evolution}, the concept of microcontent or small physics topic \autocite{peng2019application} was formulated the model; at once to validate it, a fieldwork was designed and applied to first-year engineering students studying graphic interpretation in URM as a topic..

The model helps focus on solving the problem that PI faces regarding teacher training and methodology evaluation, enabling a more efficient comparison of PI with other methodologies. The fundamental reason for this model is to provide a conceptual framework on the nature of learning in PI.

This thesis is organized as follows: the first chapter presents the problem statement; in the second chapter it is presented
the preliminary research that forms the basis of the model; the third chapter focuses on formulating the central axis of this thesis, which is the model itself; the fourth chapter showcases some virtual experiments; the sixth chapter presents two field studies, and their results are compared with the theoretical predictions of the model. Finally, in the fifth chapter, the discussion and conclusions are presented. Additionally, development material such as tests, computer programs, etc. are presented in appendices and annexes.

\chapter{Problem statement} \label{cap:prob}

Following the initial proposal raised by  \textcite{mcdermott1999resource} and subsequently supported by \textcite{bao2001concentration}, \textcite{pritchard2008mathematical}, \textcite{PhysRevSTPER.6.020105}, among others, the primary the intention of this thesis is to improve the field of educational physics by furnishing it with a substantial scientific foundation, encompassing both experimental and theoretical aspects.

However, contributing to educational physics in general from a scientific perspective that explains, predicts and agrees with observations in the classroom is totally unfeasible, given that this problem contemplates sociocultural aspects belonging to a particular historical moment, diversity in intentions government, peculiarities of the institutions, objectives of educational institutions, curricular dynamics, ideology of teachers and cognitive characteristics that favor or hinder students' learning of physics, among many others.

Based on the above, it is necessary to identify research fields within educational physics that contain a higher degree of scientific elements. These fields should unquestionably be connected to the protocols of various currently active methodologies. The aim is to investigate how to enhance or modify certain steps within these methodologies with the intention of formulating a partial theory that can be experimentally or observationally tested in the classroom.

The most suitable methodology for this purpose is peer instruction (PI). With PI, you have control over the questions, quickly ascertain the students' responses, and can measure the outcomes of both successes and failures. Furthermore, a model of the methodology is available, enabling monitoring of the interaction among peers and the actions of the instructor.

However, despite research demonstrating that PI can enhance student learning outcomes, it faces several challenges. Student resistance is one issue, as some students may not feel comfortable with PI and may prefer a more traditional lecture-based approach. Instructor training is another hurdle since the success of PI hinges largely on the quality of questions posed and the instructor's ability to facilitate productive discussions among students. Consequently, instructors require additional training, time, and resources. Assessment in PI relies heavily on multiple-choice questions to gauge students' understanding, which is useful for providing immediate feedback and stimulating discussion but may fail to capture the full depth of students' comprehension. Additionally, evaluating the effectiveness of the methodology and comparing it with other teaching approaches can be challenging \autocite{crouch2001peer,PhysRevSTPER.6.020123}.

In light of the aforementioned, it may not be reasonable to expect a scientism \footnote{In this work, scientism should be understood as an epistemological approach that advocates tolerance towards various forms of knowledge, including intuition, everyday knowledge, artistic, and technical knowledge. It does not regard science as the sole form of knowledge, nor does it idealize it as infallible. Instead, it opposes a dogmatic stance that portrays science as the sole essential pillar in society and culture \autocite{teixido2021necesidades}.} explanation for every aspect of educational physics or all its methodologies. Even in the case of peer instruction, which holds promise, we are still a long way from developing a robust theory encompassing the learning processes occurring within this methodology. A prudent step forward is to formulate a simplified model that addresses the fundamental aspects of PI as a learning mechanism and is amenable to experimental or observational validation. Rather than a mathematical model based on functions leading to differential equations, a more suitable approach involves a computational model that considers discrete and potentially random facets of learning.

Hence, the formulation of the model necessitates insights from the field of neuroscience regarding learning and techniques derived from the study of sociophysics, which examines interactions among individuals within a cultural context.

\section{Problem}

As previously discussed, this thesis addresses the challenge of imbuing a scientific dimension into the peer interaction methodology to bridge the gap between the intuitive or artistic perception of physics education and an evidence-based pedagogical perspective. Thus, the central question guiding this research is: How should we formulate, simulate, and validate a computational model that comprehensively accounts for the learning mechanisms engendered in students when they engage with PI methodology?

To answer this question, the model under development must encompass several key factors: the role of the instructor within the PI framework, the time required for the effective consolidation of a specific physics topic within the PI approach, the impact of group size on the learning process, the influence of the quality of cognitive interactions on learning outcomes, and the scope of the content to be taught.

Applying a scientific framework to the PI  methodology involves grounding it in established cognitive learning principles, scrutinizing the characteristics of the subject matter to be taught, considering the impact of instructors' teaching styles and ideologies, and comprehending the dynamics of educational institutions. Validating such a model within the classroom necessitates a deep understanding of both the students' existing disciplinary knowledge and their sociocultural attributes, values, and attitudes.

\section{Question research}

How can a computational model be formulated, simulated, and validated to comprehensively account for the learning mechanisms in students engaging with the PI methodology, and what insights does the developed model provide regarding the instructor's role, the time required for effective consolidation of specific physics topics, the influence of cognitive interactions on learning outcomes, and the relevant scope of content within the PI approach?

\section{Objectives}
\subsection{General objective}
The overarching goal is to formulate, simulate, and validate a computational model that explains the learning mechanisms occurring in students when they engage in PI methodology.

\subsection{Specific objectives}

\begin{enumerate}
    \item Conduct field studies on a selected population to ascertain their cognitive characteristics, disciplinary knowledge, and attitudes toward physics and its learning, for provide the foundation for formulating the model's assumptions, delineating its limitations, and identifying pertinent parameters.
    \item Establish an epistemological framework to define the concepts of learning and knowledge, determining the scope and depth of the subjects to be taught, with the purpose of identify key cognitive principles underpinning the model.
    \item Define the mathematical formalism supporting the model, incorporating fundamental elements of linear algebra and probability theory.
    \item Formulate the computational model elucidating the processes of physics learning, subsequently simulating and validating it.
    \item Create educational experiment tools and measurement instruments to compare the model's predictions with classroom results.
\end{enumerate}

\section{Justification and limitations of the research}

The justification for this research encompasses two primary aspects. Firstly, in alignment with McDermott's perspective, it is imperative to bridge the divide between methodologies grounded in scientific research and those rooted in intuition and empirical experience. Secondly, it holds significant value to gain insight into the learning mechanisms of students and the instructor's role within the peer instruction (PI) methodology, with the ultimate aim of bringing these insights into the classroom.

The primary limitation of this research lies in the aspects that the model does not address. These include student attitudes, conceptual interferences that may arise when students study other subjects, learning experiences, and the potential confusion students may encounter beyond the classroom.

\chapter{Unraveling the target population's state: an in-depth exploration of their initial knowledge}

In order to specify the educational system that will be modeled, an exploration of the student population to which the model refers was necessary, an exploration of the student population to which the model refers was necessary. This section presents the results of an exploratory research on conceptual aspects of mechanics in a group of engineering students during their first semesters in Bogotá  (Colombia). 

The objective of this exploratory population study was two-fold. First, to determine the conceptual understanding and attitude towards learning physics of first semesters engineering students in Bogot\'a regarding mechanics. Second, to conduct a pilot test aimed at identifying the main challenges students face when interpreting graphs of one-dimensional kinematics.

To achieve these objectives, we applied the FCI \autocite{hestenes1992force} and TUG-K \autocite{beichner1994testing} tests, as well as the Colorado test \autocite{colorado}. In addition, we designed a tutorial \autocite{taleroMUG} and developed instruments to assess students' ability to interpret graphs of URM. To conclude the exploratory study, we conducted both an informal interview with the instructors and a structured interview with the students who participated in one of the pilot tests.

\section{Conceptual understanding of newtonian mechanics}

The FCI was administered to $786$ engineering students between the ages of $16$ and $30$ from $8$ universities in Bogot\'a, they belonged to diverse social strata; the test was administered after the instructors had taught the fundamentals of mechanics, including kinematics and Newton's laws. The objective of this preliminary research was to analyze the distribution of correct answers that the students had and to identify some common errors. 

The high school physics curriculum in Bogotá covers various topics, including uniform rectilinear motion (URM), uniform accelerated rectilinear motion (URAM) and vertical free fall. The poor performance of students on questions $1$, $3$, $19$ and $20$ caught the attention of the author of this thesis and other researchers. Questions $1$ and $3$ relate to the vertical fall of bodies, while questions $19$ and $20$ concern URM and URAM. \href{https://drive.google.com/file/d/1HM9X0O69W6tsjfXOzXULF9Uw9TuI4Epd/view?usp=sharing}{Annex 1} provides additional information on the test\footnote{To view the FCI instrument used, click on the word ``Annex".}. The search for an explanation for this surprising finding served as the motivation for the development of this work.

By crossing the results of questions $1$ and $3$ in the FCI, it can be deduced that there are three predominant ideas among the students: first, an Aristotelian thought in which it is believed that heavy bodies always fall first and then light ones; second, a Galilean thought that is based on the belief that bodies always fall at the same time; and third, a Newtonian thought, characterized by the understanding of the forces that act on the body (in the case of what is formulated in the FCI), namely the weight and the force of friction \footnote{In the FCI, the buoyant force is not taken into account since it is small compared to the weight of the spheres referred to in questions $1$ and $3$.}. According to the results, it was found that approximately $7\%$ expressed an Aristotelian thought, $25\%$ a Galilean thought and only $5\%$ a Newtonian thought.

Likewise, when we compare the results of questions $19$ and $20$, it was found that the students confused position with velocity and velocity with acceleration, though we were careful as the FCI $19$ question refers to instantaneous velocity. However, as \textcite{talero2013velocidades} shows in a URAM, the instantaneous velocity evaluated in an average time $t_p =\frac{1}{2}(t_1 + t_2)$ is equal to the mean and average velocity referred to at times $t_1$ and $t_2$. Therefore, it was not possible to infer what type of velocity the student was thinking of when answering the question.

\begin{figure}[ht]
  \centering
  \includegraphics[width=0.8\textwidth]{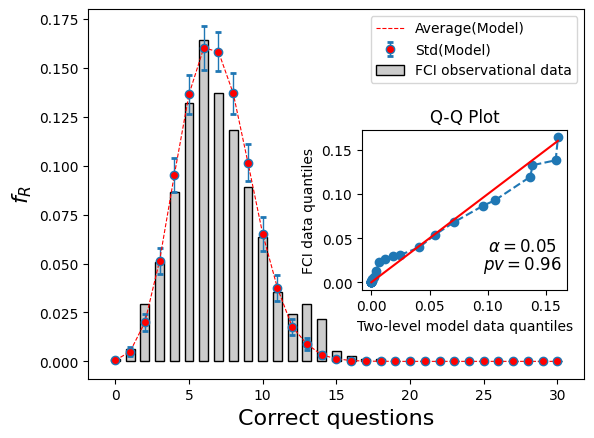}
      \caption{The bar graph corresponds to the distribution of correct answers in the FCI test, while the dashed line graph represents the fit using a two-level heuristic model. The $Q-Q$ plot explains the fit of the distributions.}
  \label{fig:fci}
\end{figure}

The bar graph shown in Fig.(\ref{fig:fci}) illustrates the distribution of the number of correct answers provided by the students. After visual analysis the graph, it appears that the distribution is similar to a binomial probability distribution with an average of around $6$, a standard deviation of approximately $2.2$ and therefore a probability of success of $\frac{1}{5}$. A more detailed study reveals that $f_R$ does not strictly follow a binomial distribution. Nevertheless, the curve suggests a certain homogeneity in the answers provided by the students. Thus, we formulated a heuristic model that assumes the presence of at least two types of students with different levels of knowledge, which are present in different proportions relative to the total number of students who answered the test.

The model consists of the following: among the $N$ students who responded to the FCI,  $N_{l}$ responded with a probability of success of $p_{l}$ in each question, while another $N_{h}$ responded with a probability of success of $p_{h}$ in each question. In other words, the model assumes that the Bogotá sample consists mainly of two types of students, each with a different level of knowledge. Thus, the model proposes that the student is in a state of constant confusion and in multiple ways. When students answer a question, they eliminate some options with a certain degree of certainty, thereby increasing their probability of answering correctly. In general, they answer all questions with the same probability, consequently each group responds with a binomial distribution with a different probability of success but an equal number of trials.

Probability theory states that two binomial distributions with the same number of trials and success probabilities $p_1$ and $p_2$ can be combined into a single binomial distribution with the same number of trials, but with probability $p$. This probability is a linear combination of $p_1$ and $p_2$ ($p=a p_1+b p_2$) which must satisfy $a+b=1$. However, in the model proposed here this last condition is not always fulfilled, since $p_l+p_h>1$ may occur in some cases. So $f_R$ in the Fig.(\ref{fig:fci}) does not necessarily have a binomial distribution.

To confront the heuristic model with the data, it was essential to simulate the students' behavior based on the dynamics mentioned above. The model was materialized using the Monte Carlo technique \autocite{gould2006introduction}. When run with the parameters shown in the Tab.(\ref{tab:dataMC}), the average relative frequency $f_R$ and its corresponding standard deviation are represented by the dashed line and error bars in Fig.(\ref{fig:fci}). The program that implements the model can be found in  \href{https://github.com/PacTal/Doctoral-thesis/blob/main/Two_Level_Model/Two_Level_Model.ipynb}{Annex 2} or in APP. (\ref{app:Twolevel}).
\begin{table}[ht]
    \centering
        \begin{tabular}{|c|c|c|c|c|c|}
        \hline
        $N$ & $\frac{N_{l}}{N}$ & $\frac{N_{h}}{N}$ & $p_{l}$ & $p_{h}$& $U$\\
        \hline\hline
        $1000$ & $0.53$ & $0.47$ & $0.2$ & $0.25$ &$100$\\
        \hline
        \end{tabular}
    \caption{Parameters used to simulate the behavior of the students when answering the FCI.}
    \label{tab:dataMC}
\end{table}
The virtual experiment, implemented in the simulation shown in annex 2 or APP.(\ref{app:Twolevel}), takes $N_l$ virtual students and makes them answer $30$ questions with a probability of $p_l$ for successfully answering each question. Simultaneously, it takes $N_h$ virtual students and makes them answer the same questions with a probability of $p_h$. It does this for $U$ virtual universities and then calculates the mean and standard deviation for each number of correct answers.

The two-tailed Kolmogorov-Smirnov test was used to validate the fit of the observed data with the data obtained from the virtual experiment. The results are shown in the quantile-quantile plot of Fig.(\ref{fig:fci}). In this test, the null hypothesis ($H_o$) is rejected if the p-value ($pv$) is less than $\alpha$. The null hypothesis affirms that the two samples come from the same distribution. The results found do not allow us to reject the null hypothesis and it must be accepted with a significance level of $\alpha=0.05$, which corresponds to a $5\%$ probability of committing a type I error: rejecting $H_o$ when it is true.

Based on the model, it can be concluded that the sample of engineering students in Bogotá who responded to the FCI consisted of two slightly different groups of students. Specifically, $53\%$ of the students responded without any knowledge by choosing one of the five options with the same probability of $\frac{1}{5}$, while $47\%$ of the students ruled out one option with certainty and responded randomly, they chose one option among the remaining four with a success probability of $\frac{1}{4}$. However, the model's predictions were not entirely accurate, as the observed number of students who answered eight questions correctly was lower than that predicted by the model, while the number of students who answered $13$ and $14$ questions correctly was higher than expected by the model.

\section{Interpretation of kinematic graphs in one dimension}

To conduct further research into the graphical interpretation of one-dimensional kinematics among engineering students from Bogotá, a specific university was selected, hereinafter referred to as University $A$. The engineering faculty of University $A$ has enrolled approximately $4000$ students from various engineering classes taking Physics I, Physics II and Physics III courses, typically in groups of $20$ students for four hours per week. In the first course, Newtonian particle mechanics is taught from a theoretical-experimental perspective using a primarily traditional methodology, with linear algebra as a prerequisite and differential calculus of one variable as a corequisite.

An inquiry instrument was developed for the study, using the TUG-K test and two questions from the FCI. The instrument was translated and adapted by the author for the target population; four faculty professors reviewed the instrument, see 
\href{https://drive.google.com/file/d/1rYZnN7DogiMH7dNOIIcjSarmv6Zbz96f/view?usp=drive_link}{Annex 3}. It was later administered to $102$ Physics I students at various times throughout the semester; finally, it was conducted with $143$ Physics I students after they had completed their study of kinematics.

\begin{figure}[ht]
  \centering
  \includegraphics[width=0.8\textwidth]{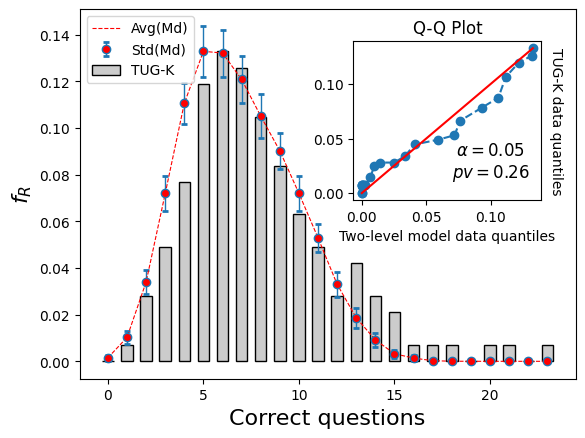}
  \caption{The bar graph corresponds to the distribution of correct answers in the TUG-K test, while the dashed line graph represents the fit using a two-level heuristic model. The $Q-Q$ plot explains the fit of the distributions.}
  \label{fig:tug-k}
\end{figure}

The bar graph in Fig.(\ref{fig:tug-k}) shows the relative frequency of correct answers given by the students. This curve was interpreted using the previously presented two-level heuristic model and the fit was evaluated using the two-tailed Kolmogorov-Smirnov test. A $pv=0.26$ was obtained, which does not allow us to reject the null hypothesis with a level of significance of $\beta=0.05$, as shown the $Q-Q$ plot in Fig.(\ref{fig:tug-k}). Therefore, it must be accepted that the two distributions come from the same distribution. This implies that the students who took the test can be divided into two groups: the first group, with a proportion of $0.6$, answered each question with a probability of $0.23$; and, a second group with a proportion of $0.4$ answered with a probability of $0.4$, see Tab.(\ref{tab:dataTug-k}).

\begin{table}[ht]
    \centering
        \begin{tabular}{|c|c|c|c|c|c|}
        \hline
        $N$ & $\frac{N_{l}}{N}$ & $\frac{N_{h}}{N}$ & $p_{l}$ & $p_{h}$& $U$\\
        \hline\hline
        $1000$ & $0.6$ & $0.4$ & $0.23$ & $0.4$ &$100$\\
        \hline
        \end{tabular}
    \caption{Parameters used to simulate the behavior of the students when answering the TUG-K.}
    \label{tab:dataTug-k}
\end{table}
Although the fact that some students answered more than half of the questions, inconsistencies were found in their answers. This is because they answered very difficult questions such as questions $1$ and $14$, whose correct answers require an accurate interpretation of the area under the curve on the graph of velocity versus time. However, they did not answer questions related to uniform motion such as $8$ or $12$, which are much simpler.

Confusion was found between the slope of a graph and its coordinates; changes in coordinates and slopes were not distinguished; the physical magnitudes on the axes were not read correctly; a correct association between the narration and the graph that represents it was not made and trajectory was confused with position, among others. These results coincided with those reported in the literature.

\section{Students' attitudes towards physics and his learning}

Given that the population at University $A$ consists of engineering students and not students of pure sciences or sciences applied to science teaching, it is not expected a priori that these students would start with a favorable attitude towards conceptual and meaningful learning of physics \autocite{colorado}. This can affect the development of instruction aimed at teaching,  to address this issue, the Colorado test was adapted, validated and administered.

Four professors belonging to University $A$ adapted the test and classified the students' answers based on expert knowledge and knowledge of the population, arranging them in order from least to most favorable scientific attitude. The Colorado test questions were classified using letters $A$ to $E$, where $A$ indicates the least favorable scientific attitude and $E$ the most favorable. The response options presented to the students were ordered according to the criteria established by the expert group (see \href{https://drive.google.com/file/d/1bWTxAwVnTZber8-bsq2DuOC_IJ4OpANi/view?usp=drive_link}{Annex 4} or APP.(\ref{app:colorado})). 

\begin{figure}[ht]
  \centering
  \includegraphics[width=0.6\textwidth]{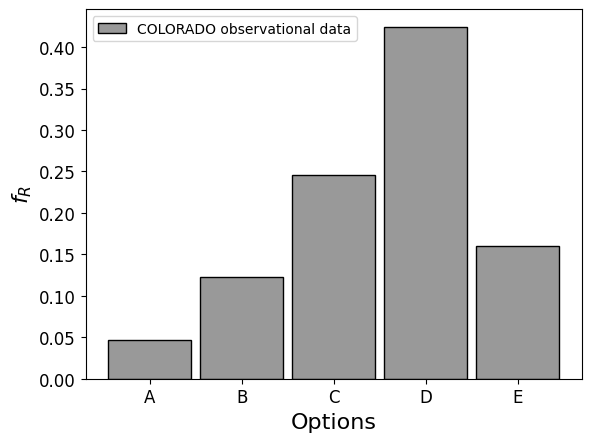}
  \caption{The options are transformed and arranged in order of increasing favorable attitude, with the least favorable option on the left and the most favorable on the right.}
  \label{fig:colorado}
\end{figure}

In line with this, a study was conducted on the scientific attitude of a sample of 106 students, using the Cronbach's alpha coefficient as an indicator of instrument validation. This coefficient measures the internal consistency of a survey and ranges between zero and one. The Cronbach's alpha coefficient is suitable when the instrument items are used additively to interpret an overall result, which requires that each item adds in the same direction. The coefficient is defined as:
\begin{equation} \label{cron}
  \alpha=\frac{N_q}{N_q-1}\left( 1-\frac{\sum_{i=1}^{N_e}\sigma^2_{i}}{\sigma^2_T}  \right).
\end{equation}
Here, $N_q$ represents the number of questions, $\sigma^2_T$ represents the variance of the total score and $\sigma^2_{i}$ represents the variance of the $i$-th. If the value of Cronbach's alpha coefficient ($\alpha$) is greater than $0.8$, the instrument is considered to be reliable; contrariwise, if the coefficient is less than $0.8$, the instrument is considered unreliable. In the present study, a Cronbach's $\alpha$ coefficient of $0.81$ was obtained, indicating that the instrument is reliable with a high degree of confidence. The results obtained are shown in the graph of Fig.(\ref{fig:colorado}), in which the items to the right represent a greater attitude.
In general, a moderately favorable attitude towards physics, its teaching and learning was found. However, a qualitative analysis shows that attitudinally, the connection with the real world, the sophistication in problem-solving, the confidence in problem-solving and the conceptual connections are unfavorable, while personal interest, making sense, applied conceptual understanding and problem-solving are favorable attitudes.

\section{Evolution of the graphical interpretation of rectilinear uniform motion}

A study was conducted with $20$ engineering students from University $A$ to track the evolution of their learning in graphic interpretation in the URM. The active learning methodology based on tutorials was utilized and an instrument consisting of eight questions was designed, validated and administered for this purpose; we will refer to this instrument in the future as TUG-URM (URM Graphics Comprehension Test), which can be seen in  \href{https://drive.google.com/file/d/1av0j72KJTBC47IdK9KoE4uh-uMyEnJPw/view?usp=drive_link}{Annex 5} or in the App.(\ref{app:TUG_URM}).

The study inquired about the kinematic interpretation of the slope, the point of cut with the vertical axis, the sign ($+,-$) in the direct reading and the area under the curve in graphs of position and velocity versus time; as well as the ability to identify a physical situation narrated in graphics.

The instructional methodology used in this research was tutorial sequences. This methodology, based on active learning, was mainly developed by Lillian McDermott and her collaborators at the University of Washington\autocite{mcdermott1999resource,mcdermott2001oersted,benegas2007tutoriales}. This type of instruction takes into account the students' characteristics, including their preconceptions and mental models. The tutorials aim to develop conceptual understanding and quantitative physical reasoning, starting from natural situations for the students based on their preconceptions. Through cognitive conflict, materialized through Socratic dialogue, the instructor guides the student to understand physics concepts. To conduct this research, the author of this thesis designed a tutorial on one-dimensional graphic interpretation (T-ODGI), which was reviewed by three professors from the engineering faculty at University $A$, see \href{https://drive.google.com/file/d/14F8bETp3JuQJTzDXQTVic-K-upTBRkM1/view?usp=drive_link}{Annex 6} or App.(\ref{app:Tutorial_MUG}). The tutorial was based on researches mentioned previously in this chapter. In addition to the URM, the tutorial covers the entire topic of graphic interpretation required by the engineering program curriculum at University $A$.

The applied research methodology consisted of the following steps: first, students read the chapter designated by the instructor before class; second, in class students meet in groups and begin to carry out the activities proposed in the tutorial; third, the instructor walks around the groups and asks pertinent questions that generate cognitive conflict, he confronts  the answers of the students and gives clues to solve the doubts; fourth, the instructor collects the general doubts and makes a magisterial clarification of these, exemplifying with situations raised in the tutorial; sixth, the TUG-URM was applied individually and without any type of notes, the TUG-URM instrument consists of $8$ questions with $5$ options and a single answer, there are $10$ different tests in order of the questions and options, respectively. See  \href{https://drive.google.com/file/d/1av0j72KJTBC47IdK9KoE4uh-uMyEnJPw/view?usp=drive_link}{Annex 5}.

\begin{figure}[ht]
  \centering
  \includegraphics[width=0.6\textwidth]{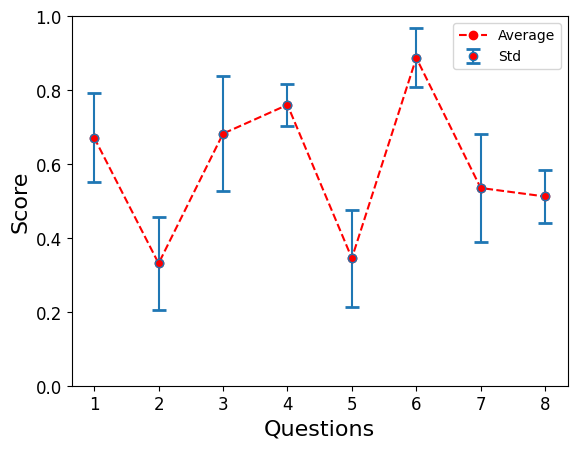}
      \caption{Course scores and standard deviation for each question.}
  \label{fig:pilot_Q}
\end{figure}

The results of the study are displayed in Fig.(\ref{fig:pilot_Q}). The horizontal axis of the figure exhibits the question number in the TUG-URM, while the vertical axis displays the score of each question along with its standard deviation. Since the TUG-URM has ten different shapes (where the order of the questions and the options varies), Tab.(\ref{tab:correspondenceTUG-URM}) illustrates the correspondence between the question number in Fig.(\ref{fig:pilot_Q}) and the question number in annex $5$; it also shows the researched concept.

\begin{table}[ht]
    \centering
        \begin{tabular}{|c|c|c|}
        \hline
        $Q$ in Fig.(\ref{fig:pilot_Q}) & Concept under research& $Q$ in Annex 5 \\
        \hline\hline
        $1$ & Velocity or position vs time graph & $2$  \\ \hline
        $2$ & Graphic identification from the narrative & $6$  \\ \hline
        $3$ & Displacement from velocity vs time graph & $8$  \\ \hline
        $4$ & Meaning of the $x(t)$ and $v(t)$ signs& $10$  \\ \hline
        $5$ & Reading the vertical axis on the velocity vs time graph & $4$  \\ \hline
        $6$ & Constant velocity & $5$  \\ \hline
        $7$ & Discern between graphs of $x(t)$ and $v(t)$ & $3$  \\ \hline
        $8$ & Reading of axes and slope in the graph of position vs time & $1$  \\ \hline
        \end{tabular}
    \caption{Concept under research in the TUG-URM}
    \label{tab:correspondenceTUG-URM}
\end{table}

According to the results shown in Fig.(\ref{fig:pilot_Q}), it is found that the students did not perform well in making a correct association between the narration of a kinematic situation and its graphic representation, either position versus time or velocity versus time. They also had a low performance when asked about the velocity reading on the velocity versus time graph. In addition, the group was very heterogeneous in these two concepts. 

On the contrary, according to the results shown in the graph of Fig.(\ref{fig:pilot_Q}), the interpretation of the signs in the graphs of position and velocity versus time and the concept of constant velocity were well understood.

\begin{figure}[ht]
  \centering
  \includegraphics[width=0.6\textwidth]{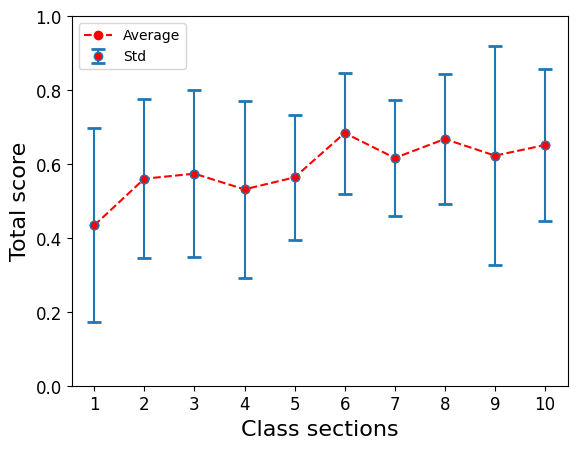}
      \caption{Evolution of the total score and standard deviation in each section.}
  \label{fig:pilot_Total_score}
\end{figure}

Although the active instruction implemented during the $10$ sections, the performance did not change significantly during the process. In addition, a high level of heterogeneity was maintained in the group, which suggests that the knowledge of each student about graphic interpretation in rectilinear motion was not acquired with certainty. The above can be inferred by observing the evolution of the average and standard deviation of the test score per section in Fig.(\ref{fig:pilot_Total_score}).

\section{The relationship between narrative and graphic interpretation in rectilinear uniform motion}

An research was conducted to examine the learning progression of the correlation between the narration of a kinematic situation and its graphical representation in the URM. The objective of the intervention was to ensure that the students acquired the necessary knowledge to interpret graphically a kinematic situation from the narrated URM. The study involved $7$ engineering students enrolled in a Physics I course at University $A$. The course lasted $8$ weeks with a weekly intensity of $12$ hours. During the research, we implemented an interactive instructional guided by tutorials with elements of collaborative learning. The T-ODGI tutorial served as the primary reference for our methodology,see \href{https://drive.google.com/file/d/14F8bETp3JuQJTzDXQTVic-K-upTBRkM1/view?usp=drive_link}{Annex 6}. Additional activities in class were addressed to reflect on the graphs shown in App.(\ref{app:Eval_Oral}).

In order to carry out the research, an inquiry instrument was designed, tested and applied, which consisted of a series of open questions that describe a daily situation. This instrument was corrected by three expert professors both in didactic and in the scolar physics and who know the population of university $A$, see \href{https://drive.google.com/file/d/1QHUMNRE4I7xwFihGbm9dETW3QA3_kvLY/view?usp=drive_link}{Annex 7} or App.(\ref{app:Tutorial_MUG}).

The research methodology involved the application of one of the $12$ problems from the instrument in each class. Those problems had the same level of difficulty and described everyday situations. To elucidate the understanding of the process of associating a narrative with its graphic representation, some conceptual actions were proposed, each with its own indicator. These conceptual actions and their indicators were as follows:

The student
\begin{enumerate} [label=\textbf{Act-\arabic*}]
    \item  identifies the physical situation, since he does a drawing that corresponds to what is expected.
    \item  establishes an adequate reference system, which is manifested in his drawing by observing the location of zero,        orientation of the axis and use of an appropriate length scale.
    \item  applies the concept of a particle, since modeling a moving body either by marking it or by considering it as a        whole.
    \item  associates the slope of a $x(t)$ graph with velocity, which is demonstrated by correctly sketching   the              inclination when drawing this type of graph.
    \item  can connect the physical situation with the kinematic graph of $x(t)$, which is demonstrated by appropriately marking the axes, using the correct scale, adjusting the cut-off points and projecting the data on the axes.
   \item can identify the direction of movement of a body in a $v(t)$ graph, which is demonstrated by correctly aligning the direction of movement expressed in the narrative with the signs in the $v(t)$ graph.
   \item identifies the URM sequences, which is demonstrated by correctly depicting the points of discontinuity of the    
         derivative in both the $x(t)$ and $v(t)$ graphs.
    
\end{enumerate}

Scoring of test was done as follows: the complete problem was observed to evaluate each of the $7$ previous criteria and an item ($A$,$B$,$C$ or $D$) was associated according to the convention shown in the Tab.(\ref{tab:asig}).

\begin{table}[!htp]
\centering
\begin{tabular}{|c|c|}
  \hline
Item  & Assignment  \\ \hline \hline
 $A$  &  Accomplished             \\ \hline
 $B$  &  Unachieved         \\ \hline
 $C$  &  It is not possible to know the answer \\ \hline
 $D$  &  There's no answer   \\ \hline
 
\end{tabular}
\caption{Item assignment}
\label{tab:asig} 
\end{table}

\begin{figure}[ht]
  \centering
  \includegraphics[width=0.7\textwidth]{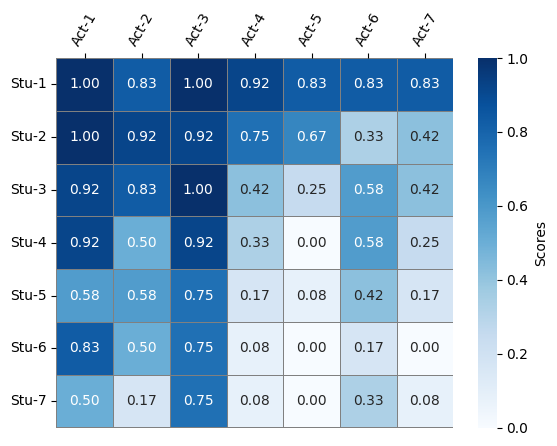}
      \caption{In this map the students (Stu) with the highest scores were located at the top of the graph and the conceptual actions (Act) were ordered from left to right.}
  \label{fig:headMappliot_two}
\end{figure}

\begin{figure}[ht]
  \centering
  \includegraphics[width=0.6\textwidth]{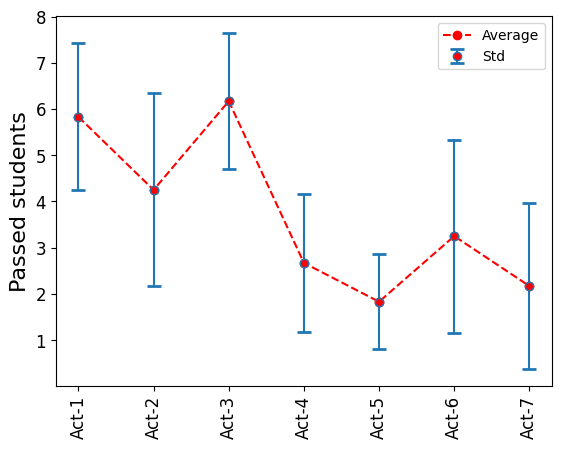}
      \caption{Number of students who completed each action during the entire process.}
  \label{fig:pilot_two}
\end{figure}

Fig.(\ref{fig:headMappliot_two}) displays a heat map indicating the average performance of each student for each conceptual action. A highly diverse group is observed: the group is very heterogeneous;  actions $1$, $2$ and $3$ are moderately achieved by students; actions $4$, $5$, $6$ and $7$ were not achieved; in particular action $5$, which corresponds to the objective of this intervention, was only moderately achieved by two students; finally,  only two students achieved moderate success in action $5$, which was central to meeting the  objective set for this intervention.\\ 
Fig.(\ref{fig:pilot_two}) depicts the average and standard deviation of the number of students who were able to complete each cognitive action throughout the process. The data reveals that: making a coherent drawing, identifying the body as a particle, and establishing a reference system; are the most successful actions, although the success rate is not uniform. Similarly, the central actions of the narrative-graphic interpretation link, such as: identifying sequences, directing movement, reading axes and slope; were only achieved by a few students at certain points in the process.

During an informal interview with the students, they were asked to provide a metacognitive reflection on the challenges they faced while learning the topic. The most relevant aspects identified were the following five:

\begin{enumerate}
    \item I would have understood better if the instructor had carried out experiments or presented practical or concrete situations of the movement that he asked me to represent graphically.
    \item The graphs are all very similar, making them difficult to distinguish. If the professor had provided more explanation when presenting them, perhaps I would have understood them better.
    \item Sometimes peers help clarify concepts, but sometimes they only confuse more.
    \item The students demonstrated a natural tendency to associate graphics with photographs or narratives of positions.
    \item There is not enough time, and concepts are easily forgotten.
\end{enumerate}

\section{Insights and conclusions from the exploration of the target population's initial knowledge} \label{sect:prelimiConclutions}

This section presents the discussion and conclusions of the study that sought to characterize the engineering population of Universidad$ A$ (among other) of Bogotá in terms of their basic knowledge of Newtonian mechanics, graphic interpretation of one-dimensional kinematics, and attitudes toward physics and its learning.

\begin{figure}[ht]
  \centering
  \includegraphics[width=0.8\textwidth]{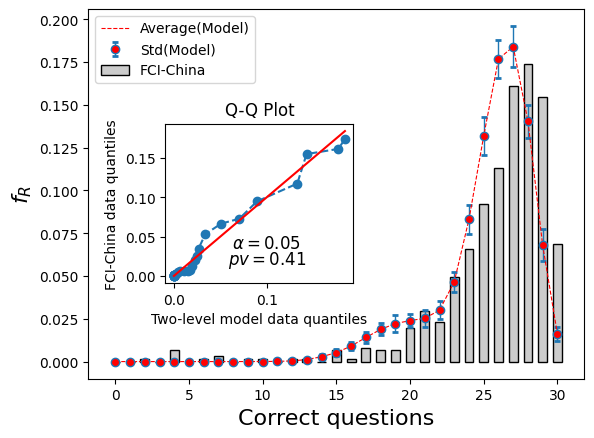}
  \caption{The bar graph corresponds to the distribution of correct answers in the FCI (China) test, while the dashed line graph represents the fit using a two-level heuristic model. The $Q-Q$ plot explains the fit of the distributions.}
  \label{fig:tug-kl}
\end{figure}

The results of the FCI and the TUG-K were interpreted using a heuristic model that postulates the existence of only two predominant types of students who answer each question with a certain probability and have different proportions in the sample. The model fits the results with a significance level of $5\%$. When comparing the results of the FCI for Bogotá with those obtained for the same test in China\autocite{bao2009learning}, similar results were found, but shifted to the right, as shown in Fig.(\ref{fig:tug-kl}).

The previous results allow us to formulate the following working hypothesis: when a student responds to a multiple-choice test with a single answer, he primarily applies their knowledge to rule out options until they attain a certain degree of certainty that enables them to select an answer. This implies that the student is always in a constant process of discernment between various physical ideas, some wrong and others correct. We will call this working hypothesis, from now on, the cognitive discernment hypothesis (CDH).

In addition to the deficiency in graphic interpretation extracted from the FCI results, it was also noted that the most students had not understood the fundamental concepts related to falling bodies during high school.

On the other hand, although an active methodology was implemented to teach the URM, the students could not discern the different aspects of graphic interpretation. This was observed in the wide heterogeneity and low score. This may have been because the students did not have the opportunity to carry out a significant metacognition process that would have allowed them to discern better.

Concerning attitude, most of the students have favorable beliefs regarding the central idea of physical science and its learning. Additionally, the students are aware of the permanent confusion, the shortage of time to develop concepts, and the lack of concrete experiments that support theoretical developments.

The results of this preliminary research serve as the basis for formulating a computational model on the evolution of learning small physical contents. This model will be formulated in the next chapter.

\chapter{Model}

The aim of this chapter is to formulate a computational model that accounts for the learning of physical micro-contents (PMC) in PI methodology. To achieve this goal, we will use the results obtained in the preliminary studies reported in the previous chapter, as well as some PER results on cognitive processes and metrics such as normalized average gain and the concentration diagram. We will also specify the meaning of the model and system, which is of fundamental importance for this work.

The chapter is organized as follows: first, some concepts present in the PER literature used in the formulation of the model are reviewed; second, the epistemological concepts necessary to conceptually locate the model are specified; and third, the model itself is formulated. The formulation of the model proceeds as follows: the system to which the model refers is defined, the basic concepts are defined, the axioms or assumptions are established, the mathematical-computational formalism is proposed and a simulation is implemented.

\section{Preliminary concepts}

We summarize some useful concepts for the formulation of the model.

\subsection{Metrics}
To construct the model or evaluate its performance, we will utilize the metrics provided underneath.
\subsubsection{Average normalized gain} \label{sub:gain}
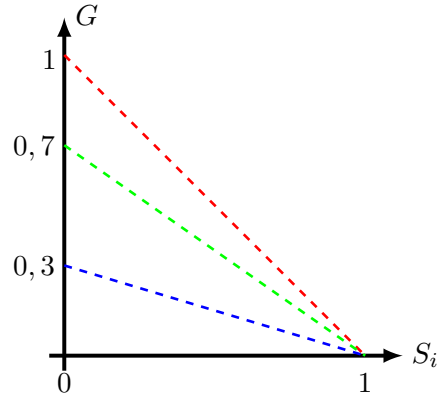
\begin{figure}[!htp]
\begin{center}
	\begin{tikzpicture}[scale=1.0]
	\draw [-latex,black,line width=1.5pt] (-0.2cm,0cm) -- (4.5cm,0cm);
	\draw [-latex,black,line width=1.5pt] (0.0cm,-0.2cm) -- (0cm,4.5cm);
	\coordinate [label=below:\textcolor{black} {$S_i$}] (x) at  (4.8cm,0.3cm);
	\coordinate [label=below:\textcolor{black} {$G$}] (x) at  (0.3cm,4.8cm);
	\coordinate [label=below:\textcolor{black} {$1$}] (x) at  (4cm,-0.1cm);
	\coordinate [label=below:\textcolor{black} {$0$}] (x) at  (0cm,-0.1cm);
	\coordinate [label=below:\textcolor{black} {$0,3$}] (x) at  (-0.4cm,1.45cm);
	\coordinate [label=below:\textcolor{black} {$0,7$}] (x) at  (-0.4cm,3.05cm);
	\coordinate [label=below:\textcolor{black} {$1$}] (x) at  (-0.2cm,4.2cm);
	\draw [red,dashed,line width=1.0pt] (0cm,4cm) -- (4cm,0cm);
	\draw [blue,dashed,line width=1.0pt] (0.0cm,1.2cm) -- (4cm,0cm);
	\draw [green,dashed,line width=1.0pt] (0.0cm,2.8cm) -- (4cm,0cm);	
	\end{tikzpicture}
\caption{Gain vs starting score.}
\label{fig:Gsi}
\end{center}
\end{figure}

The work of Hake \autocite{hake1998interactive} describes a widely used quantitative technique for measuring the learning of a group of students. This technique generally involves using the same instrument as a pre-test and post-test. The Hake's factor, also known as the normalized mean gain, is denoted by $G$ and defined as
\begin{equation}\label{ganacia}
G=\frac{ S_{f}-S_{i} }{ 1-S_{i} }.
\end{equation} 
Here, $S_i$ refers to the score of the group normalized to unity on the pre-test, while $S_f$ refers to the score on the post-test. $G$ is a measure of the effectiveness of a methodology in a particular physics course. The results are usually grouped or classified into high, medium, and low. The proficiency is considered high when $G \leq 0.7$, medium when $0.3 \leq G < 0.7$, and low when $G < 0.3$, see Fig.(\ref{fig:Gsi}).

\subsubsection{Concentration} \label{subs:concentration}
\begin{figure}[!htp]
\begin{center}
	\begin{tikzpicture}[domain=0:1,scale=5]
	\draw [-latex,black,line width=1.5pt] (-0.1cm,0cm) -- (1.1cm,0cm);
	\draw [-latex,black,line width=1.5pt] (0cm,-0.1cm) -- (0cm,1.1cm);
	\coordinate [label=below:\textcolor{black} {$u$}] (x) at  (1.12cm,0.05cm);
	\coordinate [label=below:\textcolor{black} {$C$}] (x) at  (0.06cm,1.12cm);
	\coordinate [label=below:\textcolor{black} {$1$}] (x) at  (1.0cm,0.0cm);
	\coordinate [label=below:\textcolor{black} {$1$}] (x) at  (-0.05cm,1.05cm);
	\coordinate [label=below:\textcolor{black} {$0,4$}] (x) at  (0.4cm,0.0cm);
	\coordinate [label=below:\textcolor{black} {$0,7$}] (x) at  (0.7cm,0.0cm);
	\coordinate [label=below:\textcolor{black} {$0,2$}] (x) at  (-0.1cm,0.25cm);
	\coordinate [label=below:\textcolor{black} {$0,5$}] (x) at  (-0.1cm,0.55cm);
	
     \draw [black,line width=0.5pt] (0cm,1.0cm) -- (1cm,1cm);
     \draw [black,line width=0.5pt] (1cm,1.0cm) -- (1cm,0cm);
     \draw [black,dashed,line width=0.5pt] (0cm,0.2cm) -- (1cm,0.2cm);
     \draw [black,dashed,line width=0.5pt] (0cm,0.5cm) -- (1cm,0.5cm);
     \draw [black,dashed,line width=0.5pt] (0.4cm,0.0cm) -- (0.4cm,1.0cm);
     \draw [black,dashed,line width=0.5pt] (0.7cm,0.0cm) -- (0.7cm,1.0cm);
     \draw[color=blue,line width=1pt]   plot ({\x},{2.0*sqrt(1-2.0*\x+2.0*\x*\x)-1.0}); 
	\draw[color=red,line width=1pt]   plot ({\x},{2.0*sqrt(1.0/3.0-2.0*\x/3.0+4.0*\x*\x/3.0)-1.0});
	\end{tikzpicture}
\caption{Concentration vs score.}
\label{fig:ConC}
\end{center}
\end{figure}
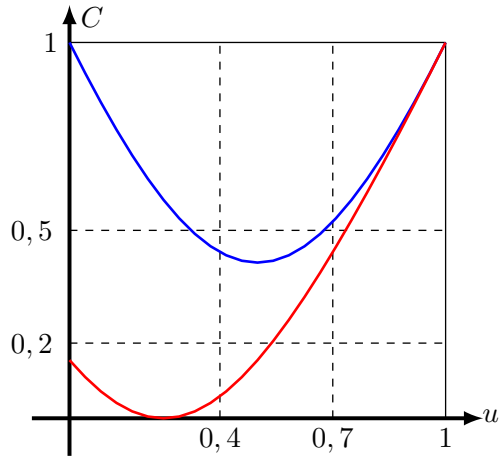

Another important and useful quantitative technique is the concentration factor or concentration analysis \autocite{bao2001concentration}, which aims to measure the distribution of $N_e$ students' responses among $m$ possible options in a multiple-choice question with a single answer. The concentration analysis is carried out by examining the relationship between the concentration $C$ and the overall test score $S$ for each of the $N_p$ questions with the same number of options. This analysis is typically performed using a single question administered to a  group of students. Concentration is defined as
\begin{equation} \label{ca}  
C = \frac{\sqrt{m}}{\sqrt{m}-1}\left(\frac{ \sqrt{\sum_{i=1}^{m}{n_i^2}}}{N_e}-\frac{1}{\sqrt{m}}       \right) ,
\end{equation}
here $n_i$ refers to the number of times the $i-th$ distractor out of $m$ is selected. To correctly interpret the results, it is necessary to restrict the domain of the ordered pair $(S, C)$ within the bounds set by the curves $C_{min}$ and $C_{max}$, which correspond to the minimum and maximum values of the concentration factor $C$ as a function of the test score $S$, respectively. See Fig.(\ref{fig:ConC}). The functions $C_{min}$ and $C_{max}$ can be expressed as follows:
\begin{equation} \label{ca3}  
  C_{min}= \frac{\sqrt{m}}{\sqrt{m}-1} \left(\frac{\sqrt{ (1-u)^{2}}}{m-1}  +u^{2}-\frac{1}{\sqrt{m}} \right)
\end{equation}
and
\begin{equation} \label{ca6}  
  C_{max}= \frac{\sqrt{m}}{\sqrt{m}-1} \left( \sqrt{(1-u)^{2} +u^{2}} -\frac{1}{\sqrt{m}} \right). 
\end{equation}
Any point $(S,C)$ outside the region bounded by the red and blue curves in the Fig.(\ref{fig:ConC}) is considered invalid and has no meaning \autocite{bao2001concentration}.\\

The concentration vs score plot shown in Fig.(\ref{fig:ConC}) is stratified vertically and horizontally.  Vertically, the performance $S$ increases from left to right. For $S<0.4$, a zone of predominantly random responses is defined, while for $0.4\leq S<0.7$, the performance is considered medium, and for $S\geq 0.7$, the performance is understood as high. Horizontally, the concentration increases from bottom to top. For $C<0.2$, the students' responses are distributed almost equally among all options, forming an almost flat histogram. For $0.2\leq C<0.5$, the distribution is predominantly over two options, while for $C\geq 0.5$, the distribution is predominantly over one option. Inferences such as high concentration with random performance, which indicates many students with a well-articulated misconception, or high concentration with high performance, which indicates many students with a well-established correct idea, can be made based on this plot.

\subsection{PI methodology}
The PI methodology was developed by Professor Eric Mazur at Harvard University \autocite{mazur1999peer,crouch2001peer,caduta}. For the purposes of this work, it is defined by the steps shown below, which were slightly adjusted in respect of the originals.
\begin{enumerate} [label=\textbf{Step-\arabic*}]
    \item Before going to class, the students have already read the theoretical course materials that were developed       
          specifically for this population through a process of didactic transposition.
    \item The instructor gives his students a test question and invites them to solve it individually. The question could be an experimental situation
    \item The instructor, the assistant, or the data acquisition system collects the answers.
    \item The instructor forms groups of two or three students and asks for a single answer to the question. The instructor,         the assistant, or the data acquisition system collects the answers given by the groups of students.
    \item The instructor explains the problem and invites his students to ask.
\end{enumerate}

The PI steps shown above are in agreement with the cognitive principles that we will show below. 

\subsection{Cognitive principles}  \label{par:cognitiveP}

According to \autocite{redish2004theoretical}, ``All cognition takes place as a result of the functioning of neurons in the individual’s brain"(p.$5$). Redis' affirm is based on neuroscience research that has shown how learning, memory, and cognitive processes are closely linked to an individual's neurons, including the intensity and number of connections between them \autocite{KANDEL2014163,BALABAN2006298}. Although understanding the intricacies of neural functioning is not necessary for the purposes of this thesis, we will highlight some cognitive principles related to the teaching and learning of physics.

According to  \autocite{redish1994implications,grote1995effect}, the principles below will be used in this work.

\begin{enumerate}
    \item A student learning a new topic must make a connection between the new knowledge and their previous knowledge.
    \item The elaboration of a student's knowledge largely depends on the context.
    \item When a student has already developed a misconceived schema, making significant changes can be very difficult.
    \item In general, students have different ways of learning.
    \item Generally, students learn most effectively through interactions with other students rather than solely through          instructions from the instructor.
    \item Spaced repetition is more effective than instruction done in a single block of work.
\end{enumerate}

\subsection{Physical micro-content, PMC}
For the purpose of this thesis, a PMC is defined as the theoretical and/or experimental adaptation of a physical concept  that is short in length and depth compared to what is typically addressed in a textbook. The PMC is designed to satisfy the specific characteristics of a particular population of students  and is in accordance with the students' initial state of knowledge \autocite{talero2013diagrama}. 

To characterize a PMC, it is subdivided into a series of conceptual attributes of increasing difficulty and dependency denoted as 
$F_0,F_1,F_2,F_3 \dots,F_{N-1}$. Each of these attributes contains levels of increasing depth, represented by $q_0,q_1,q_2,q_3 \dots,q_{m-1}$. The concatenation and interrelation of these $N$ attributes, together with their $m$ levels, define the PMC in question \autocite{axelrod1997dissemination,caduta}.The PMC is represented by a matrix, referred to as the attribute matrix, in which the conceptual attributes are arranged in increasing order from left to right, forming the columns, and the depth levels at each conceptual level increase from top to bottom, forming the rows, as shown in Tab.(\ref{tab:tabS}). Thus, each student is associated with a matrix of attributes during their learning process, which is characterized by placing a $1$ or a $0$ in each component $(i,j)$. Specifically, if a student understands attribute $F_j$ at level $q_i$ then $(i,j)\leftarrow 1$, otherwise $(i,j)\leftarrow 0$.
\begin{table}[htp]
\centering
\begin{tabular}{|c||c|c|c|c|c|c|c|}
 \hline  
  
 \textsf{} & $F_0$  & $F_1$      & $F_2$  & $\ldots$   & $F_j$  & $\ldots$  & $F_N$  \\ \hline  \hline
 $q_0$      & $0$   &       $0$  &  $0$   & $0$        & $0$    & $0$ & $1$ \\ \hline  
 $q_1$      & $0$   & $\ldots$   &        &            &        & & $\vdots$  \\ \hline  
 $q_2$      & $0$   & $\ldots$   &        &            &        & & $\vdots$    \\ \hline 
 $\vdots $  & $0$   & $\ldots$   &        & $\ddots $  &        & & $\vdots$   \\ \hline  
 $q_i$      & $0$   &  $\ldots$  &        &            &       & & $\vdots$   \\ \hline  
 $\vdots$   & $1$   &  $\ldots$  &        &            &        & & $\vdots$    \\ \hline  
 $q_m$      & $0$   & $\ldots$   &        &            &        & &  \\ \hline    
\end{tabular}
\caption{Conceptual attributes matrix of a PMC}
\label{tab:tabS} 
\end{table}

\section{Epistemological framework}

The computational model that is developed later will have a rigorous scientific perspective. This is intended to add more scientific than artistic or ideological elements to research in physics teaching, thus extending what was suggested by McDermott in $2001$, who at that time maintained that research in physics teaching contained elements of both art and science. \autocite{mcdermott2001oersted}. For this reason, it is necessary to establish a rigorous epistemological framework that allows for the analysis of the scientific validity of the model. This will help determine its effectiveness and relevance in the educational context.

The main epistemic framework on which this work is based is systemic materialism \autocite{primero2019concepto}. This philosophical system makes it possible to clearly develop each of the aspects of the model to be formulated.

\subsection{Ontology, brief comment} \label{def:ontology}
As a basis for formulating the model that is the object of this thesis, it is necessary, albeit superficially, to propose an ontological position on which to reference some epistemological aspects necessary for the proper development of this work.

Ontology, which is responsible for studying the nature of existing things in a general way, is an important part of the systemic materialist philosophy. In this philosophy, an ontology is used that is compatible with the current scientific vision and is based fundamentally on three premises. Firstly, the premise that material things exist independent of the mind. Secondly, the premise that all things are either systems or are part of a system. And thirdly, the premise that changes in the properties of existing things occur in accordance with natural or social laws.

Systemic materialism posits the existence of concrete objects with properties or modes of being, which are referred to as ``things." A thing, according to this philosophy, consists of the existing individual along with its properties. The properties of a thing can depend solely on that thing or on that thing and another; in the former case, such properties are referred to as intrinsic, while in the latter case, they are relational. The state of a thing is the collection of mathematical representations that are associated with its properties. Additionally, a thing can only have a set of states that are dictated by the natural or social law that governs the changes in its properties; these states are referred to as legal states. The patterns of natural or social laws can be universal if they apply to all existing things or derivatives if they only apply to a certain class of things. On the other hand, a thing can be classified as basic if it is not composed of other things, such as an electron; conversely, a thing can be complex if it is made up of multiple things, such as the Earth \autocite{Romero2021scientific}.

In systemic materialism, a material object is defined as a thing that has more than one element in its set of states and is thus susceptible to change. The collection of material things is referred to as matter. Additionally, an object is considered to be real in this philosophy if it is capable of interacting with another object or with itself, meaning that it can induce changes. Based on this definition, reality is the collection of all real things. Furthermore, it is postulated in systemic materialism that only material objects are real \autocite{bunge2002ser,Romero2021scientific,primero2019concepto}.

\subsection{System, emergency, and model}

We will understand by system an object whose parts are related to each other by some kind of link and surrounded by a environment. A system comprises a composition, an environment, a structure, and a mechanism \autocite{Romero2021scientific}. Composition refers to the collection  of things that are interconnected or have some kind of relationship with each other. Environment is the collection of external things that interact with internal things. Structure describes the collection of links between internal components (endostructure) and the collection of links between internal things and their environment (exostructure). The mechanism is the collection of processes that undergo the properties of the elements of the composition.

One of the most important properties of a system is emergent behavior, which refers to the fact that systems in general exhibit properties that are not present in any of their individual components \autocite{bunge2002ser}.

An example of a social system is an engineering school, since it is made up of parts such as instructors, students, classrooms, administrators, etc; it has an environment made up of warehouses, streets, other faculties, etc.; the structure, which is constituted by the relationships of learning engineering, managing, etc.; the mechanism of the faculty is the teaching-learning process \autocite{bunge2007caza}.

In this work, a factual model is understood as a conceptual representation of the mechanism of a real system. The formulation of this type of model requires four essential aspects: first, know the set of factual elements or processes to be represented in sufficient detail; second, use a clear formalism that is free of ambiguities and is as precise and exact as possible; third, relate the elements of formalism to the factual elements or processes; and fourth, define a set of axioms or assumptions and provide concrete, coherent, and reliable data \autocite{Romero2021scientific,velten2009mathematical}. It is worth noting, although this is implicit in what was stated above, that their predictions must be validated through experiments. On the other hand, since models are simplifications and idealizations of reality, they are always incomplete and limited in their scope. They are designed to help understand, explain, or predict the behavior of a system or phenomenon, but they do not capture all of the complexities and nuances of the real world. 

A computational model is a type of model that can be implemented in a computer. It is defined by a formalism that consists of an algorithm, which can be used to simulate and analyze the behavior of the system under study \autocite{pang1999introduction}. On the other hand, a mathematical model is described by a formal mathematical framework, such as differential equations or linear algebra \autocite{gershenfeld1999nature}. 

Several mathematical models of learning have been formulated \autocite{PhysRevSTPER.4.010109,PhysRevSTPER.6.020105}, but they are highly phenomenological and therefore do not take into account the individual underlying causes of each student's learning. Furthermore, while the PI methodology is discussed in \autocite{PhysRevSTPER.6.020105}, the various nuances of student-student and student-instructor interactions, as well as the importance of the content to be learned, are not investigated.

The aim of this chapter is to propose a computational and factual model that captures the learning process in the PI methodology.

\subsection{Knowledge and learning} \label{subsec:knowdge}

Following  \cite{Romero2021scientific}, we consider knowledge as a collection of cognitive processes that emerge from the neural activity of the cognising subject in response to stimuli from the subject's environment or internal to the subject. From this perspective, cognitive processes are understood to be inherent to the subject's neural network and to be stimulus-specific. In other words, knowledge is not considered as an abstract entity or separate from the subject, but as an emergent manifestation of neural activity. Without brain there is no knowledge \autocite{dieguez2022volcado}. Cognitive processes occur in the brain of the knowing subject; in very general terms, some of these processes are attention, perception, memory, planning, problem-solving, decision-making, goal-setting, self-regulation, language processing, emotion regulation, and metacognition, among others.

There are at least three types of knowledge: sensorimotor, perceptual, and conceptual. Sensorimotor knowledge occurs when the subject is required to perform physical actions in response to a stimulus. Perceptual knowledge, for its part, arises when the stimulus requires the perception and interpretation of events. Finally, conceptual knowledge refers to the ability to devise, design, check, correct, and employ similar skills that are required when the stimulus presents an intellectual challenge. These different types of knowledge are manifested in different situations and require different types of skills and responses on the part of the knowing subject \autocite{Romero2021scientific,ardila2001psicologia}. 

Here we are interested in conceptual knowledge, which is characterized, for the purposes of this thesis, by the presence of comprehension or understaing in the knowing subject. Understanding involves a series of cognitive operations whose purpose is to relate part of the pre-existing cognitive network with facts, symbols, or constructions. The main operations that lead to understanding, in order of increasing complexity, are description, subsumption, and explanation.

The description is the most basic manifestation of understanding, it consists of characterizing a fact or a concept by providing an organized set of statements. Subsumption, on the other hand, involves fitting a particular fact into a general scheme. Explanation, which is more complex than description and subsumption, requires a theoretical referent that provides the explanatory premises necessary for the subject to develop a model that fits with the world.

We understand learning as a process that leads to the acquisition of knowledge and involves changes in the structure and function of the learner's brain \autocite{dehaene2013quatre}. During the learning process, synaptic changes occur that reorganize and reinforce the neural connections related to what has been learned. As a result, the learner acquires specific abilities that they did not have before, allowing them to achieve certain objectives. A subject's level of understanding is evidence of their learning.

In the field of neuroscience, it is crucial to emphasize the close relationship between learning and knowledge. When an individual utilizes their pre-existing knowledge to respond to a particular problem, it can induce changes in the structure and function of the brain. This process, referred to as neuroplasticity, involves the strengthening or weakening of neural connections based on the experiences and behaviors of the individual. Therefore, the acquisition and application of knowledge not only shape an individual's behavior, but can also restructure the neural architecture of the brain.

\section{Model formulation}

Physics is essential to modern society as it impacts everything from technology and infrastructure to healthcare and the environment. Its principles and applications are critical in advancing knowledge, driving innovation, and solving the most pressing challenges facing humanity. In addition, physics is an essential discipline for engineering, as it provides the fundamental principles and laws that govern the behavior of the natural world. Engineers need physics to design and develop new technologies and infrastructure, make informed decisions, and drive innovation. Moreover, physics is crucial in addressing societal challenges; therefore, physics is a vital and indispensable discipline for understanding and advancing our world.

Teaching physics effectively is crucial in order to educate citizens with a fundamental understanding of the natural world. These citizens will possess innovative and creative problem-solving skills, critical thinking ability, and a deep appreciation for the world around them. By utilizing these skills in their professional lives, they can contribute to the solution of a wide range of problems that afflict humanity.

It is important to note that learning occurs in students and is sought by the teacher, who is immersed in the regulations of an educational institution. These regulations, in turn, are subject to government guidelines that reflect the values and norms of a particular culture. Seen in this way, the learning and teaching of physics are plural, with multiple edges, and their practices depend on various socio-cultural factors.

In this sense, determining whether a student has learned physics after taking a course is very difficult, mainly due to several factors. These include the plurality of content, which covers depth and breadth, among others; the heterogeneity of the students, which includes their own interests, academic history, social stratum, etc.; the heterogeneity of instructors, which involves their disciplinary and didactic training, as well as their ideology and working conditions, etc.; and the constant changes in the guidelines of the faculties that obey internal and external dynamics, including the behavior of the market.

In line with the above, it is necessary to delimit and simplify the system that will be studied in order to conduct research on the learning of physics with scientific characteristics.

\subsection{System definition} \label{subsec:def_system}

The system is composed of:

\begin{enumerate}
    \item a group of experts in physics and didactics who belong to an educational institution.
    \item a PMC.
    \item $N$ students who exhibit typical academic performance and possess a regular interest in acquiring knowledge of a specific PMC.
    \item the knowledge of each student about the PMC.
    \item the learning that each student has about the PMC. 
    \item a classroom, which includes a laboratory, equipped with electronic devices such as a video projector and computers.
    \item a set of questions presented in a test format on the PMC, which may involve experimental activities.
    \item study materials on the MCF aimed at students, which contain readings and/or videos accompanied by activities designed to enhance their 
          learning experience.
    \item cell phones that have internet connectivity, enabling real-time acquisition of answers from students.
    \item  the processes of the PI instructional methodology.
    \item gradual meetings between the instructor and the students.
    \item a assistant available that retrieves and organizes data output provided by the students to present to the instructor.
\end{enumerate}

The system's environment comprises the sociocultural, academic, and family contexts of the students, alongside the norms and activities set by the faculty and university. Moreover, it entails students' the continuum access to information via the Internet.

The endostructure of the system consists of three primary sets of links. The first set pertains to the relationships among the experts, with the objective of investigating the characteristics of the students to be instructed, producing the PMC, creating study materials for the PMC, and generating questions that foster cognitive confrontation among peers. The second set involves the continuous relationship between the instructor and the expert group to ensure the relevance of the materials. The third set refers to the relationships formed through peer interaction and instructor feedback facilitated by cognitive engagement.

It is important to note that the design of the reference material on the PMC, intended for students, as well as the design of the PMC and the questions (which serve as the measurement instrument), will be elaborated in accordance with the test design standards.

The system's exostructure also consists of three components, one for each set of internal links. Concerning the relationship among experts, the system may be influenced by the rigidity of administrative guidelines, the pressure to calibrate the study materials and instruments, among other factors. Regarding the relationship between the instructor and the group of experts, there may be communication failures or a lack of interest in optimizing the processes. In the classroom, relationships with the external environment may arise from information filtering through internet networks, emotional characteristics of students due to situations outside of class, technical failures, ineffective demonstrative experiences, poorly formulated questions, academic influence among students, absences from class, and other factors.

The mechanism of the system is made up of the set of processes that make learning possible for each student, when the PI methodology is applied within the framework of what is described here.

\subsection{Model domain}

In relation to the system defined in subsection (\ref{subsec:def_system}), the things \footnote{Things are understood in terms of what was defined in (\ref{def:ontology}).} that the model will refer to are: student knowledge and learning; the process by which a cognitive confrontation between students generates knowledge changes in each student; and, the process of interaction between the instructor and the student, which generates knowledge change in the student.

\subsection{Model assumptions} \label{sub:assumtios}
In this section, we will describe the assumptions that the model is based on. For ease of reading, we present them as a list.
\begin{enumerate} [label=\textbf{Assm-\arabic*}]
    \item The students in the system have a normal interest in the study of physics. There are few students with very good performance,           and there are also few students with low performance. This condition is achieved by adapting the PMC to the population.
    \item During the process, the students remain emotionally stable.
    \item The instructor follows the instructions of the PI methodology to the letter: promotes individual reflection, motivates students   to debate, and maintains a Socratic dialogue with students when solving each question.
    \item The process of interaction between peers and between student and instructor is considered isolated in the sense that no significant information enters from outside, either through the Web or by any other means.
    \item The most important learning of a student occurs in the cognitive confrontation with another student or with the instructor when he intends to present in public the arguments that support the answer to a question.
    \item In the interaction between students, the following is assumed:
    \begin{enumerate}
        \item Students with very similar knowledge about the PMC do not debate, and thus do not learn.
        \item Students with very different knowledge about the PCM do not debate and thus do not learn. The student with less knowledge  agrees with the wiser without making the slightest reflection.
        \item When two students with similar knowledge have a cognitive confrontation, the one with more knowledge is more probable to successfully convince the other.
    \end{enumerate}
    \item The writing, illustrations, or any other extra-disciplinary elements contained in the measurement instrument (that is, the questions) do not affect the cognitive processes of any student.
    
\end{enumerate}

\subsection{The model's mathematical formalism} \label{subsec:mathematicalFor}

In this subsection, we present the mathematical formalism upon which the model object of this thesis is formulated.
\subsubsection{Definitions}
\begin{defi}[$\mu$ set] \label{defmu}
Let $\mu$ be the set of matrices $A=\left ( a_{ij} \right )$, $B=\left ( b_{ij} \right )$ and so on,
of dimension $(M+1) \times (N+1)$ and $i=0,1,2,  \dots,M$ , $j=0,1,2,  \dots,N$; with the following properties:
 
 \begin{enumerate}
     \item  $\forall \left ( a_{ij} \right ) \in \mu$, $a_{ij}=0$  or $a_{ij}=1$ $\forall i, j$.
     \item  $\forall \left ( a_{ij} \right ) \in \mu$, $\sum_{i=0}^{M} a_{ij}$ at most is equal to $1$.
 \end{enumerate}
\end{defi}
Here are two examples of matrices in  $\mu$
 \begin{multicols}{2}
\[
\left ( a_{ij} \right ) =
\begin{pmatrix}
0 & 1 & 1  \\ 
0 & 0 & 0  \\ 
1 & 0 & 0  
\end{pmatrix}
\]


\[
\left ( b_{ij} \right ) =
\begin{pmatrix}
0 & 1 & 0 &0 \\ 
0 & 0 & 0 &0 \\ 
1 & 0 & 0 &1 
\end{pmatrix}.
\]
 \end{multicols}   

\begin{defi}[Clean matrix, $V$] \label{def:clean}
 Let $V=\left ( a_{ij} \right ) \in \mu$ with $a_{ij}=0 \; \forall i,j $; $V$ is call, matrix clean.
\end{defi}
\begin{defi}[Matrix, $Y$] \label{defM:Y}
   We denote with $Y=\left ( a_{ij} \right )$ any matrix in $\mu$ such that $a_{Mj}=1$ for every $j$, e.g.:
   \[
Y =
\begin{pmatrix}
0 & 0 & 0 &0  &0\\ 
0 & 0 & 0 &0  &0\\ 
0 & 0 & 0 &0  &0\\ 
1 & 1 & 1 &1  &1
\end{pmatrix}.
\]
\end{defi}

\begin{defi}[Discernment factor, $D(A)$] \label{defM:D}
 $\forall A= \left ( a_{ij} \right ) \in \mu$ and $m,b \in \mathbb{R}^{+}$, we define the $\eta_{ij}$ coefficient as $\eta_{ij}=m(j+1)+i+b$ and the discernment factor as
\begin{equation} \label{discer}  
D(A)=\frac{ \sum_{i=0}^{M} \sum_{j=0}^{N} \eta _{ij} a_{ij} }{\sum_{j=0}^{N} \eta_{mj}},
\end{equation}
\end{defi}
\begin{theorem}  \label{theor:dicer01}
$D(A) \in \left [0,1 \right ]$.
\end{theorem}
Here an example, we take $m=\frac{1}{2}$, $b=\frac{1}{2}$ and
 \[
    A=\left ( a_{ij} \right )=
    \begin{pmatrix}
    1 & 0\\ 
    0 & 1 \\ 
    0 & 0 
    \end{pmatrix},
\]
then the discernment factor (\ref{discer}) takes the form
$$
D(A)=\frac{\left (1 \cdot 1 +1.5 \cdot 0 \right )+\left ( 2\cdot 0+2.5\cdot 1 \right )+\left (3 \cdot0+3.5\cdot0 \right )}{3+3.5}=0.54.
$$
\begin{defi}[Level function, $F(j)$] \label{defM:Level_function}
  Let $\left ( a_{ij} \right ) \in \mu$, we define the level function as 
$$
F(j) = \left\{ \begin{array}{ll} 0\; ,  \quad \text{if} \quad a_{ij}=0\; \quad \forall i  \\ q\; , \quad \text{if} \quad  a_{qj}=1  \end{array} \right. 
$$
\end{defi}
\begin{theorem}  \label{theor:level}
$F(j) \in \left [0,M \right ]$.
\end{theorem}
\begin{defi}[Linear set] \label{defM:linear}
  Let $\mu_l \subseteq \mu$ and $\mu_l \neq V$, see definition \ref{def:clean}; if \quad $\forall \left ( a_{ij} \right ) \in \mu_l$  it is fulfilled
  $$
  \forall j \quad  F(j)\geq F(j+1),
  $$
 then $\mu_l$ is called linear set\footnote{The reason for the names will be clear in the next subsection when we interpret the model.}. An example of an element of $\mu_l$ is
 \[
    \left (b_{ij} \right )=
    \begin{pmatrix}
    0 & 0 & 0 & 1\\ 
    0 & 1 & 1 & 0\\ 
    1 & 0 & 0 & 0
    \end{pmatrix},
\]
\end{defi}
\begin{defi}[Nonlinear set] \label{defM:nonlinear}
  Let $\mu_{nl} \subseteq \mu$ and $\mu_{nl} \neq V$, if $\quad \forall \left ( a_{ij} \right ) \in \mu_{nl}$  it is fulfilled
  $$
  \exists j \mid   F(j) <  F(j+1),
  $$
 then $\mu_{nl}$ is called nonlinear set. An example of an element of $\mu_{nl}$ is
 \[
    \left (c_{ij} \right )=
    \begin{pmatrix}
    0 & 0 & 0 & 0\\ 
    0 & 0 & 0 & 0\\ 
    0 & 0 & 1 & 0\\ 
    1 & 1 & 0 & 1\\ 
    0 & 0 & 0 & 0
    \end{pmatrix},
\]
\end{defi}
\begin{defi}[Location, $L(A)$] \label{defM:location}
Let $A \in \mu$, we denote the location function as $L(A)$ and define it as
$$
L(A)= \left\{ \begin{array}{lcc}
                0 &   if  & A=V \\
                1 &  if   & A \in \mu_{nl} \\
                2 &  if  & A \in \mu_{l} \\
                3  &  if   & A=Y .
             \end{array}
   \right.
$$
\end{defi} 
\begin{defi}[Affinity, $\Omega(A,B)$] \label{defM:afinity}
Let $A=(a_{ij})$, $B=(b_{ij})$ element in $\mu$, we define the set $\Omega(A,B)$ as
$$
\Omega\left ( A,B \right )=\left \{ (x,y)\mid a_{xy}=b_{xy} \; \wedge \: b_{xy}=1  \right \}.
$$
We denote the cardinality of $\Omega(A,B)$ as $n$ and define the  affinity $p_l$, as
$$
p_l=\frac{n}{N+1},
$$
here $N+1$ is the number of columns of $A$ or $B$.

Additionally, we define a parameter $p_c$, which we call the cut parameter. His definition is
$$
p_c= \left\{ \begin{array}{lcc}
                p_l    &  if  & p_l\geq \frac{1}{2}\\
                1-p_l  &  if  & p_l <  \frac{1}{2}.
             \end{array}
   \right.
$$
An example is: 
\begin{multicols}{2}
 \[
    A=
    \begin{pmatrix}
    0 & 1 & 0 & 0\\ 
    0 & 0 & 0 & 1\\ 
    1 & 0 & 1 & 0\\ 
    0 & 0 & 0 & 0\\ 
    \end{pmatrix},
\]
 \[
    B=
    \begin{pmatrix}
    0 & 1 & 0 & 0\\ 
    0 & 0 & 0 & 1\\ 
    0 & 0 & 1 & 0\\ 
    1 & 0 & 0 & 0\\ 
    \end{pmatrix},
\]
\end{multicols}
$\Omega(A,B)=\left \{ (0,1),(1,3),(2,2) \right \}$,  $n=3$,  $p_l=\frac{3}{4}$ and $p_c=p_l$

\end{defi}
\begin{defi}[Raise, $\widehat{R}$ ] \label{defM:raise}
    Let $A=(a_{ij})$ an element in $\mu$.  $\widehat{R} A$ increases the occupancy in the leftmost column of $A$ by one unit. If $A=V$, then $RA$ does $a_{00}= 1$. An example is:
     \[
    A=
    \begin{pmatrix}
    0 & 1 & 0 & 0\\ 
    0 & 0 & 0 & 1\\ 
    0 & 0 & 1 & 0\\ 
    1 & 0 & 0 & 0\\ 
    \end{pmatrix},
\]
    \begin{multicols}{2}
 \[
    \widehat{R} A=
    \begin{pmatrix}
    0 & 0 & 0 & 0\\ 
    0 & 1 & 0 & 1\\ 
    0 & 0 & 1 & 0\\ 
    1 & 0 & 0 & 0\\ 
    \end{pmatrix},
\]
 \[
    \widehat{R}^{2} A=
    \begin{pmatrix}
    0 & 0 & 0 & 0\\ 
    0 & 0 & 0 & 1\\ 
    0 & 1 & 1 & 0\\ 
    1 & 0 & 0 & 0\\ 
    \end{pmatrix}
\]

\end{multicols}
    
\end{defi}
\begin{defi}[Copy operator, $\widehat{H}$] \label{defM:copyOperator}
    Let $A=(a_{ij}),B=(b_{ij}) \in \mu $ and $(x,y) \notin\Omega(A,B)$  we define the copy operator, denoted as $\widehat{H}$, as follows
$$
\widehat{H}\left [ A\leftarrow B, (x,y)  \right ]= \left\{ \begin{array}{lcc}
    A\left ( a_{xy}\leftarrow b_{xy} \right ) &   if  & \sum_{i=0}^{M} a_{iy}=0 \\
    a_{iy}=0 \wedge A\left ( a_{xy}\leftarrow b_{xy} \right ) &  if   & \exists \;  i \;\mid  a_{iy}=1 
             \end{array}
   \right.
$$

    $$
    \widehat{H}\left [ A\leftarrow B, (x,y)  \right ]=A\left ( a_{xy}\leftarrow b_{xy} \right ).
    $$
That is to say, the action of the operator $\widehat{H}$ is to replace only the element $a_{xy}$ of $A$ with the element $b_{xy}$ of $B$; furthermore, if there is already a $1$ in the column, it becomes zero. Note that in general $\widehat{H}\left [ A\leftarrow B, (x,y)  \right ]\neq \widehat{H}\left [ B\leftarrow A, (x,y)  \right ]$. 

\end{defi}

Given a set $\Lambda$ of $N_e$ matrices belonging to $\mu$, which depend on the parameter $t \in \mathbb{N}$, we are making the following definitions:

\begin{defi}[Concentration of states, $C$] \label{defM:concetration}
    For the purposes of this work, we are adapting the concentration factor defined in subsec.(\ref{subs:concentration}) to account for the concentration of the states of the $\Lambda$ elements. Extending the correspondence with the original interpretation, we have:
    \begin{enumerate}
        \item Item $A$, incorrect $\leftarrow $  $V$ state.
        \item Item $B$, incorrect $\leftarrow $  Noninear state.
        \item Item $C$, incorrect $\leftarrow $  Linear state.
        \item Item $D$, correct $\leftarrow $   $Y$ state.
    \end{enumerate}
 We will keep the definition of the score as the number of states $Y$.
\end{defi}

\begin{defi}[Insight gain, $G$] \label{defM:gainD}
    We adapt the gain defined in subsec.(\ref{sub:gain}) to $t$ to account for the evolution of the states in $\Lambda$. Extending the correspondence with the original interpretation, we have:
        \begin{enumerate}
        \item $\bar{S}$   $\leftarrow$  $\bar{D}$
        \item $G=\frac{S_t-S_o}{1-S_o}$ to  $t$   $\leftarrow$   $G=\frac{D_t-D_o}{1-D_o}$ to  $t$, on average and individually. 
    \end{enumerate}
    
\end{defi}

\subsection{Model  interpretation}
In this section, we will connect some relevant system concepts discussed in sec.(\ref{subsec:def_system}), with the mathematical definitions shown in sec.(\ref{subsec:mathematicalFor}).
\subsubsection{The mathematical formulation of the knowledge of a PMC}

According to the definition of knowledge given in sec.(\ref{subsec:knowdge}), we represent the knowledge that a student has about a certain PMC by means of an element of $\mu$. The various elements of $\mu$ represent the diverse connections in the neural network of the brain of the student who knows the PMC. Therefore, different elements of $\mu$ correspond to distinct neuronal configurations. This type of simple representation does not require explicitly stating any specific biological structure. So, we represent a PMC by means of a matrix that belongs to the set $\mu$; his columns represent the conceptual attribute, while his rows represent the level of each conceptual attribute, which is represented by the function $F(j)$, see def(\ref{defM:Level_function}).

The clean Matrix $V$, see def.(\ref{def:clean}), represents the particular case when a student doesn't have any knowledge about the PMC, and the matrix $Y$, see def.(\ref{defM:Y}), represents the particular case when the student has perfect knowledge about the PMC.

We understand the state of knowledge of the PMC of a particular student as a specific configuration of his neural network, represented as an element of $\mu$. Agree with this, there are four main sets of states: clean, nonlinear, linear, and outliers\footnote{Despite the fact that clear, nonlinear, linear, and outliers each represents  a set of states, we will refer to them simply as ``states'' for the sake of simplicity.}. The clean state refers to a state where the student has no knowledge of the PMC. The nonlinear state indicates that the student possesses some knowledge about the PMC but exhibits understanding of complex attributes rather than simple ones, resulting in a lack of coherence in their knowledge; in other words, the student is predominantly in a descriptive stage of knowledge. Students who are able to subsume the PMC but still do not explain are in a linear state free from incoherence. Students in the outliers state have complete knowledge of the PMC, they explain. We represent the clean state with the clean matrix, the nonlinear state with an element of $\mu_{nl}$, the linear state with an element of $\mu_l$, and finally, the outliers state is represented by the $Y$ matrix. Depending on the status a student occupies, has a more or less correct knowledge of the PMC, the state occupation is quantified by the Location function $L(A)$, see def.(\ref{defM:location}).

According to the results of the preliminary investigation reported in sec.(\ref{sect:prelimiConclutions}), students are consistently confused. Specifically, they struggle to correctly discern the components of a concept. In the context of PMC, this confusion is represented by the discernment index $D$, which quantifies the level of correct knowledge regarding a PMC, see def.(\ref{defM:D}). Therefore, $D$ serves as a representation of the accurate understanding of a PMC.

\subsubsection{Learning of PMC, formal definition}

We define the learning of a PMC as the change in the state of knowledge about a PMC with an increase in discrimination $D$ or location $L$, def.(\ref{defM:location}). Thus, a student can increase their discernment while remaining in the same location. That is to say, a student learns when, prompted by external or internal stimuli, he solve problems involving operations on components and  processes related to the PMC; these problem-solving efforts are expressed through description, subsumption, or explanation; which are already contemplated in the conceptual attributes of the PMC.

Let's consider an example:  Given a PMC with three conceptual attributes and three levels, a student starts in state $A_1$ and, after a stimulus, transitions to state $A_2$. The states are displayed below. Did the student learn anything about the PMC?

\[
 A_1=
\begin{pmatrix}
$0$ & $1$ & $0$ \\
$0$ & $0$ & $0$ \\
$1$ & $0$ & $0$ \\
\end{pmatrix}
\rightarrow
A_2=
\begin{pmatrix}
$0$ & $0$ & $0$ \\
$0$ & $1$ & $0$ \\
$1$ & $0$ & $0$ \\
\end{pmatrix}
\]
We first calculate the location, these are $L(A_1)=2$ and $L(A_2)=2$; there is no main state change, before and after the stimulus it continues in the linear state. Let us now calculate the discernment, $D(A_1)=\frac{4.5}{10.5}=0.43$ and $D(A_2)=\frac{5.5}{10.5}=0.52$, therefore the student did learn something about the PMC.

In this thesis, we consider the fundamental learning stimulus for the cognitive confrontation that occurs when a student discusses their position on a problematic situation involving the PMC with another person, either student or instructor. 

In this thesis, we consider the cognitive confrontation between students and between a student and the instructor as the main learning stimulus.  Thus, when two students in states of knowledge $A$ and $B$ of the PMC face each other in a debate about a question in the context of the PI methodology, it is possible for one of them to partially adopt, as a result of their analysis, the position of their opponent. This process is represented by the Copy operator, as shown in def.(\ref{defM:copyOperator}). 

The debate process occurs in terms of the affinity between students see def.(\ref{defM:afinity}). 

If the students are identical, then no cognitive effort is expected, and there will be no change in the state of PMC knowledge. If the students have different knowledge, then the one in the higher location is more likely to win the debate, and in that scenario, the loser copies the depth of one of the winner's conceptual attributes, see def.(\ref{defM:location}). In any case, the loser will be the one who changes and has the chance to learn.

We have a particular interest in comprehending the evolution of the learning process within the collective group of students while subjected to an PI methodology. This representation is effectively conveyed through defs.(\ref{defM:concetration}) and (\ref{defM:gainD}).

\subsubsection{Learning evolution of a PMC} \label{sub:LearnigEvolution}

Within the PI methodology, there is a moment when the instructor explains the answer to the question, capturing the attention of the students for a few minutes. According to source \cite{hake1998interactive,mcdermott2001oersted,benegas2007tutoriales}, this instruction becomes more effective when the instructor use active learning during explanation.

Our objective is to quantify the effectiveness of this brief interaction. To achieve this, we will employ a $p_m \in \mathbb{R}$ number, ranging from $0$ to $1$, where $1$ denotes the most active instruction and $0$  indicates the least active. We characterize the active processes according to the following criteria, each with the same weight:
\begin{enumerate}
    \item The instructor has a general understanding of the knowledge level that their students possess regarding the PMC.
    \item Throughout the development of the solution to the question, the instructor fosters participation and promotes         Socratic dialogue.
\end{enumerate}
 $p_m$ is taken the same for all students because it fundamentally depends on the instructor and his environment. On the other hand, this interaction is important in the learning process in PI because it allows to understand the importance of the instructor.

 Similarly, when experts define the PMC, it becomes imperative to have a comprehensive understanding of the population. To achieve this, various types of instruments are often used. In light of this, the instructor designs and administers an entrance test that provides an overview of students' initial knowledge of the PMC early in the process. The general state of initial knowledge is characterized by a number $p_o\in \left [0,1\right ]$, which is interpreted as follows: when a student is faced individually for the first time with a question referring to the PMC, he has a probability $p_o$ of occupying a given level for each conceptual attribute. 
 
 Regarding the interaction between students, specifically the cognitive confrontation or debate about the correct answer to a question in the PI methodology, we describe it as follows:
 
 \begin{enumerate}
     \item  If students are identical or completely different, an argument-rich debate is not expected. In the first case, total agreement is expected, and in the second, the least knowledgeable will adhere to the most knowledgeable without analysis. Therefore, this type of interaction is not relevant.
     \item If the above is not fulfilled, the students have a significant debate and there is a change in the knowledge of the PMC in the loser of the debate, the student with the lowest knowledge has a greater probability of losing. Obviously, the winner can also change their knowledge state, but this is considered very unlikely and is disregarded. This is because, according to preliminary studies, the type of population to which this model responds does not have significant metacognition capacity. Based on the def.(\ref{defM:afinity}), when two students, in states represented by matrices $A$ and $B$ interac,  the exchange can occur in four cases:

\begin{enumerate} [label=\textbf{Case-\arabic*}]
         \item  If $L(B)>L(A) \wedge p_l\, >\frac{1}{2} $ is fulfilled, then $p_c=p_l$ is done and an $x_r \in \left [0,1 \right )$ is generated; if $x_r<p_c$, then $\widehat{H}\left [ A\leftarrow B, (x,y)  \right ]$ is done; else $\widehat{H}\left [ B\leftarrow A, (x,y)  \right ]$ is done.
         \item  If $L(B)>L(A) \wedge p_l\, < \frac{1}{2} $ is fulfilled, then $p_c=1-p_l$ is done and an $x_r \in \left [0,1 \right )$ is generated; if $x_r<p_c$, then $\widehat{H}\left [ A\leftarrow B, (x,y)  \right ]$ is done; else $\widehat{H}\left [ B\leftarrow A, (x,y)  \right ]$ is done.
         \item  If $L(B)<L(A)$, then  $A$ is exchanged for $B$, and $B$ for $A$ and step $1$ or $2$ is performed according to the value of $p_l$.
         \item  If $L(A)=L(B)$ is fulfilled, then $p_c=p_l$ is done and an $x_r \in \left [0,1 \right )$ is generated; if $x_r<p_c$, then $\widehat{H}\left [ A\leftarrow B, (x,y)  \right ]$ is done; else $\widehat{H}\left [ B\leftarrow A, (x,y)  \right ]$ is done.
 \end{enumerate}
 \end{enumerate}
Besides, in the interaction between the instructor and the student $A$, only the student can change. We will represent this interaction with $\widehat{H}\left [ A\leftarrow Y, (x,y)  \right ]$. 

It is important to remember that the previous list of options corresponds to a single interaction, but in a class, the quality of the interaction is related to the exchange of ideas during cognitive confrontations. Thus, in a single PI cycle, there can be several confrontations that represent the quality of the interaction. Interactions can also occur when you hear other peers talking.

The simulation of the model is presented \href{https://github.com/PacTal/Doctoral-thesis/blob/main/Model/ModeloPhD_2023.ipynb}{Annex 8} and can also be seen in App.(\ref{app:pricipalProgram}).

\chapter{Virtual experiments}

Now that the model was formulated and simulated, we can see the different evolutions of student learning under different parameters  and  initial conditions. We will call these implementations virtual experiments (VEX) \autocite{exvphysics}. The objective of this section is to design a variety of VEX that account for the different possibilities that can occur in the context of the PI methodology. For this we will use the program shown in \href{https://github.com/PacTal/Doctoral-thesis/blob/main/Model/ModeloPhD_2023.ipynb}{Annex 8}.

\section{VEX 1: instructional quality}

Discernment ($D$) allows us to know how the group of students is going through the four states of knowledge presented above, due to the PI methodology. The score allows us to know the proportion of students from the Outliers state who already know the PMC in depth. We ask ourselves, how are these two metrics related when the quality of the interaction with the instructor increases in the PI methodology, that is, if it goes from traditional to highly active? Also, how does this relationship depend on the students' initial knowledge?

\begin{figure}[ht]
  \centering
  \includegraphics[width=0.7\textwidth]{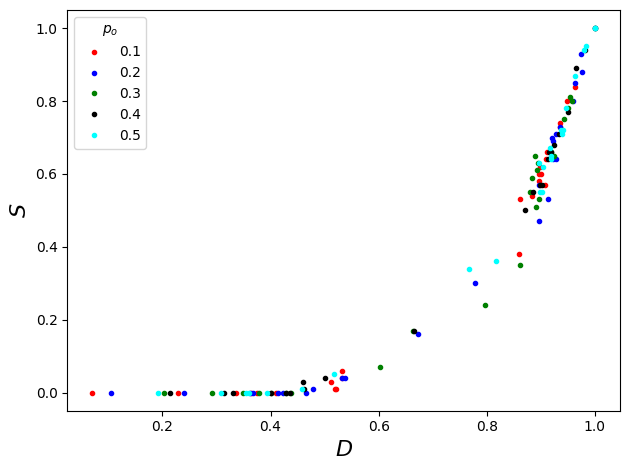}
      \caption{Relationship between discernment and score as the quality of instruction changes.}  
  \label{fig:score_discern}
\end{figure}

\begin{figure}[ht]
  \centering
  \includegraphics[width=0.7\textwidth]{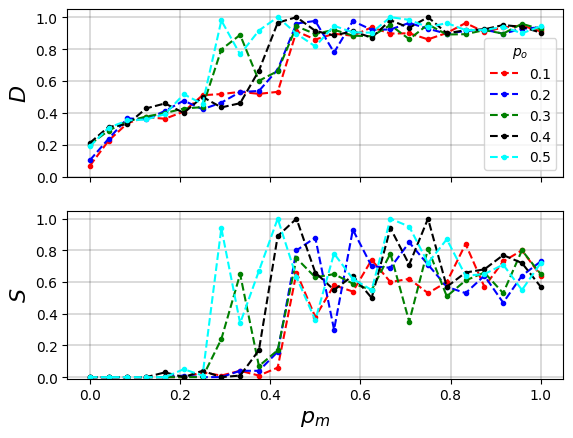}
      \caption{Above, there is discernment versus instructional quality; while below, there is a comparison of score versus the quality of instruction.} 
  \label{fig:score_discern_pm}
\end{figure}

A VEX is proposed with a PMC consisting of three conceptual attributes of three levels, which will be denoted as $\mu_{3\times 3}$ from now on. The VEX involves $100$ students and $102$ PI cycles, see Tab.(\ref{tab:Ins_quality}). This situation is typical of large enrollment courses. The VEX projects what would happen if the PI instruction methodology were applied. The graphs in Figs.(\ref{fig:score_discern}) and (\ref{fig:score_discern_pm}) show the results of the VEX.
\begin{table}[ht]
    \centering
    \begin{tabular}{|c|c|c|}
        \hline
        $\mu_{MN}$  & $n_e$ & $n_c$ \\
        \hline
        $\mu_{3\times 3}$ & $100$ & $102$  \\
        \hline
    \end{tabular}
    \caption{Instructional quality}
    \label{tab:Ins_quality}
\end{table}

The VEX $1$ considers multiple scenarios in which an equal number of students encounter the learning of the same PMC, subjected to the same number of PI cycles, while the quality of active instruction by the instructor improves. The varying factor among these systems is the initial knowledge of the students. Naturally, changes occur within each cycle, such as students' attention, dynamics of interaction among peers, and other aspects; these factors are taken into account through the random component of the model.

The distribution of points that relates discernment and the score is well defined, as shown in the graph of Fig.(\ref{fig:score_discern}). It is observed that discernment increases without the score, at all levels of initial knowledge and without a higher quality of defined instruction. In addition, there is a sparsely populated area between approximately $0.5$ and $0.9$, where the score increases. After $0.9$, there is a heavily populated area with high scores. This means that if students are not discerning well on the PMC, then they will not perform well, but if they are discerning well, then they will score high, as expected.

In Fig.(\ref{fig:score_discern_pm}), it can be observed that with minimal active instruction, the score is practically zero and the discernment does not exceed $0.5$. With active instruction at $0.4$, both score and discernment increase, somewhat earlier in groups with higher initial knowledge. After the quality of active instruction has exceeded about $0.5$, insight is pretty much the same for all groups, regardless of their initial knowledge. Something similar occurs with the score, although there is less variability, and no significant differences are observed between the groups with different levels of initial knowledge.

Note that, despite the large number of cycles and the presence of groups with high initial performance levels, it is not always possible to achieve a total group score. Sometimes the desired score is attained before completing all the cycles, only to experience a subsequent decline in performance. However, towards the end of the process, a consistent and high level of discernment is observed.

It is possible to achieve the conditions for VEX $1$, $100$  students, a PMC  of $\mu_{3\times 3}$ and $100$ cycles are required. The first two are easy to get; however, the third is a bit more difficult, but it can be achieved if $15$-minute cycles are designed, therefore $8$ cycles can be achieved in one class and $4$ more with extra-class monitoring activities, so in a week $20$ will be reached and in $5$ weeks the process is completed.

Finally, it is important to note that when the PMC is of greater dimension, the distribution of the points in the graph of Fig.(\ref{fig:score_discern_pm}) is not clearly defined, and the same occurs when the number of students or the number of cycles decreases. In addition, the discernment grows a little more linear and the score is more heterogeneous.

\section{VEX 2: evolution between states}

In this VEX, a system characterized by the set of parameters shown in Tab.(\ref{tab:evolution_P}) is being considered. We are focusing on observing the distribution of students' knowledge about a PMC in states Clean, Nonlinear, Linear and Outlier; both at the beginning of the process and at the end. We are also studying the evolution of the score and concentration ($S$ vs $C$) in each PI cycle of the system and the gain of insight, $G$.

\begin{figure}[ht]
     \centering
     \begin{subfigure}[b]{0.49\textwidth}
         \centering
         \includegraphics[width=\textwidth]{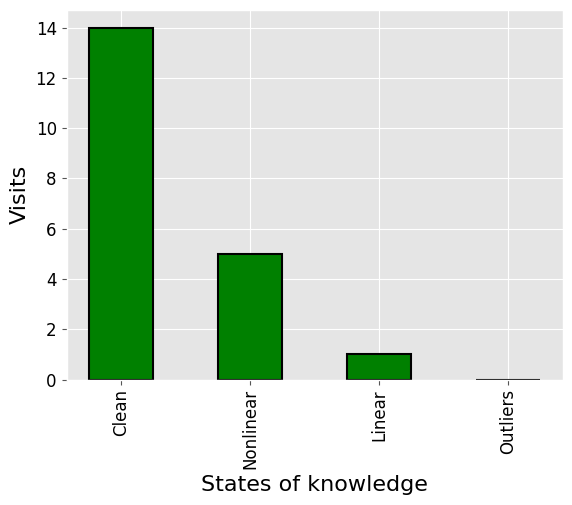}
         \caption{Initial distribution.}
         \label{fig:histo_two_1}
     \end{subfigure}
     \begin{subfigure}[b]{0.49\textwidth}
         \centering
         \includegraphics[width=\textwidth]{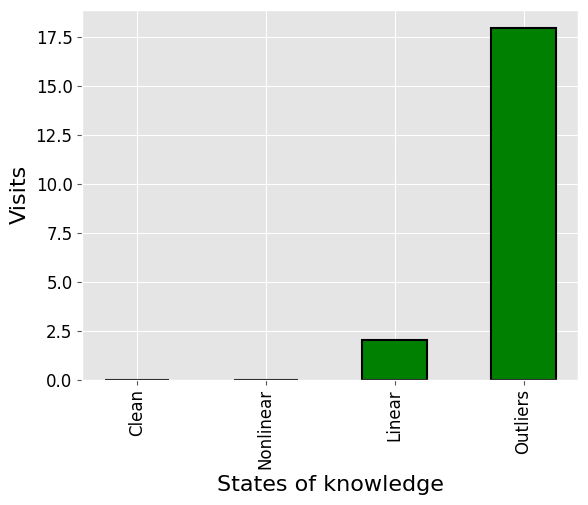}
         \caption{Final distribution.}
         \label{fig:histo_two_2}
     \end{subfigure}
\caption{Distribution of students' knowledge just at the beginning and at the end of all cycles of instruction in the PI methodology.}
\label{fig:histo_two}
\end{figure}

\begin{table}[ht]
    \centering
    \begin{tabular}{|c|c|c|c|c|}
        \hline
        $\mu_{MN}$ & $p_o$ & $p_m$ & $n_e$ & $n_c$ \\
        \hline
        $\mu_{3\times 3}$ & $0.1$ & $0.8$ & $20$ & $60$ \\
        \hline
    \end{tabular}
    \caption{Parameters for the evolution of a PMC with three conceptual attributes and three levels}
    \label{tab:evolution_P}
\end{table}
Fig.(\ref{fig:histo_two}) displays histograms representing the visits of the knowledge states at the beginning and the end of the aforementioned VEX. Fig.(\ref{fig:histo_two_1}) illustrates the initial distribution with the majority of students in a clean state, while Fig.(\ref{fig:histo_two_2}) portrays a distribution where all students are in the Outlier and  Linear states. It is evident that not all students did achieve high levels of proficiency. Nevertheless, all students have progressed to a state of knowledge involving substitution and explanation, represented by the states of greater discernment.


\begin{figure}[ht]
     \centering
     \begin{subfigure}[b]{0.49\textwidth}
         \centering
         \includegraphics[width=\textwidth]{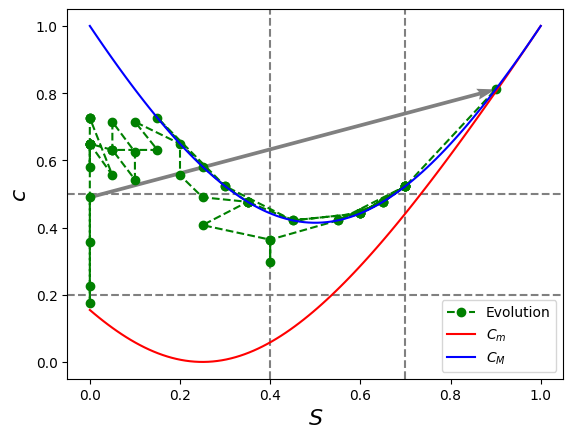}
         \caption{Evolution of concentration vs score.}
         \label{fig:CvsS}
     \end{subfigure}
     \begin{subfigure}[b]{0.49\textwidth}
         \centering
         \includegraphics[width=\textwidth]{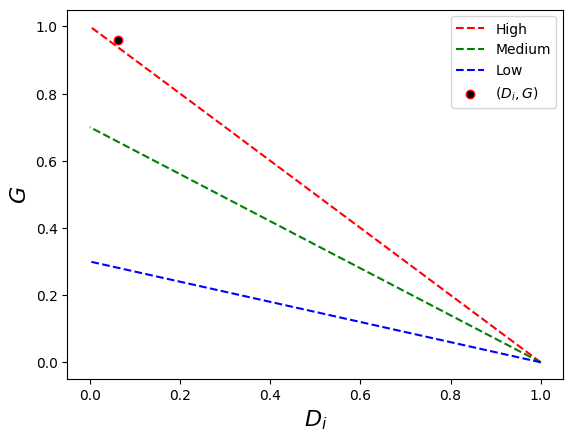}
         \caption{Gain of the discern.}
         \label{fig:gananciaD}
     \end{subfigure}
\caption{Concentration  analysis and gain.}
\label{fig:graphs_C_s_D}
\end{figure}

Fig.(\ref{fig:CvsS}) shows the evolution of the relationship between concentration and score as the number of PI cycles increases. The starting and ending points are connected by a light gray arrow and are surrounded by a series of dots that represent the $C$ vs $S$ relationship in each cycle. Note how, in several consecutive instances, the concentration reaches its maximum for a given score, and there is a setback in the score. This implies that, in this scenario, overall, students with lesser knowledge of the PMC persuade those with more knowledge, and the system undergoes a process of ``unlearning'' \footnote{The word is used to indicate that some students undergo a change in neural structure that implies a decrease in knowledge about the PMC.}.

The acquired gain in discernment throughout the entire learning process can be observed in Fig.(\ref{fig:gananciaD}). The gain is high and above the $G=1-D_i$ line, indicating a very high performance. However, it is important to highlight that in this work, the gain does not pertain to the score but rather to the discernment, which serves as a more appropriate indicator.

In general terms, the graphs in Fig.(\ref{fig:graphs_C_s_D}) depict the implementation of a system with the parameters illustrated in Tab.(\ref{tab:evolution_P}), which could plausibly occur in alternative scenarios. Obtaining an overview of the system's behavior will be accomplished by materializing multiple systems and averaging their results. This topic will be addressed in the upcoming virtual experiments.

\section{VEX 3: Monte Carlo}

The Monte Carlo method has immense importance in the computational simulation of learning models in the PI methodology. This powerful technique provides a statistical approach to estimate the general behavior of the system under study through random sampling. In the context of this work, the Monte Carlo method enables the simulation of the scoring and discernment behavior of the system under specific parameters; without this, it would be very difficult to conduct an analysis. By generating random samples within specified constraints, it becomes possible to approximate the behavior and results of the learning processes of a PMC. This application of the Monte Carlo method allows for a more comprehensive understanding of the system's performance and facilitates the evaluation of different scenarios and parameter settings.

Below, we illustrate the Monte Carlo method with two virtual experiments that enable us to statistically understand the behavior of discernment and the score in virtual student systems learning a PMC under the PI methodology.

\subsection{First virtual Monte Carlo experiment}

\begin{figure}[ht]
     \centering
     \begin{subfigure}[b]{0.49\textwidth}
         \centering
         \includegraphics[width=\textwidth]{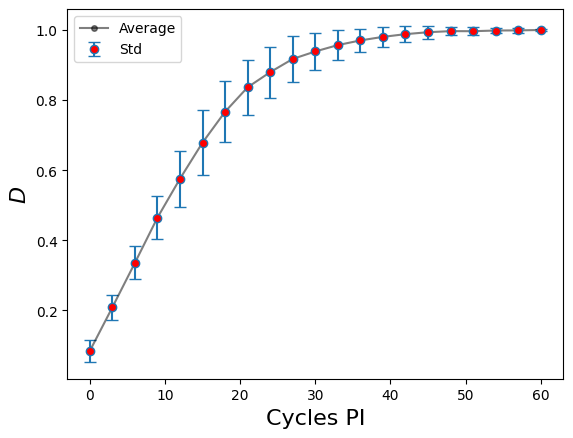}
         \caption{Statistical behavior of the evolution of discernment in the first virtual experiment.}
         \label{fig:Statistical_Dicer}
     \end{subfigure}
     \begin{subfigure}[b]{0.49\textwidth}
         \centering
         \includegraphics[width=\textwidth]{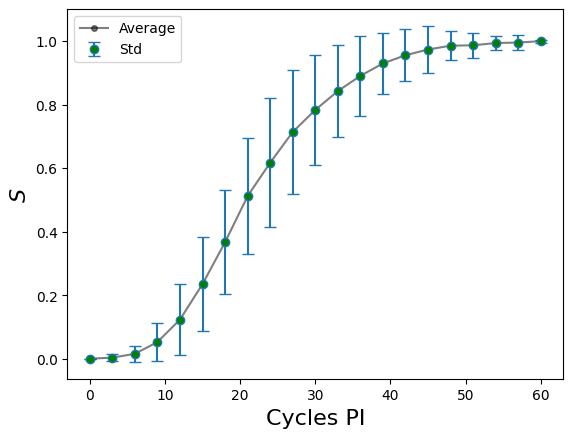}
         \caption{Statistical behavior of the evolution of the learning score in the first experimental experiment.}
         \label{fig:Statistical_score}
     \end{subfigure}
\caption{Score and discernment in virtual experiment $1$.}
\label{fig:MonteUno}
\end{figure}

This virtual experiment consists of a system composed of the parameters described in Tab.(\ref{tab:Monte1}). $100$ samples are taken, and the mean and standard deviation are calculated for both discernment and score. The results are shown in the graphs of Fig.(\ref{fig:MonteUno}).

\begin{table}[ht]
    \centering
    \begin{tabular}{|c|c|c|c|c|}
        \hline
        $\mu_{MN}$ & $p_o$ & $p_m$ & $n_e$ & $n_c$ \\
        \hline
        $\mu_{2\times 3}$ & $0.1$ & $0.6$ & $24$ & $60$ \\
        \hline
    \end{tabular}
    \caption{Virtual Monte Carlo experiment $1$ with $100$ samples.}
    \label{tab:Monte1}
\end{table}

In Fig.(\ref{fig:Statistical_score}), the asymptotic growth of discernment is observed with a standard deviation that decreases towards the last PI cycles. Discernment behaves with a linear trend in the first cycles, but then it becomes asymptotic with a higher standard deviation when most of the students come out of the PMC confusion and their discernment approaches unity.

Also, in Fig.(\ref{fig:Statistical_Dicer}) shows that the grading starts and ends with a near-zero standard deviation, which means that the vast majority of virtual learners have learned the PMC. Between the first and the last cycle, the system is quite heterogeneous, as indicated by the large value of the standard deviation in the middle zone of the cycles.

Comparing the evolution of the score and the discernment, we see that the system is much more homogeneous in the discernment than in the score. This is to be expected since the score is very demanding and binary, indicating correct or incorrect answers. In contrast, the discernment captures more nuances, accounting for the cognitive movement of the student through the conceptual attributes and their levels in the PMC.

\subsection{Second virtual Monte Carlo experiment}

\begin{figure}[ht]
     \centering
     \begin{subfigure}[b]{0.49\textwidth}
         \centering
         \includegraphics[width=\textwidth]{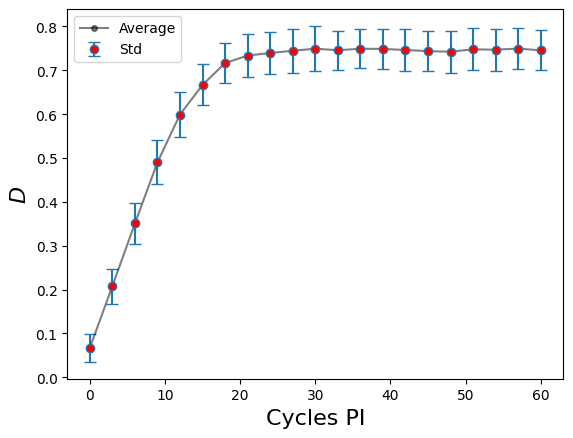}
         \caption{Statistical behavior of the evolution of discernment in the second virtual experiment.}
         \label{fig:y equals x}
     \end{subfigure}
     \begin{subfigure}[b]{0.49\textwidth}
         \centering
         \includegraphics[width=\textwidth]{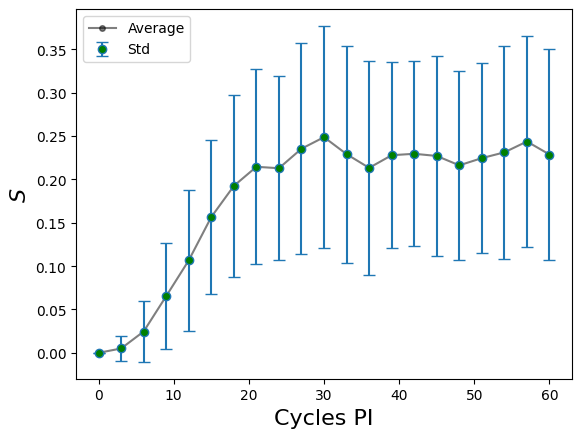}
         \caption{Statistical behavior of the evolution of the learning score in the first experimental experiment.}
         \label{fig:three sin x}
     \end{subfigure}
\caption{Score and discernment in virtual experiment $2$.}
\label{fig:MonteDos}
\end{figure}

In the first virtual experiment, PMC $\mu_{2\times 3}$ was taken  and in the second experiment, $\mu_{3\times 2}$ was taken. The system is composed of the parameters described in Tab.(\ref{tab:MonteDos}), $100$ samples are taken, and the mean and standard deviation are calculated for both discernment and score. The results are shown in the graphs of Fig.(\ref{fig:MonteDos}). 

\begin{table}[ht]
    \centering
    \begin{tabular}{|c|c|c|c|c|}
        \hline
        $\mu_{MN}$ & $p_o$ & $p_m$ & $n_e$ & $n_c$ \\
        \hline
        $\mu_{3\times 2}$ & $0.1$ & $0.6$ & $24$ & $60$ \\
        \hline
    \end{tabular}
    \caption{Parameters of the second virtual Monte Carlo experiment.}
    \label{tab:MonteDos}
\end{table}
In Fig.(\ref{fig:MonteDos}), it can be observed that the system exhibits greater heterogeneity in discernment compared to the first virtual experiment. Similarly, the score demonstrates a higher standard deviation, indicating a more pronounced heterogeneity than that observed in the initial experiment. Since all the parameters and the number of samples remain the same in both virtual experiments, the significant difference is attributed to the variation in the number of attributes and their depth. In other words, it is concluded that the learning of a PMC is more efficient when a PMC is designed with a greater number of conceptual attributes and shallow depth, rather than having a few conceptual attributes with substantial depth.

Based on the results presented in these virtual experiments, it is inferred that discernment is the most relevant indicator, or metric, to consider when quantifying the a real system  learning evolution of students exposed to the learning of a PMC using the PI methodology. 

In the upcoming chapter, this indicator will be primarily utilized to compare the model's results with the experimental findings from learning research.

\chapter{Experimental confrontation of the model}

In this chapter, the results of two experimental studies are presented and compared with the computational model that was developed and presented previously. The first study corresponds to the teaching of a fully theoretical PMC on the graphic interpretation of the URM, while the second study is both theoretical and experimental, focusing on a PMC that pertains to the fall of bodies in the air. We will be comparing two experimental studies with the results predicted by the model.

To quantitatively verify, a theory of partial truth by correspondence will be utilized \autocite{Romero2021scientific}. Based on the results of the virtual experiments conducted in the previous section, discernment emerges as the most relevant metric in the model. Therefore, it will serve as the variable to compare with the actual experience.

Experimentally, there exist $n_c$ PI cycles, and within each of these cycles, an average discernment $D$ and its corresponding  error $\varepsilon$ are obtained. The simulation is adjusted to set the experimental conditions and is executed to obtain a mean discernment $D_s$, while the accompanying error is considered irrelevant, it is expected that $\left | D_s-D \right | < \varepsilon $. More precisely, the $i-th$ cycle is verified with a truth value of $1-\varepsilon_r\: _i $, where $\varepsilon_r=\left | \frac{D-D_s}{D_s} \right |$ is the relative error. For all cycles the truth value $V$ of the verification is given by

\begin{equation} \label{eq:true_value}
    V=\frac{1}{n_c} \sum_{i=1}^{n_c} \left (1- \varepsilon_r\: _i\right ),
\end{equation}
where $V$ ranges between $0$ and $1$. If the check is true, the value is 1, and if it is zero, the value is false. Any value in between will provide a partial truth value. That is to say,  the closer to $1$, the truer, and the closer to $0$, the falser.

\section{Graphic interpretation in URM}

The general aspects of investigating the graphic interpretation of rectilinear motion were presented in chapter $1$; however, there is a scarcity of more precise research on the graphic interpretation of uniform rectilinear motion (URM). Three recent works, in particular, have been highlighted
\autocite{delgadorodriguez,talero2015paradoja,talero2013experimentos}.

The work carried out by \textcite{delgadorodriguez} to focused on designing and evaluating an educational scenario that uses Geogebra dynamic geometry software as a strategy to address challenges in interpreting graphs that represent the relationship between position and time in classical kinematics. The objective was to establish connections between objects and phenomena in order to explain and predict these phenomena effectively. The designed scenario effectively fostered these connections by incorporating a Geogebra software-based simulation, thereby creating an intermediary realm. The results of the research indicate the significance of incorporating this software in physics classes, particularly in the study of uniform rectilinear motion and its graphical representation of $x(t)$. 

\textcite{talero2015paradoja}'s research focuses on demonstrating how the graphical interpretation of uniform rectilinear motion and geometric series can elucidate the well-known paradox of Achilles and the tortoise. Furthermore, the study shows how this finding can be used as a teaching tool in elementary mechanics courses to impart knowledge about one-dimensional kinematics.

The work presented by \textcite{talero2013experimentos} involved extending the traditional problem that addresses the distance traveled by a fly at a constant speed while moving between two trains in uniform rectilinear motion with a head-on collision. The study explored both the trivial solution and the solution attributed to Neumann, utilizing series and simulations. It demonstrates how this problem can serve as valuable teaching material in elementary physics lessons, illustrating the relationship between physics and mathematics and its connection to the graphical interpretation of uniform rectilinear motion.

The first work showcased an investigation that employed simulation as a tool, while the second work presented an exercise that can be generalized and simulated to calculate the terms of a series. Lastly, the third work demonstrated virtual experiments focused on addressing a disciplinary problem related to uniform rectilinear motion (URM). These works differ from the present study as this work aims to simulate the learning process of students rather than simulating specific experiments or calculations.

\subsection{Population, instrument and PMC}

The study was conducted with a population of students aged between $18$ and $22$ years, who were enrolled in the Faculty of Engineering at $X$ University  in Bogotá, Colombia. The students were solely focused on their studies and did not hold any employment.

The instrument used for investigating the learning of graphic interpretation was a questionnaire comprising $9$ questions and $6$ versions, which can be found in \href{https://drive.google.com/file/d/1bWpdHpD0lahGRvwY7CuDTRfljfoDhYn0/view?usp=drive_link}{Annex 9} or App.(\ref{app:6test}) We employed a modified version of the PI methodology, which involves utilizing a comprehensive questionnaire instead of a single question, while keeping all other aspects unchanged. The modified approach we employed in the study involved the use of a complete questionnaire, differing from the standard practice of utilizing only one question. The theoretical reference provided to the students was chapters $3$ and $4$ of the manual previously developed and shown in \href{https://drive.google.com/file/d/14F8bETp3JuQJTzDXQTVic-K-upTBRkM1/view?usp=drive_link}{Annex 6}.

The PMC consists of three conceptual attributes and three levels of depth, what is shown in the Tabs (\ref{Tab:definePMC3x3_MUR}) and (\ref{Tab:def_PMC_MUR}). It is necessary to highlight that questions $1$, $2$, and $3$ of the instrument pertaining to level $q_0$, while questions $4$, $5$, and $6$ correspond to level $q_1$ and $7$, $8$, and $9$ are associated with level $q_2$.

\begin{table}[ht]
  \centering
  \begin{minipage}{0.3\textwidth}
    \centering
        \begin{tabular}{|c||c|c|c|}
         \hline  
         \textsf{} & $F_0$  & $F_1$  & $F_2$  \\ \hline  \hline
         
         $q_0$      &  &  &  \\ \hline  
         $q_1$      &  &  &  \\ \hline  
         $q_2$      &  &  &   \\ \hline 
        \end{tabular}
        \caption{$\mu_{3\times 3}$ of the UMR}
        \label{Tab:definePMC3x3_MUR} 
  \end{minipage}%
  \hfill
  \begin{minipage}{0.7\textwidth}
    \centering
    \begin{tabular}{|c||c|}
     \hline  
               & Definitions    \\ \hline  \hline
     $F_0$     & Physically read the axes and the area under the curve.   \\ \hline  
     $F_1$     & Physically identify the slope.  \\ \hline  
     $F_2$     & Infers a graph from text or another graph.  \\ \hline \hline
     $q_0$     & Answer one question out of three possible.  \\ \hline  
     $q_1$     & Answer two question out of three possible.  \\ \hline  
     $q_2$     & Answer three question out of three possible.  \\ \hline
    \end{tabular}
    \caption{Definition of the PMC: graphical interpretation of the UMR.}
    \label{Tab:def_PMC_MUR} 
  \end{minipage}
\end{table}
The experiment was conducted in eleven lessons, with each lesson implementing an PI main cycle, followed by the continuation of the planned theme in the normal course. The number of students varied during each class, ranging from a minimum of $16$ to a maximum of $20$, and they were not the same students; at times, one would be missing, while at other times, others would be missing. For this reason it was not possible to analyze the concentration diagrams, and we focused on the observation of the distribution of students in the states of knowledge and on the comparison of theoretical and experimental discernment. The raw experimental data can be seen in \href{https://docs.google.com/spreadsheets/d/1shGNhmcMZrlg7gHhM22AUvDuQdR4A-Hekpgxv4UVPic/edit?usp=drive_link}{Annex 10} or in App.(\ref{app:data_PI}).\\

\begin{figure}[ht]
    \centering
    \includegraphics[width=0.6\textwidth]{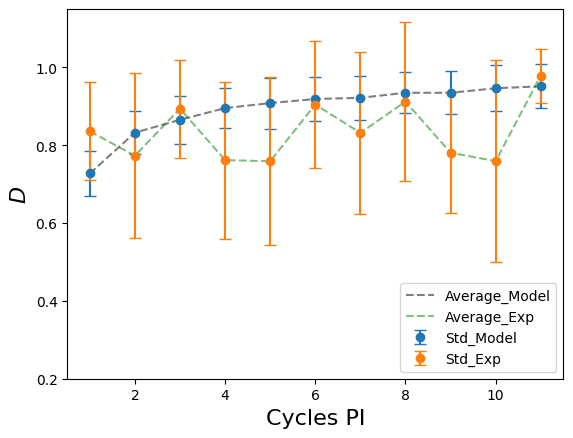}
    \caption{Comparison of experimental discernment $D$ with theoretical discernment $D_s$ extracted from the simulation.}
    \label{fig:MUR_exp_vs_theo}
\end{figure}

Fig.(\ref{fig:MUR_exp_vs_theo}) displays the experimental curve depicting the behavior of discernment $S$ as a function of the number of global PI cycles. The experimental error bars are related to the standard deviation of the discernment and account for the heterogeneity of the group, while the simulation error bars account for the heterogeneity generated by the model. Clearly, the heterogeneity is higher than what was predicted by the model.
The virtual experiment that fits closest to the experimental result has the parameters shown in Tab(\ref{tab:MUR_UC}).
\begin{table}[ht]
    \centering
    \begin{tabular}{|c|c|c|c|c|c|}
        \hline
        $\mu_{MN}$ & $p_o$ & $p_m$ & $n_e$ & $n_c$ & $n_{ccc}$\\
        \hline
        $\mu_{3\times 3}$ & $0.6$ & $0.9$ & $20$ & $11$ & $10$\\
        \hline
    \end{tabular}
    \caption{The parameters of the VEX that seek to adjust to the experimental results shown in Fig.(\ref{fig:occupation_Exp_MUR}).}
    \label{tab:MUR_UC}
\end{table}
According to the model, $n_{ccc}$ is the number of cognitive confrontations per PI cycle, which is estimated from the characteristics of the group in general, and obeys the foundations developed in the formulation of the model, here it was estimated at $10$.

On the other hand, we use the Eq.(\ref{eq:true_value})  to estimate the degree of trueness of the fit and find $V=0.9$.

\begin{figure}[ht]
     \centering
     \begin{subfigure}[b]{0.53\textwidth}
         \centering
         \includegraphics[width=\textwidth]{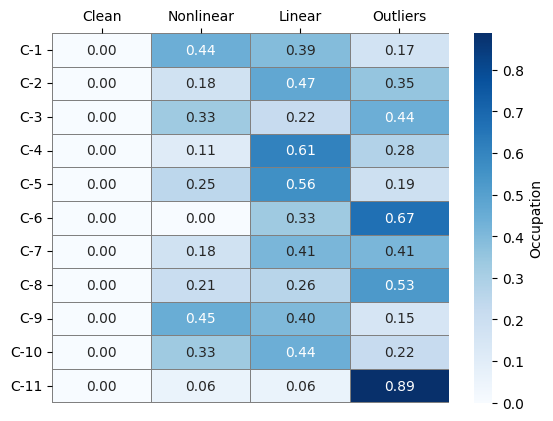}
         \caption{Occupation of the states}
         \label{fig:ocupation}
     \end{subfigure}
     \begin{subfigure}[b]{0.45\textwidth}
         \centering
         \includegraphics[width=\textwidth]{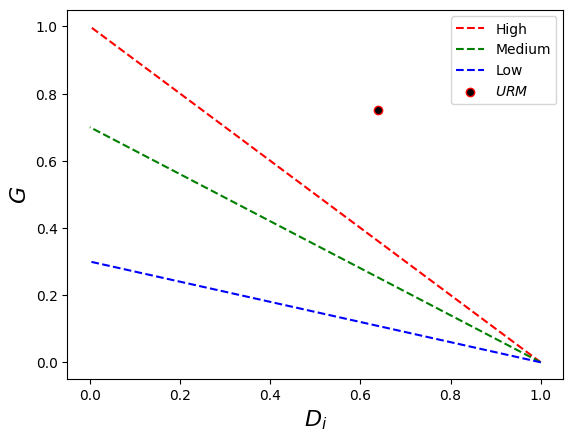}
         \caption{Discernment gain}
         \label{fig:gain_D}
     \end{subfigure} 
     \caption{Occupied states and gain.}
     \label{fig:occupation_Exp_MUR}
\end{figure}


The occupation of states Clean, Nonlinear, Linear and Outliers throughout the entire learning process of graphical interpretation in the UMR is depicted in Fig.(\ref{fig:ocupation}), while in Fig.(\ref{fig:gain_D}) the gain of discernment is shown. In the first graph  illustrates the occupation of the knowledge states during the cycles $c-1$, $c-2$, etc., conducted in each lecture. This shows the following: state Clean is not empty from the beginning, indicating that students commence with some prior knowledge; state Outliers experiences fluctuations with consistent population growth and reaches a maximum precisely in the last PI cycle or class section, this indicates that students have doubts about their answers during the process, suggesting that their knowledge is not yet fully consolidated; lastly, the occupation of states Nonlinear and Linear indicates that, during the process, students tended to be slightly more coherent than incoherent. Therefore, the students do not yet have consolidated knowledge, as there is no predominant tendency in the occupation of states Linear and Outliers. Further PI cycles may be necessary to achieve this consolidation. Finally, the second graph illustrates the gain of discernment throughout the experimental process. 

\section{Fall of bodies in the air}

The literature on the research about the learning of falling bodies is abundant, being approached from different perspectives, including historical analysis \autocite{alvarez2012fenomeno}, simulations \autocite{talero2013experimentos}, and experimental methodologies \autocite{cruz2015movimiento,diaz1989comprension}, among others. 

The field work developed and presented in this section adopts a different perspective, applying the PI methodology within the conceptual framework of the previously developed model \autocite{caduta}. In this section, the relevant results will be presented to validate the developed model.

\subsection{Student population}

The experiment of learning took place during the second semester of $2015$ in two universities located in Bogota, Colombia. Henceforth, these universities will be referred to as $X$ and $Y$. The students at University $X$ were studying engineering, while those at University $Y$ were undergoing training as chemistry teachers.

At University $X$, the experiment was conducted within two groups, $g_A$ y $gB$, of a physics course held in the evening. Each group consisted of $13$ men and $3$ women, aged $18$ to $40$. Most of the students held daytime jobs. On the other hand, at University Y, the experiment was carried out with a single group $g_C$ of a daytime physics course comprising $8$ men and $8$ women, aged between $16$ and $18$. None of the students had employment. In both cases, the classes had approximately $20$ students, but only $16$ were selected based on attendance criteria and their interest in learning; 

\subsection{Definition of the PMC}

In the literature related to the investigation of the teaching-learning process of falling bodies, two main conceptual errors have been identified. On the one hand, some students hold the mistaken idea that when two bodies are dropped simultaneously from the same height, the heavier body will hit the ground first. On the other hand, there are students who believe that all bodies always fall at the same time. In this work, we will refer to the first answer as ``Aristotelian thought'' and the second answer as ``Galilean thought''. In general, both answers are not in agreement with the experimental results \footnote{Of course, Aristotelian and Galilean thought have multiple nuances, but in this work, we will only refer to the two aspects mentioned above.}. However, research has shown that some students adhere to Newtonian mechanics, which aligns with experimental observations. We will refer to these students as those who possess Newtonian thought.

To define the PMC, three conceptual attributes have been established: Aristotelian thoughts represented by $F_A$, Galilean thoughts represented by $F_G$, and Newtonian thoughts represented by $F_N$. Each conceptual attribute is associated with four levels of discernment, ranging from the smallest to the largest, represented by $q_0$, $q_1$, and $q_2$. See Tab.(\ref{Tab:tS}).

\begin{table}[htp]
\centering
\begin{tabular}{|c||c|c|c|}
 \hline  
 \textsf{} & $F_A$  & $F_G$  & $F_N$  \\ \hline  \hline
 
 $q_0$      &  &  &  \\ \hline  
 $q_1$      &  &  &  \\ \hline  
 $q_2$      &  &  &   \\ \hline 
 $q_3$      &  &  &   \\ \hline 
\end{tabular}
\caption{Definition of the PMC $\mu_{4\times3}$ of the falling bodies.}
\label{Tab:tS} 
\end{table}

\subsection{Definition of an instrument to measure the learning of the PMC about falling bodies}

The measurement instrument is defined and tested according to the criteria established in section $4$. It consists of $5$ home experiments named $\alpha$, $\beta$, $\gamma$, $\delta$, and $\epsilon$; each with three true (T) or false (F) questions ($\alpha_A$, $\alpha_G$ and $\alpha_N$) that gradually increase in difficulty that evaluate each thought conceptual attribute.  The $\mu$ state is set to three digits according to the convention shown in Tab.(\ref{Tab:Combention}) and an example in Tab.(\ref{Tab:tS_example}). The Instrument can be seen in the \href{https://drive.google.com/file/d/15LzAjUd4-vo1Bdq85OLS-L0mIB0f8_3M/view?usp=drive_link}{Annex 11} or in App.(\ref{app:indagCaida}). Likewise, an instrument is defined as a theoretical reference on the fall of bodies that can be seen in \href{https://drive.google.com/file/d/15LzAjUd4-vo1Bdq85OLS-L0mIB0f8_3M/view?usp=drive_link}{Annex 12} or in App.(\ref{app:rtd_caida}). This instrument contemplates a numerical solution for the fall of a sphere in the air. Although analytical solutions may exist, this approach is chosen for its relevance and coherence with the characteristics of the population. It is important to note that falling bodies are not always spheres; however, this presentation serves as a starting point for discussion. On the other hand, the data of $g_A$, $g_B$, and $g_C$ is shown in the
\href{https://drive.google.com/file/d/1A4_IxLHQ6vOuaK8WiHwcQzPgLVdL83ab/view?usp=drive_link}{Annex 13} and in App.(\ref{app:rEx_caida}).

\begin{table}[htbp]
  \centering
  \begin{minipage}{0.45\textwidth}
    \centering
        \begin{tabular}{|c||c|}
         \hline  
          Digit  & Convention   \\ \hline \hline
         $0$ & $q_0$ is busy       \\ \hline  
         $1$ & $q_1$ is busy  \\ \hline  
         $2$ & $q_2$ is busy  \\ \hline 
         $3$ & $q_3$ is busy  \\ \hline  
        \end{tabular}
        \caption{Level assignment.}
        \label{Tab:Combention} 
  \end{minipage}%
  \hfill
  \begin{minipage}{0.45\textwidth}
    \centering
    \begin{tabular}{|c||c|c|c|}
     \hline  
     $320$ & $F_A$  & $F_G$  & $F_N$  \\ \hline  \hline
     $q_0$     & $0$ & $0$ & $1$ \\ \hline  
     $q_1$     & $0$ &$0$  & $0$ \\ \hline  
     $q_2$     & $0$ & $1$ &  $0$ \\ \hline
     $q_3$     & $1$ & $0$ &  $0$ \\ \hline
    \end{tabular}
    \caption{Example, $320$.}
    \label{Tab:tS_example} 
  \end{minipage}
\end{table}
The experimental parameters of Group $A$ are shown in Tab.(\ref{tab:gA}), where the active methodology $p_m$ is characterized by the two criteria defined in the subsec.(\ref{sub:LearnigEvolution}). In addition, the model uses $50$ Monte Carlo samples to simulate the system.

\begin{figure}[ht]
     \centering
     \begin{subfigure}[b]{0.48\textwidth}
         \centering
         \includegraphics[width=\textwidth]{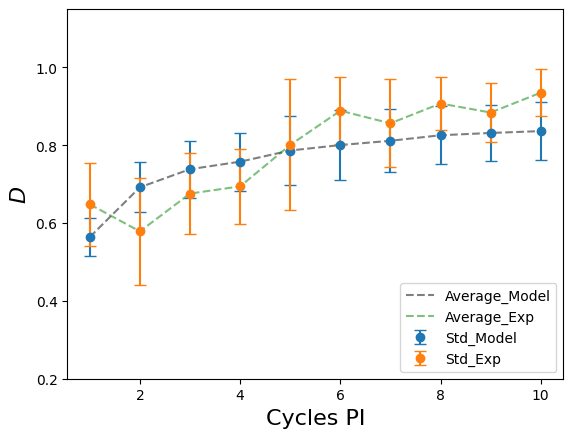}
         \caption{Group $g_A$}
         \label{fig:gA}
     \end{subfigure}
     \begin{subfigure}[b]{0.48\textwidth}
         \centering
         \includegraphics[width=\textwidth]{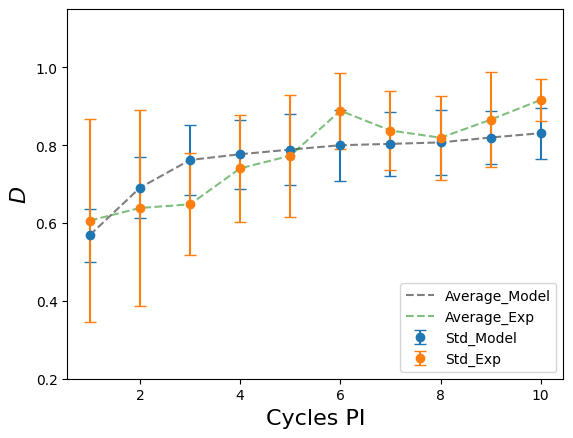}
         \caption{Group $g_B$}
         \label{fig:gB}
     \end{subfigure}
     \begin{subfigure}[b]{0.5\textwidth}
         \centering
         \includegraphics[width=\textwidth]{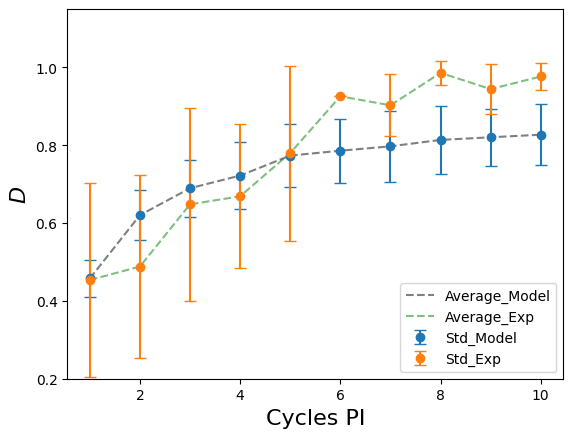}
         \caption{Group $g_C$}
         \label{fig:gC}
     \end{subfigure}
\caption{Experimental confrontation of $D$ with the model.}
\label{fig:graphs_C_vs_D}
\end{figure}

\begin{table}[ht]
    \centering
    \begin{tabular}{|c|c|c|c|c|c|}
        \hline
        $\mu_{MN}$ & $p_o$ & $p_m$ & $n_e$ & $n_c$ & $n_{ccc}$\\
        \hline
        $\mu_{3\times 4}$ & $0.63$ & $0.95$ & $16$ & $10$ & $8$ \\
        \hline
    \end{tabular}
    \caption{Parameters for group $A$.}
    \label{tab:gA}
\end{table}
In Fig.(\ref{fig:gA}), we are showing the behavior of $D$ associated with group $A$, both experimental and theoretical. A greater heterogeneity is observed in the experimental group than in the theoretical group, as well as an agreement in the first $5$ cycles and a distancing in the subsequent cycles. Applying the veracity criterion of the validation of the model defined by Eq.(\ref{eq:true_value}), $vg_A= 0.91$ is found.

\begin{table}[htbp]
  \centering
  \begin{minipage}{0.48\textwidth}
      \centering
    \begin{tabular}{|c|c|c|c|c|c|}
        \hline
        $\mu_{MN}$ & $p_o$ & $p_m$ & $n_e$ & $n_c$ & $n_{ccc}$ \\
        \hline
        $\mu_{3\times 4}$ & $0.6$ & $0.95$ & $16$ & $10$ & $8$\\
        \hline
    \end{tabular}
    \caption{Parameters for group $B$.}
    \label{tab:gB}
  \end{minipage}%
  \hfill
  \begin{minipage}{0.48\textwidth}
    \    \centering
    \begin{tabular}{|c|c|c|c|c|c|}
        \hline
        $\mu_{MN}$ & $p_o$ & $p_m$ & $n_e$ & $n_c$ & $n_{ccc}$\\
        \hline
        $\mu_{3\times 4}$ & $0.45$ & $0.95$ & $16$ & $10$ & $8$\\
        \hline
    \end{tabular}
    \caption{Parameters for group $C$.}
    \label{tab:gC}
  \end{minipage}
\end{table}

The experimental parameters of Group $B$ are shown in Tab.(\ref{tab:gB}), the model also uses $50$ Monte Carlo samples to simulate the system. In Fig.(\ref{fig:gB}), we are showing the behavior of $D$ associated with group $B$, both experimental and theoretical. Experimentally, there is great confusion at the beginning of the process, but the group becomes more homogeneous as they discern. Likewise, the results of the model and the experiment overlap their error bars in $8$ out of $10$ cycles and we found $Vg_B=0.93$.

The experimental parameters of Group $C$ are shown in Tab.(\ref{tab:gC}). The instructional quality parameter $p_m$ was estimated (as in groups $gA$ and $gB$) by the author of this research and the instructor, furthermore we find $vg_C=0.88$.\\

It is important to highlight that, according to the virtual experiments carried out and shown in the previous chapter, due to the nature of the model, the most important and susceptible metric to be confronted experimentally isthe discernment $D$ and not the is the score $S$. This is because majority occupation of state Outliers requires many cycles PI to achieve consolidation, which is not the case in classroom experience.

It is also necessary to remember that the score has been reinterpreted in this work. Here, the correct answer of the traditional concentration analysis has been replaced by the number of students that occupy the Outliers state, and the incorrect answers correspond to the other knowledge states Linear, Nonlinear, and Clean. In this sense, it is important to know the evolution of the score and concentration in order to observe the global coherence of the model and its agreement with experience.

\begin{figure}[ht]
     \centering
     \begin{subfigure}[b]{0.49\textwidth}
         \centering
         \includegraphics[width=\textwidth]{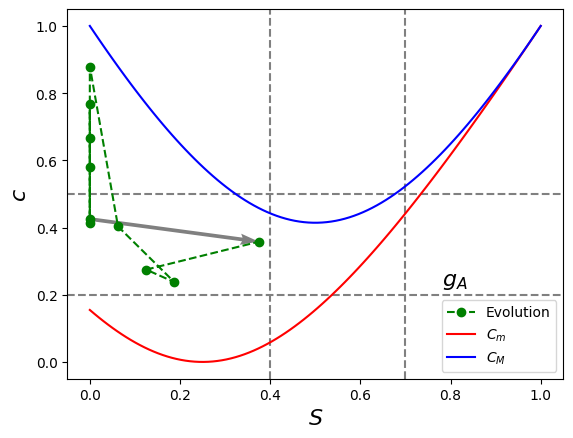}
         \caption{Concentration analysis $g_A$.}
         \label{fig:conc_gA}
     \end{subfigure}
     \hfill
     \begin{subfigure}[b]{0.49\textwidth}
         \centering
         \includegraphics[width=\textwidth]{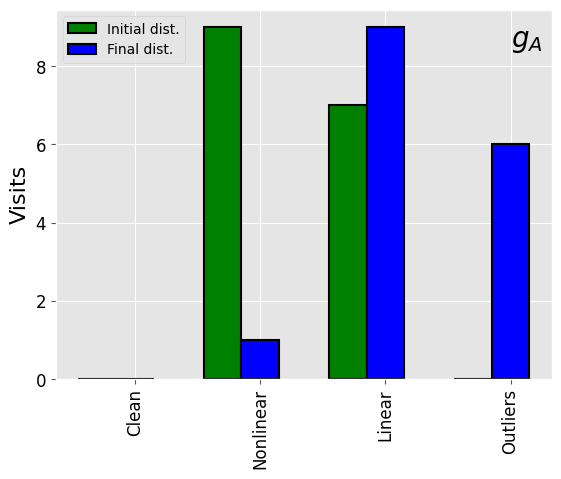}
         \caption{Occupation $g_A$}
         \label{fig:Occup_gA}
     \end{subfigure}
    \caption{Analysis of concentration and occupied states in the $g_A$.}
    \label{fig:conc_and_occu-gA}
\end{figure}

Below, we show the experimental results using Bao's concentration analysis and display the distribution at the beginning and at the end of the students in their states of knowledge, see Fig.(\ref{fig:conc_and_occu-gA}) for group $g_A$.\\

In the concentration analysis shown in Fig.(\ref{fig:conc_gA}), the green dots joined with dashed lines indicate the different values of score and concentration during the evolution process. The gray arrow goes from the starting point to the endpoint in the process. The order of the points is not important since, in most cases, it can be inferred from the graph and does not provide relevant information. The crucial aspect lies in the distribution of the points: if the line that joins the points is discontinuous, only one path is expected, otherwise there could be more. In this case, group $g_A$, the concentration analysis shows that the knowledge was not correct until the last lessons, with only a small part of the population achieving it.

On the other hand, the graph in Fig.(\ref{fig:Occup_gA}) shows the distribution of students in their states of knowledge at the beginning and at the end of the process. It is observed, in accordance with the concentration analysis, that only a small part migrates to the Outliers state. Additionally, at the beginning, the students already have an idea of the fall of the bodies and they already occupy states of nonlinear and linear knowledge; however, the system starts and ends in a low Bi-Modal region  \footnote{To be precise, it should be called Bi-state, but in order not to overload the vocabulary, we will continue to use the terms suggested by \textcite{bao2001concentration} }. It is also observed that there is a migration from the nonlinear to linear state during the process, indicating a more coherent knowledge.

\begin{figure}[ht]
     \centering
     \begin{subfigure}[b]{0.49\textwidth}
         \centering
         \includegraphics[width=\textwidth]{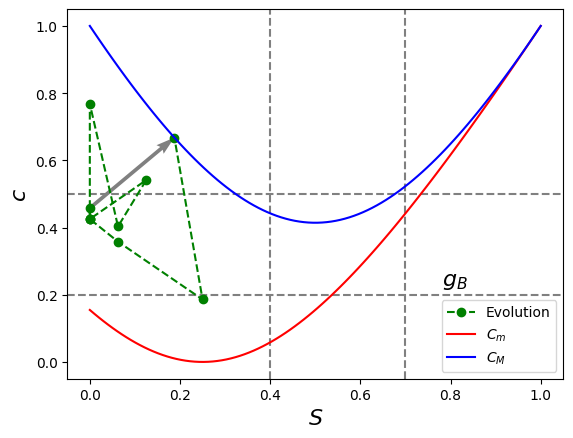}
         \caption{Concentration analysis $g_B$.}
         \label{fig:conc_gB}
     \end{subfigure}
     \hfill
     \begin{subfigure}[b]{0.49\textwidth}
         \centering
         \includegraphics[width=\textwidth]{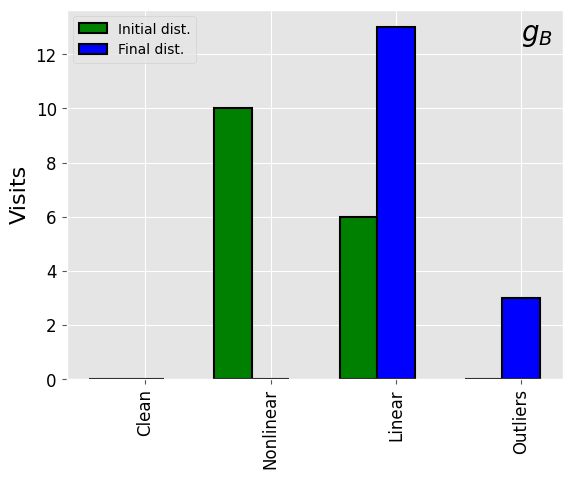}
         \caption{Occupation $g_B$}
         \label{fig:Occup_gB}
     \end{subfigure}
    \caption{Analysis of concentration and occupied states in the $g_B$.}
    \label{fig:conc_and_occu-gB}
\end{figure}
The results of group $B$ are observed in the graph of Fig.(\ref{fig:conc_and_occu-gB}). The system starts in a Bi-Modal region and ends in a One-Modal region, which is seen directly in Fig.(\ref{fig:Occup_gB}). The initial distribution does not include any students in the Clean state, and there are already some who possess coherent knowledge of the PMC. As the process progresses, there is a shift to the right, resulting in the vacating of the Clean and Nonlinear states, while the students occupy the Linear and Outliers states.

\begin{figure}[ht]
     \centering
     \begin{subfigure}[b]{0.49\textwidth}
         \centering
         \includegraphics[width=\textwidth]{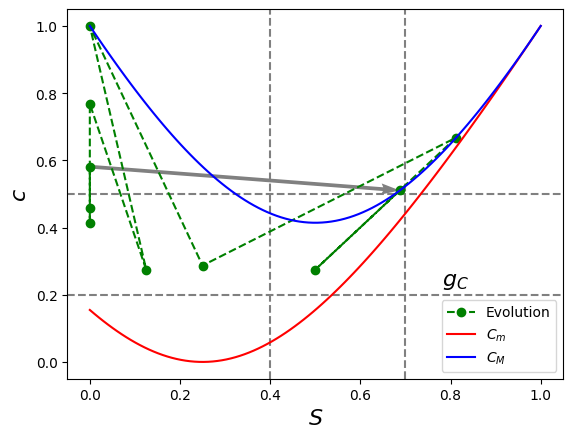}
         \caption{Concentration analysis $g_C$.}
         \label{fig:conc_gC}
     \end{subfigure}
     \hfill
     \begin{subfigure}[b]{0.49\textwidth}
         \centering
         \includegraphics[width=\textwidth]{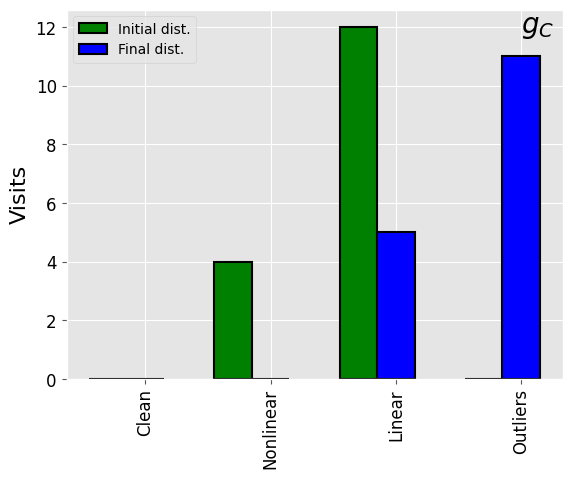}
         \caption{Occupation $g_C$}
         \label{fig:Occup_gC}
     \end{subfigure}
    \caption{Analysis of concentration and occupied states in the $g_C$.}
    \label{fig:conc_and_occu-gC}
\end{figure}

The cluster in Fig.(\ref{fig:conc_and_occu-gC}) begins in a low region of One-Model and evolves to a mid region of One-Model. What is interesting about the evolution of this group is the instability in the occupation of the Outliers state. Before the last section, the maximum occupation was reached, and in the last one, as expected. This reflects the lack of knowledge consolidation during the process, as demonstrated by the VeX developed in the previous chapter. Finally, similar to the two previous groups, there is a shift to the right in the occupation of knowledge states, see Fig.(\ref{fig:Occup_gC}).

\begin{figure}[ht]
    \centering
    \includegraphics[width=0.6\textwidth]{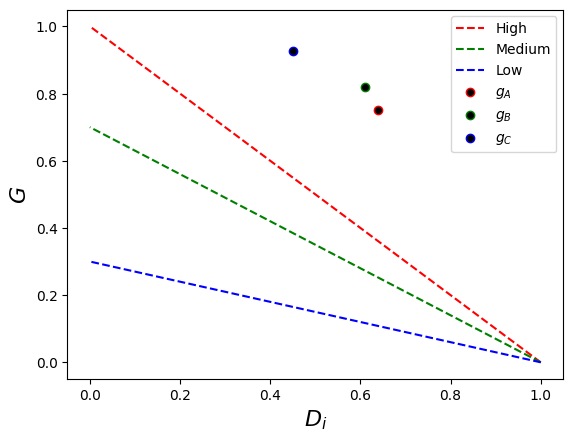}
    \caption{Experimental gain.}
    \label{fig:gain_cuer}
\end{figure}

Hake's gain of the three groups is shown in Fig.(\ref{fig:gain_cuer}). It is essential to highlight that the gain in this thesis has been redefined to reflect discernment rather than the score. This adjustment is more suitable for the nature of the model. The results are satisfactory and indicate a favorable evolution of knowledge that ensures internal coherence among the majority of students. However, it does not guarantee high consolidation or complete mastery of knowledge about the PMC.

Groups $A$ and $B$ experienced a more significant gain compared to Group $C$, primarily due to their higher initial level of insight. In this condition, students face cognitive challenges as they try to give meaning to their knowledge. On the other hand, Group C, which starts with a lower initial condition, continues to evolve within a state of incoherence. Despite making limited initial progress in discernment, a high gain does not necessarily move them out of the Nonlinear state.

When comparing the experimental results with the theoretical predictions from the model, we find a good agreement as long as the parameters are adjusted cautiously.

\chapter{Discussion and conclusions}

Previous works have documented the importance of developing learning models to improve the quality of physics teaching and learning \autocite{PhysRevSTPER.6.020105,PhysRevSTPER.4.010109,pritchard2008mathematical,redish2004theoretical}. In particular, \textcite{PhysRevSTPER.6.020105} proposed a mathematical phenomenological model of learning in PI based on a master equation, from which he derives consequences that agree with the class data obtained in class. \textcite{PhysRevSTPER.6.020105} base his model on the impossibility of individually tracking the learning of each student but focused on the group as a whole, consequently unable to access the cognitive details or psychological interactions during the PI interaction process. Also his model describes short-term learning in PI.

Nitta's  model does not take into account the characteristics of the population and its link with the content to be taught, which is of vital importance for the process: it is necessary to know the population to understand what and how to teach. It also does not differentiate in PI between the interactions among students and those between the teacher and student. Additionally, it does not provide any insights into the student's knowledge process in its different stages.

In this thesis, a computational model that represents the evolution of learning in PI was formulated, simulated, and validated. The model is based on some principles of cognitive neuroscience and some results of the PER. It relies on a PMC that strongly depends on the characteristics of the population and the transitions that students undergo through four states of knowledge of the PMC in various PI cycles.

The model proposes a learning mechanism in the interaction between students and the teacher, as well as the quality of the teacher's instruction and the discussions between students, allowing for a discussion on the consolidation of long-term learning. The model was validated with two classroom experiments and showed good agreement.

The formal structure of the model consists of a set of matrices that represent the state of knowledge of each student in the PMC. It evolves according to the learning rules and is materialized through the Monte Carlo method.

The premises and results of the model developed in this work highlight the importance of the instructor. Without the instructor's guidance, the students would wander irregularly between states without stopping at any state, despite a large number of PI cycles. In addition, the model highlights the permanent confusion of students regarding the knowledge of a PMC and the significance of the insight they achieve through the PI that goes beyond reviewing the traditional score or gain. In other words, what is essential is the change in the insight of a PMC. The model is formulated to account for the evolution of learning, allowing the observation of the process of forgetting and consolidating knowledge in the students, and answering the question: until when is it possible to have consolidated knowledge? The model also allows us to see how it becomes more challenging to transition between increasingly coherent states.

On the other hand, if the model were implemented in the classroom, it would allow the generation of alarms about the students' performance during the process. The transition through the different states of knowledge would enable the detection of students with some type of learning problem.

The model developed in this work has limitations and points for improvement. An important limitation is the lack of objective criteria to establish two parameters: the quality of the teacher's instruction in explaining the correct answer in a PI cycle and the quality of cognitive confrontation between students. Another limitation is the failure to unveil the sociological and psychological dynamics that occur during the cognitive confrontation between students.

This work generates some perspectives and suggestions for future work, and the four main ones are highlighted. First, investigating how to objectify the parameters described in the previous paragraph. Second, designing non-invasive methodologies to understand in detail the dynamics of cognitive confrontation, such as determining if debates occur, and if they do, under what terms they occur, to test the interaction hypothesis proposed in this model. Third, developing questionnaires directly related to the laboratory experience, which would reduce cultural bias and make the questionnaires more objective. Fourth, investigating methodologies to develop individual activities tailored to each student, allowing for proposing a cognitive evolution based on their particular knowledge.

In conclusion, based on field studies, a computational model was formulated, simulated, and validated to explain the evolution of learning of a PMC in the PI learning methodology applied to first-semester Engineering students. The model proposes a learning mechanism during a cognitive confrontation that elucidates how students clarify their understanding of the knowledge within the PMC.

It was found that the role of the instructor is crucial in the PI methodology, as learning is not possible without their guidance. The model incorporates learning mechanisms based on neuroscience. Additionally, the study revealed that consolidating the knowledge of a PMC requires multiple PI cycles.

Moreover, the research indicated that the size of the group influences the rate of learning, as it enhances the quality of cognitive confrontation. Furthermore, the quality of cognitive confrontation was also identified as a crucial factor.

Lastly, the study demonstrated that the complexity and depth of a PMC impact its learnability. Shallow and simple PMCs are easier to learn, whereas PMCs that grow in depth and complexity become more challenging to comprehend.


\printbibliography[heading=subbibliography]

\appendix
\clearpage 
\addappheadtotoc
\appendixpage
\chapter{Two-level heuristic model simulation} \label{app:Twolevel}

\section{Simulation implementation in python}

\begin{lstlisting}[language=Python,style=mystyle] 

import numpy as np
from scipy import stats
import matplotlib.pyplot as plt
import matplotlib.patches as patches
from seaborn._core.properties import LineWidth
#----FCI data
distrubutionFCI=[0.001272264631, 0.006361323155, 0.02926208651,
0.05089058524,0.08651399491, 0.1323155216, 0.1641221374, 
0.1374045802, 0.1183206107,0.08905852417, 0.06361323155, 0.03562340967,
0.02417302799, 0.02926208651,0.02162849873, 0.005089058524, 
0.002544529262, 0.001272264631, 0.001272264631,
                 0, 0, 0, 0, 0, 0, 0, 0, 0, 0, 0, 0]
#---------Tug-K
distrubutionTugK=[0.0,0.006993006993, 0.02797202797, 0.04895104895, 
0.07692307692, 0.1188811189,0.1328671329, 0.1258741259, 0.1048951049,
0.08391608392, 0.06293706294,0.04895104895, 0.02797202797, 0.04195804196,
0.02797202797, 0.02097902098,0.006993006993, 0.006993006993, 
0.006993006993, 0, 0.006993006993,0.006993006993, 0, 0.006993006993]
#----------- FCI china
distrubutionFCI_china=[0.000000,0.000000, 0.001642, 0.000000, 
0.006568, 0.000000, 0.001642,0.003284, 0.000000, 0.001642,
0.001642, 0.000000, 0.001642,0.001642, 0.000000, 0.006568, 
0.001642, 0.008210, 0.006568,0.006568, 0.019704, 0.029557,
0.022989, 0.049261, 0.065681, 0.091954, 0.113300, 0.160920,
0.174056, 0.154351, 0.068966]
#-------------
#print("Normalization test=",np.sum(distrubutionTugK))
print("Normalization test=",np.sum(distrubutionFCI_china))
#----
def Count_essays(A,n): #A is the number student, n number of questions
  acum=[0 for i in range(n+1)]
  for k in range(n+1):
    for j in A:
      if k==j:
        acum[k]=acum[k]+1
  return acum
#---Parameters for the FCI
#n,p,q=30,0.2,0.41 # n is the number of correct correct questions,p is pl, q is ph
#N1,N2,U=470,530,100  #N1 is Nl, N2 is Nh and U is the university virtual number
#----------------
#---Parameters for the Tug-K
#n,p,q=23,0.23,0.4 # n is the number of correct correct questions,p is pl, q is ph
#N1,N2,U=600,400,100  #N1 is Nl, N2 is Nh and U is the university virtual number
#----------------Parameters for the FCI-China
n,p,q=30,0.65,0.876 # n is the number of correct correct questions,p is pl, q is ph
N1,N2,U=150,850,100  #N1 is Nl, N2 is Nh and U is the university virtual number
#-----------------------
esasys=list(range(n+1))
Rus,Rus2=[],[]
for k in range(U):
  arr=np.random.binomial(n,p,N1).tolist() # Answer of student l in the iniversity U
  arr2=np.random.binomial(n,q,N2).tolist() # Answer of student h in the iniversity U
  Rus.append(Count_essays(arr,n))
  Rus2.append(Count_essays(arr2,n))
#---
for k in range(U):
  for j in range(n+1):
    Rus[k][j]=Rus[k][j]/(N1+N2)
    Rus2[k][j]=Rus2[k][j]/(N1+N2)
#---
row_vis_com_uni1=np.transpose(Rus)
row_vis_com_uni2=np.transpose(Rus2)
row_vis_com_uni=row_vis_com_uni1+row_vis_com_uni2
aveg,dstd=[],[]
for q in range(n+1):
  aveg.append(np.mean(row_vis_com_uni[q]))
  dstd.append(np.std(row_vis_com_uni[q]))
print("Normal-Monte-Carlo=",np.sum(aveg))
#-----Perform the Kolmogorov-Smirnov test
#statistic, pvalue = stats.ks_2samp(aveg,distrubutionFCI)#, alternative='less'
#statistic, pvalue = stats.ks_2samp(aveg,distrubutionTugK)#, alternative='less'
statistic, pvalue = stats.ks_2samp(aveg,distrubutionFCI_china)#, alternative='less'
print(f"Test statistic: {statistic:.4f}")
print(f"P-value: {pvalue:.4f}")
pv=pvalue
alpha=0.05
if pvalue < alpha:
    print("The null hypothesis can be rejected at the", alpha, "level of significance.")
else:
    print("The null hypothesis cannot be rejected at the", alpha, "level of significance.")
#-----
# Calculate empirical quantiles for each sample
n1 = len(aveg)
#n2 = len(distrubutionFCI)
#n2 = len(distrubutionTugK)
n2 = len(distrubutionFCI_china)
empirical_quantiles1 = np.arange(1, n1 + 1) / n1
empirical_quantiles2 = np.arange(1, n2 + 1) / n2
#--xxxxxxxxxxxxxxxxxxxxx----The graphics--------xxxxxxxxxxxxxxxxxxxxxxxxxxxxxxx
fig, ax = plt.subplots()
ax.errorbar(x=esasys, y=aveg, yerr=dstd, capsize=2, capthick=2, marker='o',linestyle='none',
             linewidth=1.0,markerfacecolor='red',label='Std(Model)')
ax.plot(esasys, aveg,'--',color='red',linewidth=0.8,label='Average(Model)')
#ax.bar(esasys,distrubutionFCI,color=(0.8, 0.8, 0.8),width=0.6,align='center',
#        edgecolor='black',linewidth=1.0,label='FCI observational data')
#ax.bar(esasys,distrubutionTugK,color=(0.8, 0.8, 0.8),width=0.6,align='center',
#        edgecolor='black',linewidth=1.0,label='TUG-K')
ax.bar(esasys,distrubutionFCI_china,color=(0.8, 0.8, 0.8),width=0.6,align='center',
        edgecolor='black',linewidth=1.0,label='FCI-China')
ax.set_xlabel('Correct questions',fontsize=16,color='black')
ax.set_ylabel('
',fontsize=16,color='black')
ax.grid(False)
ax.legend(loc='upper left')
#ax.legend(loc='upper left', bbox_to_anchor=(0.005,0.14))
# insert--------------
ax1 = fig.add_subplot(5, 5, 16)
# Set the position of the subplot relative to the background plot
left, bottom, width, height = ax1.get_position().bounds
ax1.set_position([left + 0.9 * width, bottom + 0.3 * height, 2.2 * width, 2.5 * height])
#ax1.plot(np.quantile(aveg, empirical_quantiles1), np.quantile(distrubutionFCI, empirical_quantiles2), 'o--')
#ax1.plot(np.quantile(aveg, empirical_quantiles1), np.quantile(distrubutionTugK, empirical_quantiles2), 'o--')
ax1.plot(np.quantile(aveg, empirical_quantiles1), np.quantile(distrubutionFCI_china, empirical_quantiles2), 'o--')
ax1.plot([np.min(aveg), np.max(aveg)], [np.min(aveg), np.max(aveg)], 'r-')
ax1.set_xlabel('Two-level model data quantiles')
#ax1.set_ylabel('FCI data quantiles')
#ax1.set_ylabel('TUG-K data quantiles', labelpad=-170, rotation=-90)
ax1.set_ylabel('FCI-China data quantiles')
ax1.set_title('Q-Q Plot')
#ax1.text(0.14, 0.035, r"
", fontsize=12, ha="center")
ax1.text(0.1, 0.035, r"
".format(alpha), fontsize=12, ha="center")
ax1.text(0.1, 0.015, r"
".format(pv), fontsize=12, ha="center")
#----
plt.show()

\end{lstlisting}
\chapter{Colorado test adaptation} \label{app:colorado}

\section*{Survey presented to students}

Instrucciones. A continuación leerás una serie de afirmaciones  que pueden o no describir lo que piensas sobre el aprender física. Marca tu hoja de respuestas de acuerdo con el siguiente convenio:
\begin{description}
\item [A] Completamente en desacuerdo
\item [B] Desacuerdo
\item [C] Neutral
\item [D] De acuerdo
\item [E] Completamente de acuerdo
\end{description}

\begin{enumerate}

\item El problema para aprender física no es tener que memorizar toda la información expuesta en clase, consiste en no entender los principios.
\item	Cuando resuelvo un problema de física, trato de pensar  cual sería un valor lógico para la respuesta.
\item	Yo considero que la física es útil  en mi vida cotidiana.
\item	Para mí es bueno hacer muchísimos problemas para aprender física.
\item	Después de estudiar un tema de física y creo entenderlo, tengo dificultades para resolver problemas del mismo tema.
\item	Las formulas de los diferentes temas de física generalmente están relacionadas.
\item	Cuanto más entienden los físicos, se comprueba que la mayoría de las ideas actuales de la física son incorrectas. 
\item	Cuando resuelvo un problema de física, lo único que hago es buscar  una fórmula  para utilizar  las variables dadas en el problema y reemplazo los valores.
\item	Una buena manera de aprender física para mi es leer el libro de texto en detalle. 
\item	Normalmente no solo hay un método correcto para resolver un problema de física. 
\item	Hasta que no entiendo por qué y cómo funcionan las cosas no estoy satisfecho 
\item	Puedo aprender física aunque el profesor no explique bien el tema en la clase.
\item	Siempre espero que las formulas de la física me ayuden a comprender las ideas y no solo usarlas para hacer cálculos.
\item	Estudio física para adquirir conocimiento que me será útil en mi vida afuera de la universidad.
\item	Si no encuentro como resolver algún problema de física en mi primer intento, trato de resolverlo de otra manera.
\item	Casi toda persona que se esfuerza es capaz  de entender física.
\item	Entender  física consiste básicamente en poder recordar algo que hayas leído o que hayas visto.
\item	Podrían resultar dos diferentes valores correctos para la respuesta de un problema en física si se utilizan dos enfoques distintos.
\item	Para entender física, la discuto con mis amigos y compañeros.
\item	Si en cinco minutos no  llego a la respuesta de un ejercicio de física pido  ayuda a otra persona o lo dejo para después.
\item	Si no recuerdo la  fórmula necesaria para resolver un problema en el examen,  busco otra  manera (¡sin hacer  copia!)  para resolverlo.
\item	Para resolver un problema de física busco otro similar que ya haya resuelto.
\item	Cuando hago un problema de física  y mis cálculos dan resultados diferentes a los que yo esperaba, no confío en mis cálculos y reviso el problema otra vez.
\item	En física es importante para mí entender las fórmulas antes de poder usarlas correctamente.
\item	Disfruto resolviendo problemas de física.
\item	En la física, las formulas matemáticas expresan relaciones  importantes  entre las cantidades medidas.
\item	Es importante que el gobierno apruebe nuevas ideas científicas antes de que puedan ser ampliamente aceptadas.
\item	Aprender física cambia mis ideas acerca de cómo funcionan las cosas.
\item	Para aprender física, solo tengo que memorizar las soluciones de los problemas parecidos.
\item	Las habilidades lógicas que utilizo para entender física me son útiles en mi vida diaria.
\item	Utilizamos esta pregunta para eliminar las encuestas de las personas que no están leyendo las preguntas. Por favor selecciona la opción D (de acuerdo) para preservar tus respuestas.
\item	Detenerse para entender de dónde vienen las fórmulas en física es bueno, aunque   se quite tiempo para hacer ejercicios.
\item	Siento que entender a detalle sólo algunos problemas de física es un buen método para aprender.
\item	Normalmente se me ocurre una forma de resolver los problemas de física.
\item	Lo que se trata en física tiene relación con lo que se experimenta en el mundo real.
\item	Hay veces en que resuelvo problemas de física en más de una manera para mejorar mi entendimiento.
\item	Para entender física, a veces relaciono mis experiencias personales con lo  que estoy analizando.
\item	Es posible explicar física sin necesidad de fórmulas matemáticas.
\item	Cuando resuelvo problemas de física, explícitamente pienso en los principios físicos que aplican al problema.
\item	Si me confundo  en un problema de física, leo el tema por mi cuenta o busco asesoría.
\item	Es posible que los físicos hagan cuidadosamente un experimento y que obtengan dos resultados muy diferentes que ambos sean correctos.
\item	Cuando estudio física, trato de relacionar la información importante con lo que ya conozco en lugar de memorizarla en la forma en la que me la presentan.
 
\end{enumerate}

\chapter{Test URM} \label{app:TUG_URM}

The instrument was presented to the students as shown below.

\section*{Cuestionario rápido sobre MUR} 
Este cuestionario busca evaluar sus conocimientos en  cinemática unidimensional.  El cuestionario no debe extraerse de su protección plástica, no se debe escribir sobre él y debe ser devuelto al terminar la prueba. Este cuestionario  es de selección múltiple con única respuesta la cual se consigna  únicamente en su hoja de respuestas, recuerde  que tachones,  enmendaduras o marcar más de una respuesta por pregunta  anulan irremediablemente el punto. No olvide escribir en su hoja de respuestas su nombre y el código del examen, el cual se  encuentra en el pie de página de este.\\ 
\textbf{Profesor: Paco Talero}

\begin{figure*}[ht]
\begin{center}
\includegraphics[width=16.0cm,height=21.0cm]{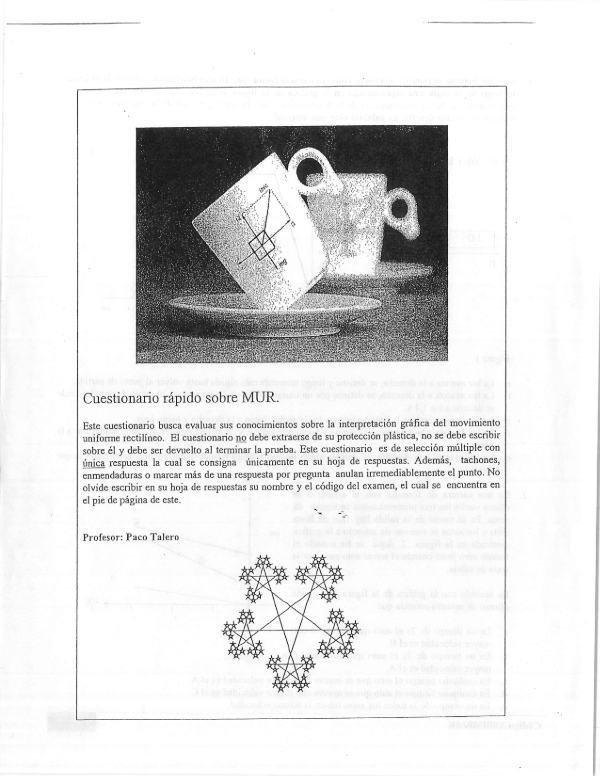}
\end{center}
\end{figure*}
\begin{figure*}[ht]
\includegraphics[width=16.0cm,height=21.0cm]{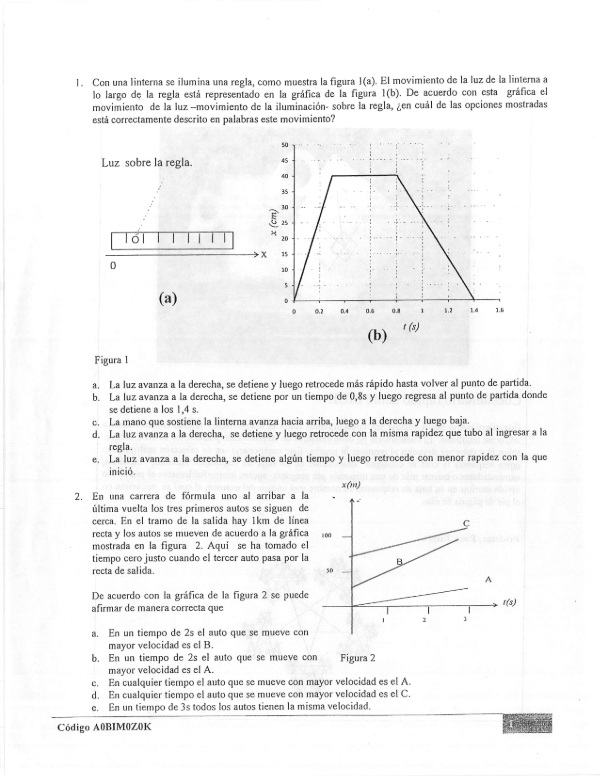}
\end{figure*}
\begin{figure*}[ht]
\includegraphics[width=16.0cm,height=21.0cm]{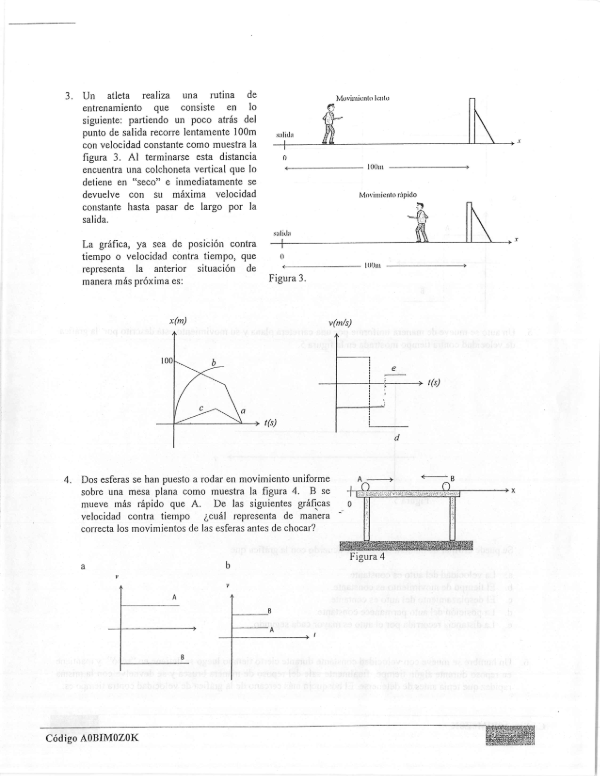}
\end{figure*}
\begin{figure*}[ht]
\includegraphics[width=16.0cm,height=21.0cm]{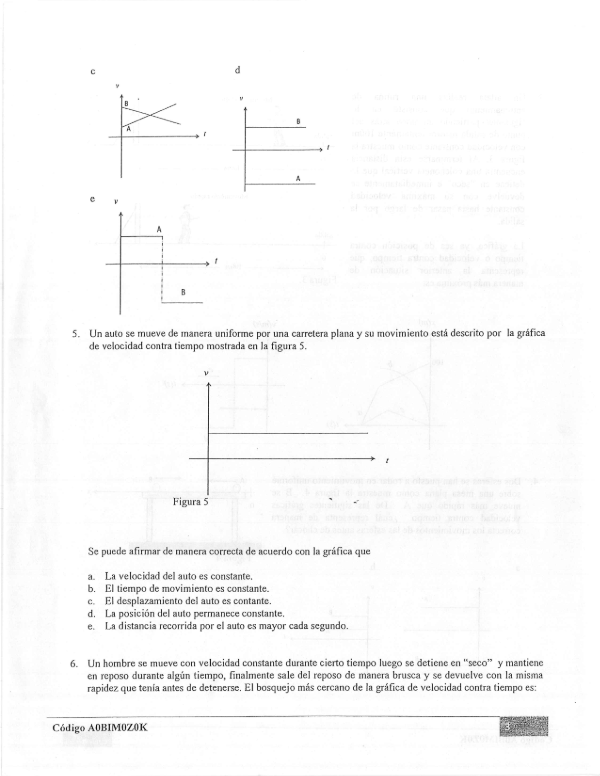}
\end{figure*}
\begin{figure*}[ht]
\includegraphics[width=16.0cm,height=21.0cm]{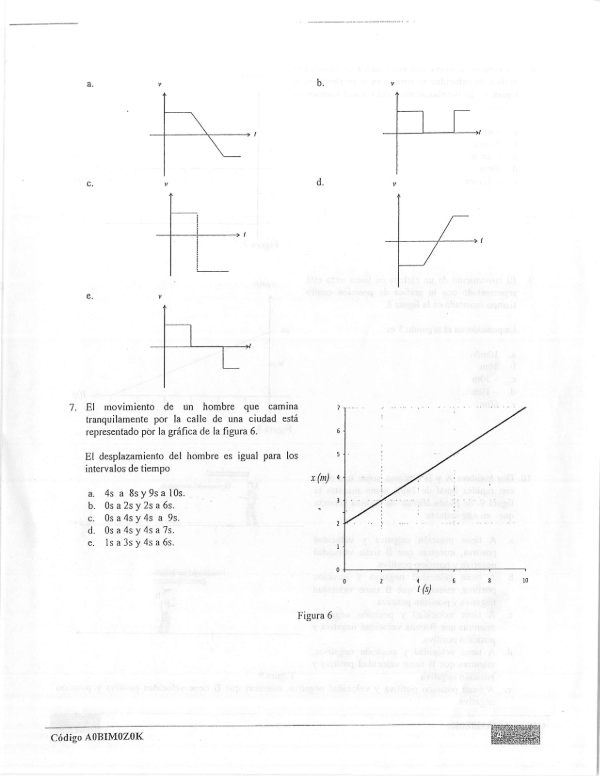}
\end{figure*}
\begin{figure*}[ht]
\includegraphics[width=16.0cm,height=21.0cm]{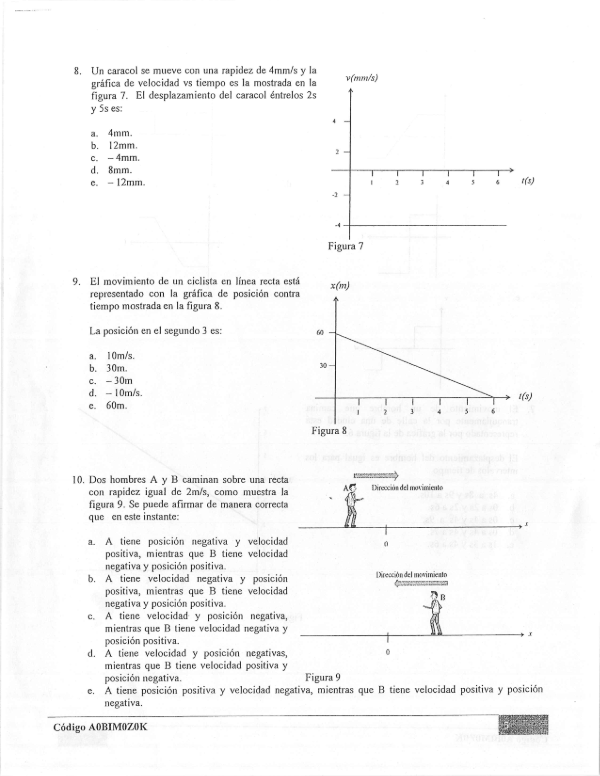}
\end{figure*}

\chapter{Handbook about the graphical interpretation of accelerated uniform motion} \label{app:Tutorial_MUG}

This manual was developed as part of the research work aimed at developing curriculum-immersed content. Its primary objective is to serve as a basis for researching how students from the population described in this study approach the fundamental concepts of graphical interpretation of one-dimensional accelerated uniform motion. This appendix only shows a part of it, to see it in full see \href{https://drive.google.com/file/d/14F8bETp3JuQJTzDXQTVic-K-upTBRkM1/view?usp=drive_link}{Annex 6}.

\begin{figure*}[ht]
\begin{center}
\includegraphics[width=16.0cm,height=21.0cm]{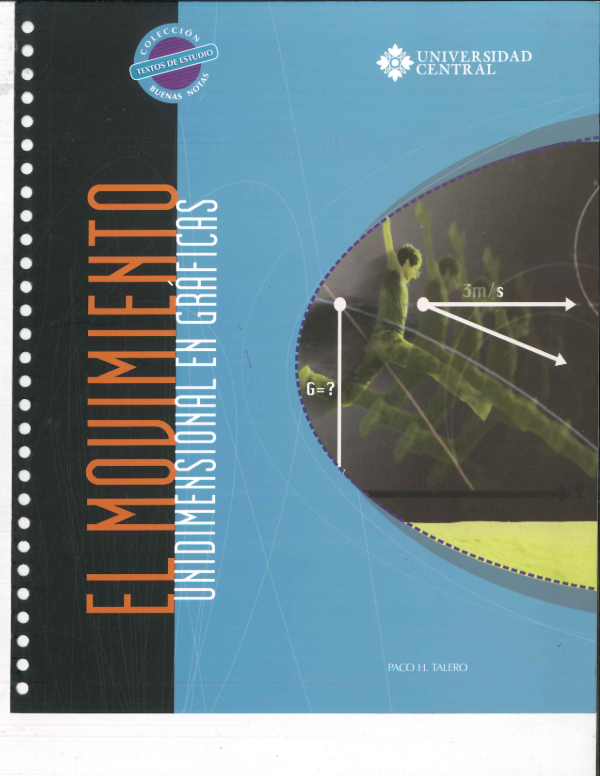}
\end{center}
\end{figure*}
\begin{figure*}[ht]
\includegraphics[width=16.0cm,height=21.0cm]{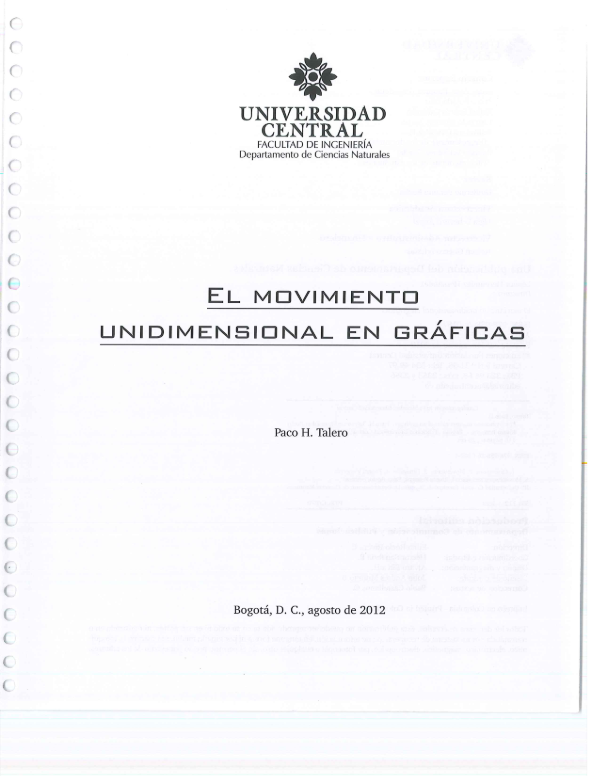}
\end{figure*}
\begin{figure*}[ht]
\includegraphics[width=16.0cm,height=21.0cm]{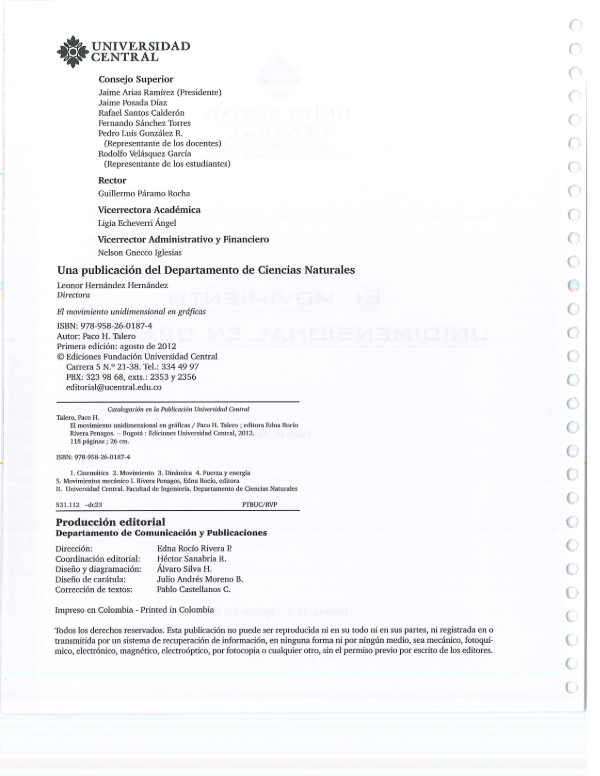}
\end{figure*}
\begin{figure*}[ht]
\includegraphics[width=16.0cm,height=21.0cm]{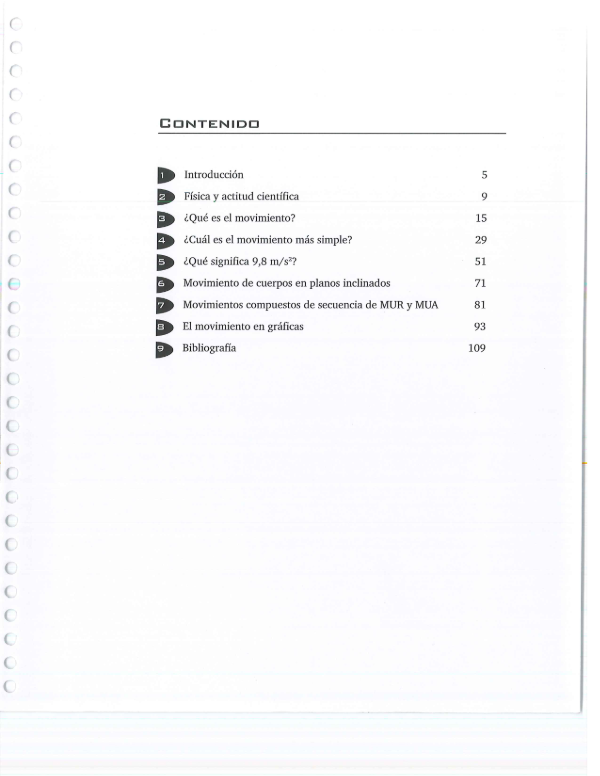}
\end{figure*}
\chapter{Instrument on narrative}

\section{Model problem}

Un auto se mueve por una carretera recta con una rapidez constante de $80km/h$ durante $15$ minutos, de repente el conductor observa a lo lejos un obstáculo en la vía y baja de manera inmediata la rapidez de su auto a $50km/h$, luego continúa con esta rapidez durante $3$ minutos más en dirección al obstáculo. 
\begin{enumerate}
\item Realice un dibujo que ilustre toda la situación.
\item Ubique un sistema de referencia adecuado para estudiar el movimiento del auto.
\item Identifique la partícula que representa el auto.
\item Pinte la función de posición del auto con respecto al tiempo.
\item Pinte la función de velocidad del auto con respecto al tiempo.
\end{enumerate}

\section{12 equivalent problems}

\begin{enumerate}

\item Un auto se mueve por una carretera recta con una rapidez constante de $80km/h$ durante $15$ minutos, de repente el conductor observa a lo lejos un obstáculo en la vía y baja de manera inmediata la rapidez de su auto a $50km/h$, luego continúa con esta rapidez durante $3$ minutos más en dirección al obstáculo. 

\item Un motociclista se mueve por una carretera llana con una rapidez constante de $120km/h$ durante $5$ minutos, se detiene súbitamente durante otros $5$ minutos y finalmente regresa con una rapidez de $60km/h$ durante $10$ minutos por la misma vía. 

\item Un auto se mueve por una carretera recta con una rapidez constante de $60km/h$ durante $15$ minutos, de repente el conductor observa a lo lejos un peaje en la vía y baja de manera inmediata la rapidez de su auto a $30km/h$, luego continúa con esta rapidez durante 3 minutos más en dirección al peaje. 

\item Un atleta  corre por una pista recta con una rapidez constante de $6m/s$ durante $10$ segundos, hasta que encuentra una colchoneta vertical y de manera inmediata se regresa con una rapidez de  $4m/s$, luego continúa con esta rapidez durante $15$ segundos más hasta alcanzar el final de la pista donde se detiene súbitamente. 

\item En una competencia de natación un competidor se encuentra dentro de la piscina en el punto de salida, al escuchar el disparo de partida comienza a nadar  por su carril recto con una rapidez constante de $2m/s$  hacia la otra  orilla, al llegar allí  gira rápidamente sin cambiar de manera apreciable su rapidez y regresa al punto de partida. Esto lo realiza hasta que retorna tres veces al punto de salida. Si la piscina mide $50m$ de largo:   

\item En un entrenamiento de atletismo un atleta pasa por la salida justo cuando el entrenador activa su cronómetro, recorre una distancia de $50$ metros  en $10$ segundos, aquí baja su rapidez a la mitad y se mueve en la misma dirección durante $20$ segundos más  que es justo cuando el entrenador pita y el atleta se detiene en seco.

\item Un hombre camina en línea recta por un parque plano y largo. Recorre 10 metros en 5 segundos y se detiene bruscamente, se mantiene así durante 4 segundos y luego súbitamente  se regresa hacia el punto de salida con una rapidez de $3m/s$ durante $4s$, al cabo del cual se detiene en "seco" y se mantiene en reposo durante $5s$ más.  

\item En un juego mecánico una esfera se mueve sobre una mesa plana en línea recta, partiendo del centro recorre una distancia de $20cm$ en $4$ segundos con rapidez constante, allí se detiene bruscamente y se mantiene así durante $4$ segundos,  luego súbitamente  se regresa hacia el punto de salida con una rapidez de $4cm/s$ durante $2s$, al cabo del cual se detiene en "seco" y se mantiene en reposo durante $4s$ más.  

\item En una prueba de motociclismo acrobático un motociclista  se mueve sobre una pista plana en línea recta, partiendo  del frente de la puerta principal  recorre una distancia de $100m$ en $4$ segundos con rapidez constante, allí se detiene bruscamente y se mantiene así durante un minuto,  luego súbitamente   regresa hacia el punto de salida con una rapidez de $40m/s$ durante $20s$.  

\item En una carrera de $100m$ lisos un atleta al escuchar el disparo se queda quieto durante $2s$, pero luego parte súbitamente recorriendo $50m$ en $5s$, luego súbitamente baja su rapidez a $4m/s$ y continua así pasando de largo por la meta.  

\item  En una carrera de autos hay un tramo largo y recto de pista que contiene el punto de salida, un auto pasa por el punto de salida con una velocidad de $120km/h$, se mantiene así durante dos minutos luego  súbitamente reduce su velocidad a $90km/h$, continúa así durante un minuto y luego reduce su velocidad a $60km/h$, continúa así durante dos minutos y luego súbitamente reduce su velocidad a $30km/h$.  

\item Un hombre  camina por un parque a través de un sendero recto. Pasa por un puente sobre el riachuelo del parque con una rapidez de $1m/s$ hacia el norte y camina con este ritmo durante dos minutos al final de los cuales se detiene bruscamente, permanece en reposo durante un minuto, luego  inicia una rápida marcha hacia el sur pasando por el puente después de transcurridos $20s$ de haber salido del reposo.
 
\end{enumerate}

\chapter{Exercisers for discussion about G URM} \label{app:Eval_Oral}
\section{Graphic analysis of position and velocity vs time}
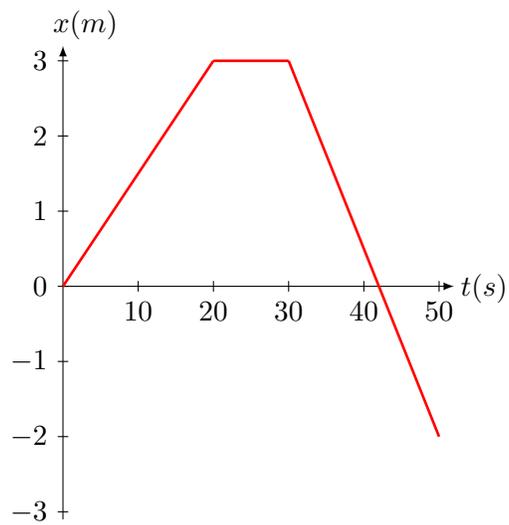
\begin{figure}[!htp]
\begin{center}
	\begin{tikzpicture}
	\draw [-latex] (0cm,0cm) -- (5.2cm,0cm);
	\draw [-latex] (0cm,-3.1cm) -- (0cm,3.2cm);
	\coordinate [label=below:\textcolor{black} {$t(s)$}] (x) at  (5.6cm,0.3cm);
	\coordinate [label=below:\textcolor{black} {$x(m)$}] (x) at  (0.3cm,3.8cm);
     \foreach \x/\xtext in {1/10, 2/20, 3/30, 4/40, 5/50}  
     \draw[shift={(\x,0)}] (0pt,2pt)--(0pt,-2pt) node[below] {$\xtext$}; 
     \foreach \y/\ytext in {-3/-3, -2/-2, -1/-1, 0/0, 1/1,2/2,3/3}  
     \draw[shift={(0.0,\y)}] (2pt,0pt)--(-2pt,0pt) node[left] {$\ytext$}; 
	\draw [red,line width=1.0pt] (0cm,0cm) -- (2cm,3cm);
	\draw [red,line width=1.0pt] (2cm,3cm) -- (3cm,3cm);
	\draw [red,line width=1.0pt] (3cm,3cm) -- (5cm,-2cm);	
	\end{tikzpicture}
\caption{Question $1$, position vs time.}
\end{center}
\end{figure}
\begin{figure}[!htp]
\begin{center}
\begin{tikzpicture}
	\draw [-latex] (0cm,0cm) -- (5.2cm,0cm);
	\draw [-latex] (0cm,-3.1cm) -- (0cm,3.2cm);
	\coordinate [label=below:\textcolor{black} {$t(s)$}] (x) at  (5.6cm,0.3cm);
	\coordinate [label=below:\textcolor{black} {$x(m)$}] (x) at  (0.3cm,3.8cm);
     \foreach \x/\xtext in {1/10, 2/20, 3/30, 4/40, 5/50}  
     \draw[shift={(\x,0)}] (0pt,2pt)--(0pt,-2pt) node[below] {$\xtext$}; 
     \foreach \y/\ytext in {-3/-3, -2/-2, -1/-1, 0/0, 1/1,2/2,3/3}  
     \draw[shift={(0.0,\y)}] (2pt,0pt)--(-2pt,0pt) node[left] {$\ytext$}; 
	\draw [red,line width=1.0pt] (0cm,0cm) -- (2cm,-3cm);
	\draw [red,line width=1.0pt] (2cm,-3cm) -- (3cm,-3cm);
	\draw [red,line width=1.0pt] (3cm,-3cm) -- (5cm,2cm);	
	\end{tikzpicture}
	\caption{Question $2$, position vs time.}
\end{center}
\end{figure}
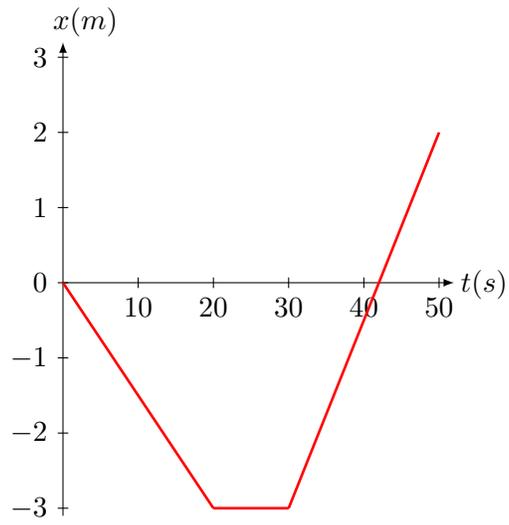
\begin{figure}[!htp]
\begin{center}

\begin{tikzpicture}
	\draw [-latex] (0cm,0cm) -- (5.2cm,0cm);
	\draw [-latex] (0cm,-3.1cm) -- (0cm,3.2cm);
	\coordinate [label=below:\textcolor{black} {$t(s)$}] (x) at  (5.6cm,0.3cm);
	\coordinate [label=below:\textcolor{black} {$x(m)$}] (x) at  (0.3cm,3.8cm);
     \foreach \x/\xtext in {1/10, 2/20, 3/30, 4/40, 5/50}  
     \draw[shift={(\x,0)}] (0pt,2pt)--(0pt,-2pt) node[below] {$\xtext$}; 
     \foreach \y/\ytext in {-3/-3, -2/-2, -1/-1, 0/0, 1/1,2/2,3/3}  
     \draw[shift={(0.0,\y)}] (2pt,0pt)--(-2pt,0pt) node[left] {$\ytext$}; 
	\draw [red,line width=1.0pt] (0cm,0cm) -- (1cm,1cm);
	\draw [red,line width=1.0pt] (1cm,1cm) -- (4cm,-3cm);
	\draw [red,line width=1.0pt] (4cm,-3cm) -- (5cm,-3cm);	
	\end{tikzpicture}
\caption{Question $3$, position vs time.}
\end{center}
\end{figure}
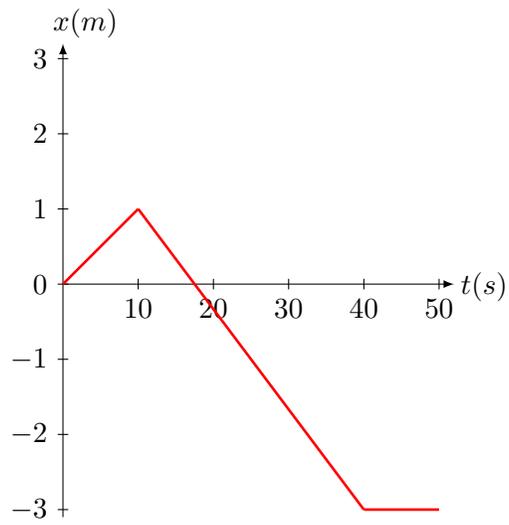

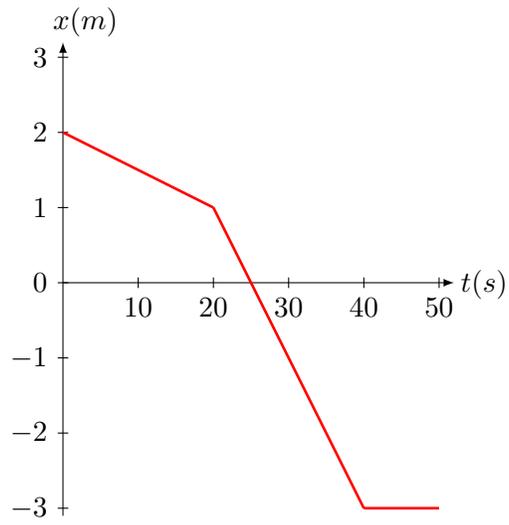
\begin{figure}[!htp]
\begin{center}
\begin{tikzpicture}
	\draw [-latex] (0cm,0cm) -- (5.2cm,0cm);
	\draw [-latex] (0cm,-3.1cm) -- (0cm,3.2cm);
	\coordinate [label=below:\textcolor{black} {$t(s)$}] (x) at  (5.6cm,0.3cm);
	\coordinate [label=below:\textcolor{black} {$x(m)$}] (x) at  (0.3cm,3.8cm);
     \foreach \x/\xtext in {1/10, 2/20, 3/30, 4/40, 5/50}  
     \draw[shift={(\x,0)}] (0pt,2pt)--(0pt,-2pt) node[below] {$\xtext$}; 
     \foreach \y/\ytext in {-3/-3, -2/-2, -1/-1, 0/0, 1/1,2/2,3/3}  
     \draw[shift={(0.0,\y)}] (2pt,0pt)--(-2pt,0pt) node[left] {$\ytext$}; 
	\draw [red,line width=1.0pt] (0cm,2cm) -- (2cm,1cm);
	\draw [red,line width=1.0pt] (2cm,1cm) -- (4cm,-3cm);
	\draw [red,line width=1.0pt] (4cm,-3cm) -- (5cm,-3cm);	
	\end{tikzpicture}
\caption{Question $4$, position vs time.}
\end{center}
\end{figure}
\begin{figure}[!htp]
\begin{center}

\begin{tikzpicture}
	\draw [-latex] (0cm,0cm) -- (5.2cm,0cm);
	\draw [-latex] (0cm,-3.1cm) -- (0cm,3.2cm);
	\coordinate [label=below:\textcolor{black} {$t(s)$}] (x) at  (5.6cm,0.3cm);
	\coordinate [label=below:\textcolor{black} {$x(m)$}] (x) at  (0.3cm,3.8cm);
     \foreach \x/\xtext in {1/10, 2/20, 3/30, 4/40, 5/50}  
     \draw[shift={(\x,0)}] (0pt,2pt)--(0pt,-2pt) node[below] {$\xtext$}; 
     \foreach \y/\ytext in {-3/-3, -2/-2, -1/-1, 0/0, 1/1,2/2,3/3}  
     \draw[shift={(0.0,\y)}] (2pt,0pt)--(-2pt,0pt) node[left] {$\ytext$}; 
	\draw [red,line width=1.0pt] (0cm,-3cm) -- (2cm,0cm);
	\draw [red,line width=1.0pt] (2cm,0cm) -- (3cm,3cm);
	\draw [red,line width=1.0pt] (3cm,3cm) -- (5cm,3cm);	
	\end{tikzpicture}
\caption{Question $5$, position vs time.}
\end{center}
\end{figure}
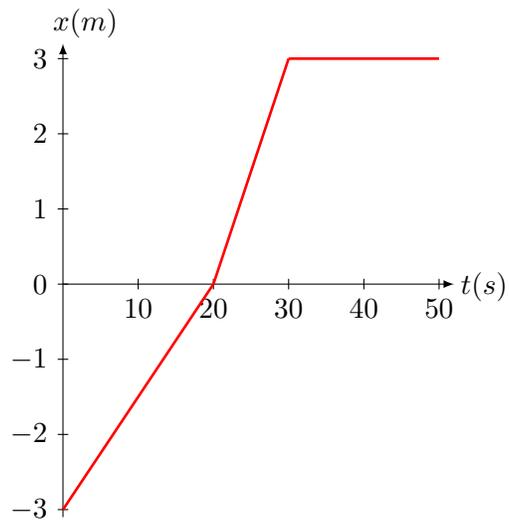

\begin{figure}[!htp]
\begin{center}

\begin{tikzpicture}
	\draw [-latex] (0cm,0cm) -- (5.2cm,0cm);
	\draw [-latex] (0cm,-3.1cm) -- (0cm,3.2cm);
	\coordinate [label=below:\textcolor{black} {$t(s)$}] (x) at  (5.6cm,0.3cm);
	\coordinate [label=below:\textcolor{black} {$v(m/s)$}] (x) at  (0.3cm,3.8cm);
     \foreach \x/\xtext in {1/10, 2/20, 3/30, 4/40, 5/50}  
     \draw[shift={(\x,0)}] (0pt,2pt)--(0pt,-2pt) node[below] {$\xtext$}; 
     \foreach \y/\ytext in {-3/-3, -2/-2, -1/-1, 0/0, 1/1,2/2,3/3}  
     \draw[shift={(0.0,\y)}] (2pt,0pt)--(-2pt,0pt) node[left] {$\ytext$}; 
	\draw [blue,line width=1.0pt] (0cm,3cm) -- (2.5cm,3cm);
	\draw [blue,line width=1.0pt] (2.5cm,0cm) -- (4.0cm,0cm);
	\draw [blue,line width=1.0pt] (4cm,-3cm) -- (5cm,-3cm);
	\draw [dashed,blue,line width=0.5pt] (2.5cm,0cm) -- (2.5cm,3cm);
	\draw [dashed,blue,line width=0.5pt] (4cm,0cm) -- (4cm,-3cm);		
	\end{tikzpicture}
	\caption{Question $1$, velocity vs time.}
\end{center}
\end{figure}
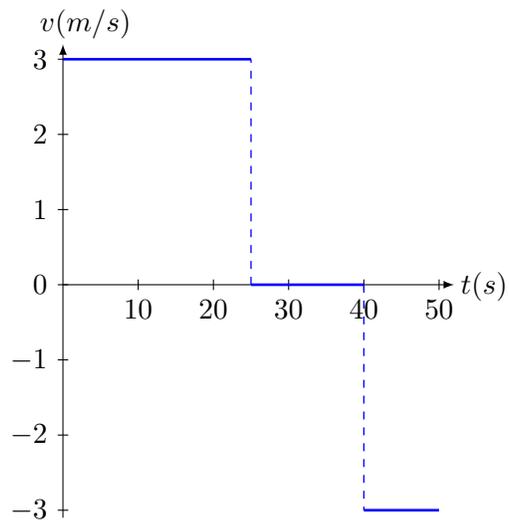

\begin{figure}[!htp]
\begin{center}

\begin{tikzpicture}
	\draw [-latex] (0cm,0cm) -- (5.2cm,0cm);
	\draw [-latex] (0cm,-3.1cm) -- (0cm,3.2cm);
	\coordinate [label=below:\textcolor{black} {$t(s)$}] (x) at  (5.6cm,0.3cm);
	\coordinate [label=below:\textcolor{black} {$v(m/s)$}] (x) at  (0.3cm,3.8cm);
     \foreach \x/\xtext in {1/10, 2/20, 3/30, 4/40, 5/50}  
     \draw[shift={(\x,0)}] (0pt,2pt)--(0pt,-2pt) node[below] {$\xtext$}; 
     \foreach \y/\ytext in {-3/-3, -2/-2, -1/-1, 0/0, 1/1,2/2,3/3}  
     \draw[shift={(0.0,\y)}] (2pt,0pt)--(-2pt,0pt) node[left] {$\ytext$}; 
	\draw [blue,line width=1.0pt] (0cm,-3cm) -- (2.5cm,-3cm);
	\draw [blue,line width=1.0pt] (2.5cm,1cm) -- (4cm,1cm);
	\draw [blue,line width=1.0pt] (4cm,2cm) -- (5cm,2cm);
	\draw [dashed,blue,line width=0.5pt] (2.5cm,-3cm) -- (2.5cm,1cm);
	\draw [dashed,blue,line width=0.5pt] (4cm,1cm) -- (4cm,2cm);		
	\end{tikzpicture}
\caption{Question $2$, velocity vs time.}
\end{center}
\end{figure}
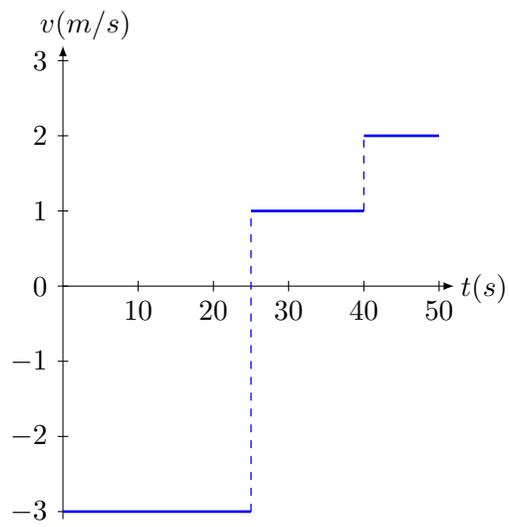

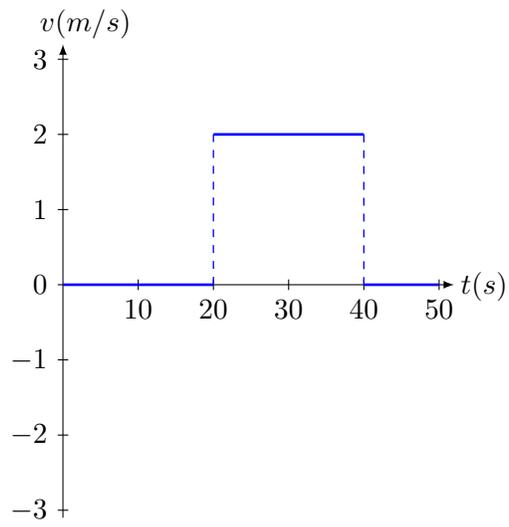
\begin{figure}[!htp]
\begin{center}

\begin{tikzpicture}
	\draw [-latex] (0cm,0cm) -- (5.2cm,0cm);
	\draw [-latex] (0cm,-3.1cm) -- (0cm,3.2cm);
	\coordinate [label=below:\textcolor{black} {$t(s)$}] (x) at  (5.6cm,0.3cm);
	\coordinate [label=below:\textcolor{black} {$v(m/s)$}] (x) at  (0.3cm,3.8cm);
     \foreach \x/\xtext in {1/10, 2/20, 3/30, 4/40, 5/50}  
     \draw[shift={(\x,0)}] (0pt,2pt)--(0pt,-2pt) node[below] {$\xtext$}; 
     \foreach \y/\ytext in {-3/-3, -2/-2, -1/-1, 0/0, 1/1,2/2,3/3}  
     \draw[shift={(0.0,\y)}] (2pt,0pt)--(-2pt,0pt) node[left] {$\ytext$}; 
	\draw [blue,line width=1.0pt] (0cm,0cm) -- (2cm,0cm);
	\draw [blue,line width=1.0pt] (2cm,2cm) -- (4cm,2cm);
	\draw [blue,line width=1.0pt] (4cm,0cm) -- (5cm,0cm);
	\draw [dashed,blue,line width=0.5pt] (2cm,0cm) -- (2cm,2cm);
	\draw [dashed,blue,line width=0.5pt] (4cm,2cm) -- (4cm,0cm);		
	\end{tikzpicture}
\caption{Question $3$, velocity vs time.}
\end{center}
\end{figure}

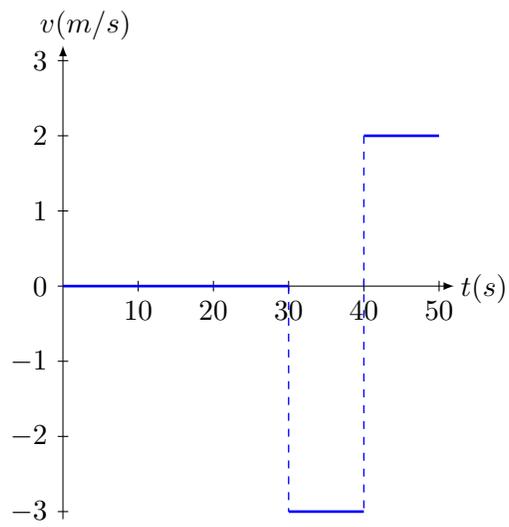
\begin{figure}[!htp]
\begin{center}

\begin{tikzpicture}
	\draw [-latex] (0cm,0cm) -- (5.2cm,0cm);
	\draw [-latex] (0cm,-3.1cm) -- (0cm,3.2cm);
	\coordinate [label=below:\textcolor{black} {$t(s)$}] (x) at  (5.6cm,0.3cm);
	\coordinate [label=below:\textcolor{black} {$v(m/s)$}] (x) at  (0.3cm,3.8cm);
     \foreach \x/\xtext in {1/10, 2/20, 3/30, 4/40, 5/50}  
     \draw[shift={(\x,0)}] (0pt,2pt)--(0pt,-2pt) node[below] {$\xtext$}; 
     \foreach \y/\ytext in {-3/-3, -2/-2, -1/-1, 0/0, 1/1,2/2,3/3}  
     \draw[shift={(0.0,\y)}] (2pt,0pt)--(-2pt,0pt) node[left] {$\ytext$}; 
	\draw [blue,line width=1.0pt] (0cm,0cm) -- (3cm,0cm);
	\draw [blue,line width=1.0pt] (3cm,-3cm) -- (4cm,-3cm);
	\draw [blue,line width=1.0pt] (4cm,2cm) -- (5cm,2cm);
	\draw [dashed,blue,line width=0.5pt] (3cm,0cm) -- (3cm,-3cm);
	\draw [dashed,blue,line width=0.5pt] (4cm,2cm) -- (4cm,-3cm);		
	\end{tikzpicture}
	\caption{Question $4$, velocity vs time.}
\end{center}
\end{figure}
\begin{figure}[!htp]
\begin{center}

\begin{tikzpicture}
	\draw [-latex] (0cm,0cm) -- (5.2cm,0cm);
	\draw [-latex] (0cm,-3.1cm) -- (0cm,3.2cm);
	\coordinate [label=below:\textcolor{black} {$t(s)$}] (x) at  (5.6cm,0.3cm);
	\coordinate [label=below:\textcolor{black} {$v(m/s)$}] (x) at  (0.3cm,3.8cm);
     \foreach \x/\xtext in {1/10, 2/20, 3/30, 4/40, 5/50}  
     \draw[shift={(\x,0)}] (0pt,2pt)--(0pt,-2pt) node[below] {$\xtext$}; 
     \foreach \y/\ytext in {-3/-3, -2/-2, -1/-1, 0/0, 1/1,2/2,3/3}  
     \draw[shift={(0.0,\y)}] (2pt,0pt)--(-2pt,0pt) node[left] {$\ytext$}; 
	\draw [blue,line width=1.0pt] (0cm,1cm) -- (1cm,1cm);
	\draw [blue,line width=1.0pt] (1cm,3cm) -- (3cm,3cm);
	\draw [blue,line width=1.0pt] (3cm,-2cm) -- (5cm,-2cm);
	\draw [dashed,blue,line width=0.5pt] (1cm,1cm) -- (1cm,3cm);
	\draw [dashed,blue,line width=0.5pt] (3cm,-2cm) -- (3cm,3cm);		
	\end{tikzpicture}
	\caption{Question $5$, velocity vs time.}
\end{center}
\end{figure}
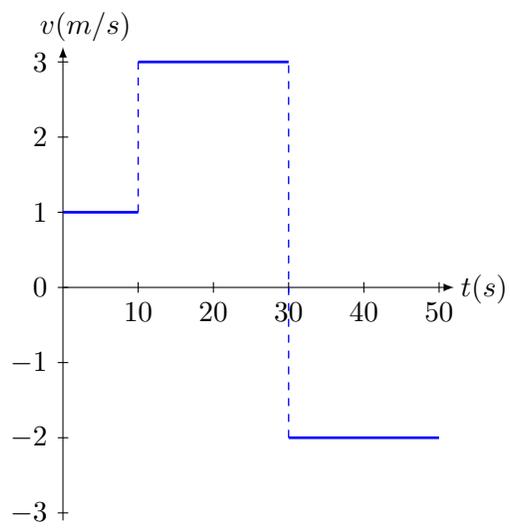
\chapter{Model simulation} \label{app:pricipalProgram}

\section{Implementation of a simulation in Python}

\begin{lstlisting}[language=Python,style=mystyle] 

import numpy as np
import matplotlib.pyplot as plt
import matplotlib.style as style
import seaborn as sns
import pandas as pd
#----Combinations without repeats
def combinations_without_repeats(n): # Number of student (n) is even
    lst=range(n)
    pairs = []
    Index = np.random.permutation(len(lst)) #This line does the permutation of the indexes of the student list (student system)
    for i in range(0, len(lst), 2): #--0,2,4,6,8...n-1
        pair = [lst[Index[i]], lst[Index[i+1]]] #--Make classmates
        pairs.append(pair) #--In this line is make the list of classmates
    return pairs
#--
print(len(combinations_without_repeats(12)))
combinations_without_repeats(12)
#---Decreasing elements
def is_increasing(lst):   #--Does the list have elements in descending order?
  cont=0
  for i in range(len(lst) - 1):
    if lst[i] >= lst[i + 1]:
      cont+=1
  if cont==len(lst)-1:return True
  else: return False
#--Concentration
R=[10,0,0,0,0]
Ne=10
def C(Ne,R):
  sum_op,m=0,len(R)
  for i in range(m):
    sum_op=sum_op+R[i]**2
  c= (np.sqrt(m)/(np.sqrt(m)-1))*(np.sqrt(sum_op)/Ne -1/np.sqrt(m))
  return c
#---
print("C=",round(C(10,R),3))
#----Cmin y Cmax
def cmM(u,Ne,R):
  m=len(R)
  c_min=(np.sqrt(m)/(np.sqrt(m)-1))*(np.sqrt( (1-u)**2/(m-1)+u**2) -1/np.sqrt(m))
  c_max=(np.sqrt(m)/(np.sqrt(m)-1))*(  np.sqrt(u**2+(1-u)**2) -1/np.sqrt(m)  )
  return c_min,c_max
#--Table
u=np.linspace(0,1,20) #--lista lineal
c_m,c_M=[],[]
for x in u:
  c_m.append(cmM(x,Ne,R)[0])
  c_M.append(cmM(x,Ne,R)[1])
#-----
plt.plot(u,c_m)
plt.plot(u,c_M)
plt.show()
#---A student
import numpy as np
class student:
    def __init__(self,M,N,po):
      self.M=M #number of levels M, rows
      self.N=N #number of attributes N, columns
      self.po=po #initial knowledge, ability to discern
      self._m =0.5 #slope of the linear growth function of the attribute
      self._b=0.5  #point of the linear growth function of the attribute
      self.s=np.zeros((M,N))# occupancy state, is filled with zeros
      self.v=np.zeros((M,N))# values matrix, increments by _m each column
      self.M_score=np.zeros((M,N))# score matrix
      #---
      for j in range(self.N):     # We are filling the state of occupation matrix, in accordance
        i=np.random.randint(0, self.M) # with po, in each attribute there is only one occupied level.
        r_n=np.random.random()
        if r_n<=po:
          self.s[i][j]=1
      #--We assign a value according to the difficulty of the attribute and the level
      for j in range(self.N):          # We are filling the value matrix, in
        for i in range(self.M):        # accordance with according to _ m and _b.
          self.v[i][j]=self._m*(j+1)+i+self._b

    #--The methods
    #--calculate score or discernment
    def score(self):
      sum_s,smax=0,0
      for j in range(self.N):
        smax=smax+self.v[self.M-1][j]
      #---
      for i in range(self.M):
        for j in range(self.N):
          self.M_score[i][j]=self.s[i][j]*self.v[i][j] #Multiplies the values in array
          sum_s=sum_s+self.M_score[i][j]  #a by the values in array value and adds all values
      return sum_s/smax # 10.5
      #----
    def change(self,q,F,occupy): # We are filling the state matrix from the outside.
      self.s[q][F]=occupy
    #---
    def VER(self):  # We can see the state matrix and the value matrix.
      return self.s,self.v
#---End of Studt class
#---Evolution
#---Global parameters
N=3    #--N the conceptual attributes
M=3    #--M are of level
po=0.8 #---initial background
p_mt=0.8  # Instructional quality
n=20 #Number of students is a even number
number_of_lessons=11
s_sys=[]
for k in range(n):
  s_sys.append(student(M,N,po))
SystemS=s_sys #--Fixed student system
#----Sampling

class samplig:
    def __init__(self,SystemS):
      self.SystemS=SystemS #
      self.n=len(SystemS)
    #--Discernment Index
    def D(self):
      sumaD=0
      for k in range(self.n):
        sumaD=sumaD+self.SystemS[k].score()
      return sumaD/self.n
    #--Clean state
    def Clean_States(self):
      cleanS=0
      for k in range(self.n):
        if self.SystemS[k].score()==0:
          cleanS=cleanS+1
      return cleanS
    #---Highest discernment state
    def OutlierS_States(self):
      Hd_S=0
      for k in range(self.n):
        if np.sum(self.SystemS[k].VER()[0][M-1])==N:
          Hd_S=Hd_S+1
      return Hd_S
    #---Linear and nonlinear states of learning
    def L_nl_states(self):
      l_st,nl_st=0,0
      for k in range(self.n):
        if self.SystemS[k].score()!=0 and np.sum(self.SystemS[k].VER()[0][M-1])!=N:
          stz=self.SystemS[k].VER()[0].T
          idx=[]
          for j in range(N):
            ztz=list(stz[j])
            if 1 in ztz:
              id=ztz.index(1)
              idx.append(id)
            else:idx.append(-1)
          if is_increasing(idx)==True:
            l_st+=1
          else:nl_st+=1
      return l_st,nl_st
    #----score and concentration
    def s_c(self):
      Ne=self.n
      Visits=[self.Clean_States(),self.L_nl_states()[1],self.L_nl_states()[0],self.OutlierS_States()]
      Concentration=C(Ne,Visits)
      ScoRe=self.OutlierS_States()/Ne
      return ScoRe,Concentration
     #--System microcontent
    def SysMic(self):
      grupS=student(M,N,0)#---Group of students as empty table
      matrix_occupation=np.zeros((M,N))# score matrix
      for k in range(self.n):
        matrix_occupation+=self.SystemS[k].VER()[0]
      #matrix_occupation=matrix_occupation/(self.n*N)
      for i in range(M):
        for j in range(N):
          NewV=matrix_occupation[i][j]
          grupS.change(i,j,NewV)
      return grupS,np.sum(matrix_occupation/(self.n*N))
     #----Histogram
    def Show_Histogram(self):
       Visits=[self.Clean_States(),self.L_nl_states()[1],self.L_nl_states()[0],self.OutlierS_States()]
       Learnig_states=['Clean','Nonlinear','Linear','Outliers']
       plt.style.use('ggplot')
       plt.figure(edgecolor='red') #navy,gray edgecolor facecolor='gray'
       plt.bar(Learnig_states,Visits,color='green',width=0.5,align='center',edgecolor='black',linewidth=1.5)
       plt.ylabel('Visits',fontsize=16,color='black')
       plt.xlabel('States of knowledge',fontsize=16,color='black')
       plt.grid(True)
       plt.xticks(rotation=90,color='black',fontsize=12)
       plt.yticks(color='black',fontsize=12)
       #plt.title("Histogram of visits", color='black')
       plt.show()
#-*_*_*_*_*_*_*_*_*_*_*_*_*_* Performance
fis1=samplig(SystemS)
print("Discernment Index=",fis1.D())
print("Clean states=",fis1.Clean_States())
print("Highest discernment state=",fis1.OutlierS_States())
print("Linear and nonlinear=",fis1.L_nl_states()[0])
print("states of learning",fis1.L_nl_states()[1])
print("s=",fis1.s_c()[0])
print("c=",fis1.s_c()[1])
#print("stado_sistema=",fis1.SysMic()[0])
print("ocupacion_stado_sistema=",fis1.SysMic()[1])
fis1.Show_Histogram()

#---Microcontent status
Laura=student(3,4,0.7) #--A particular student
def Microcontent(Laura):
  number_of_levels=len(Laura.VER()[0])#--Number of levels, rows
  n_C_a=len(Laura.VER()[0][0])        #--Number of conceptual attributes,columns
  #---
  LeveS,AttributeS=[],[]
  for i in range(number_of_levels):  #--M
    a='q'+str(i)
    LeveS.append(a)
  for j in range(n_C_a):    #--N
    b='F'+str(j)
    AttributeS.append(b)
  #---
  Data={}
  for r in range(len(AttributeS)):
    Data[AttributeS[r]]=Laura.VER()[0].T[r]
  #----
  df = pd.DataFrame(Data, index=LeveS) # Convert dictionary to Pandas DataFrame
  #---The heatmap
  ax = sns.heatmap(df, annot=True,fmt='.1f', cmap='Greens',linewidth=0.5,linecolor='black')
  ax.xaxis.tick_top() # Move x-axis labels to top
  plt.title("Microcontent status")
  plt.xlabel("Conceptual attributes",color='black')
  plt.ylabel("Levels")
  plt.show()
#-**_*_*_*_*
Microcontent(fis1.SysMic()[0])

#---Peer Interaction
class peer_instruction:
  def __init__(self,SystemS):
    self.SystemS=SystemS #
    self.n=len(SystemS)
  #--This function calculates nc and pA
  def transfer_probability(self,r,s):  #--- They are two r,s wiht r<s
    nc=0
    for i in range(M):
      for j in range(N):
        if self.SystemS[r].VER()[0][i][j]==1 and self.SystemS[s].VER()[0][i][j]==1:
          nc=nc+1
    pl=nc/N
    return pl
  # Classification of states
  def ClassState(self,r):
    #possible_states=0'clean',1'nonlinear',2'linear',3'outliers'
    if self.SystemS[r].score()==0: m_state=0
    else:
      if np.sum(self.SystemS[r].VER()[0][M-1])==N: m_state=3
      else:
        stz1=self.SystemS[r].VER()[0].T
        idx1=[]
        for j in range(N):
          ztz1=list(stz1[j])
          if 1 in ztz1:
            id1=ztz1.index(1)
            idx1.append(id1)
          else:
            idx1.append(-1)
        if is_increasing(idx1)==True:
          m_state=2
        else:
          m_state=1
    return m_state
  #--left (--r--) copy right (--s--) only used inside function Exchange---
  def copy(self,r,s):
    ab=[]
    for i in range(M):
      for j in range(N):
        if self.SystemS[r].VER()[0][i][j]!= self.SystemS[s].VER()[0][i][j]:
          ab.append((i,j))
    #----
    if ab==[]:
      Ar=self.SystemS[r]
    else:
      if len(ab)==1:
        qi,Fj=ab[0][0],ab[0][1]
        self.SystemS[r].change(qi,Fj,self.SystemS[s].VER()[0][qi][Fj])
        for i in range(M):
          if i!=qi:
            self.SystemS[r].change(i,Fj,0)
        Ar=self.SystemS[r]
      else:
        IndX=np.random.randint(0,len(ab)-1)
        qi,Fj=ab[IndX][0],ab[IndX][1]
        self.SystemS[r].change(qi,Fj,self.SystemS[s].VER()[0][qi][Fj])
        for i in range(M):
          if i!=qi:
            self.SystemS[r].change(i,Fj,0)
        Ar=self.SystemS[r]
    return Ar
  #----Exchange---
  def exchange(self,r,s):
    new_r,new_s=self.SystemS[r],self.SystemS[s]
    r_c=self.ClassState(r)
    s_c=self.ClassState(s) #--student status
    #If the students are completely different or completely the same, they do not interact.
    p_l=self.transfer_probability(r,s)
    if 0.0<p_l<1.0:
      #-----
      if r_c<s_c:     #--In the work they are L(A) and L(B)
        if p_l>=0.5:pc=p_l
        else: pc=1.-p_l
        if np.random.random()<pc:
          self.SystemS[r]=self.copy(r,s) #--Loser is r, r copy to s
        else:
          self.SystemS[s]=self.copy(s,r) #--winner is s, s copy to r the loser
      #--------
      elif r_c>s_c:
        if p_l>=0.5:pc=p_l
        else: pc=1.-p_l
        if np.random.random()<pc:
          self.SystemS[s]=self.copy(s,r) #--Loser is s, s copy to r
        else:
          self.SystemS[r]=self.copy(r,s) #--winner is r, r copy to s the loser
      #----
      elif r_c==s_c:
        if np.random.random()<0.5:
          self.SystemS[r]=self.copy(r,s) #--Loser is r, r copy to s
        else:
          self.SystemS[s]=self.copy(s,r) #--Loser is s, s copy to r
      #----------
#----
r,s=0,1
Sara_antes=SystemS[r].VER()[0]
Alice_antes=SystemS[s].VER()[0]

print("Sara antes\n",Sara_antes)
print("Alice antes\n",Alice_antes)
biologia=peer_instruction(SystemS)
print("microestado de r=",biologia.ClassState(r))
print("microestado de s=",biologia.ClassState(s))
print("pl=",biologia.transfer_probability(r,s))
#--Ejecutar
biologia.exchange(r,s)
Sara_despues=SystemS[r].VER()[0]
Alice_despues=SystemS[s].VER()[0]
print("Sara_despues\n",Sara_despues)
print("Alice_despues\n",Alice_despues)
#-----
#---redefinition
def Int_Inst(r):
  if np.random.random()<p_mt: #--If the instruction is appropriate, the student learns
    columnj=[]
    if SystemS[r].score()!=0:
      for j in range(N):
        if SystemS[r].VER()[0][M-1][j]!=1:
          list_j_column=list(SystemS[r].VER()[0].T[j])
          list_sum = sum(list_j_column) #sums the elements of the j column, must be 0 or 1
          if list_sum!=0:
            columnj.append(j)
      if columnj!=[]:
        min_columj = min(columnj) #chosen column number
        list_i_row=list(SystemS[r].VER()[0].T[min_columj])
        ind_row=list_i_row.index(1)
        #-----we make the exchange
        SystemS[r].change(ind_row,min_columj,0)
        SystemS[r].change(ind_row+1,min_columj,1)
      else:
        for j in range(N):
          list_j_column2=list(SystemS[r].VER()[0].T[j])
          list_sum2 = sum(list_j_column2) #sums the elements of the j column, must be 0 or 1
          if list_sum2==0:
            SystemS[r].change(0,j,1)
            break
    else:
      SystemS[r].change(0,0,1)
#----
r=0
Sara_antes=SystemS[r].VER()[0]
print("Sara antes\n",Sara_antes)
Int_Inst(r)
Sara_d=SystemS[r].VER()[0]

#---The student system
fis1=samplig(SystemS)
Fis=peer_instruction(SystemS)#--system
lectures,discernment,sCr,coc=[],[],[],[]
for nl in range(number_of_lessons):
  sCr.append(fis1.s_c()[0])
  coc.append(fis1.s_c()[1])
  lectures.append(nl)
  discernment.append(fis1.D())
  #-peer instruction
  couples=combinations_without_repeats(n)
  for n_collisions in couples:
    r,s=n_collisions[0],n_collisions[1]
    Fis.exchange(r,s) #--Peer Instruction
    Int_Inst(r)       #--Interaction with the instructor
#----
fis1.Show_Histogram()
Microcontent(fis1.SysMic()[0])
#---
#--Table
u=np.linspace(0,1,100) #--lista lineal
R=[fis1.Clean_States(),fis1.L_nl_states()[1],fis1.L_nl_states()[0],fis1.OutlierS_States()]
c_m,c_M=[],[]
for x in u:
  c_m.append(cmM(x,n,R)[0])
  c_M.append(cmM(x,n,R)[1])

#plt.style.use('ggplot')
plt.style.use('default')
plt.plot(lectures,discernment,'o--',color="red",markersize=4, alpha=1.0,label='D')
plt.plot(lectures,sCr,'o--',color="blue",markersize=4, alpha=1.0,label='score')
plt.grid(False)
plt.legend(loc='upper left')  #---lower center, lower right, upper left
plt.ylabel('
',fontsize=16)
plt.xlabel('
',fontsize=16)
plt.ylim(0, 1.1)
plt.title('Discernment Index')
#plt.axis('equal')
plt.show()
#---
xoo,yoo=sCr[0],coc[0]
xff,yff=sCr[-1]-sCr[0],coc[-1]-coc[0]
#-----
plt.axhline(y=0.2, color='grey', linestyle='--')
plt.axhline(y=0.5, color='grey', linestyle='--')
plt.axvline(x=0.4, color='grey', linestyle='--')
plt.axvline(x=0.7, color='grey', linestyle='--')
plt.quiver(xoo,yoo,xff,yff, angles='xy', scale_units='xy', scale=1, color='gray',
          linestyle='dashed',linewidth=2)
plt.plot(sCr,coc,'o--',color="green",label='Evolution')
plt.plot(u,c_m,color="red",label='
')
plt.plot(u,c_M,color="blue",label='
')
plt.grid(False)
plt.legend(loc='lower right')  #---lower center, lower right, upper left
plt.ylabel('
',fontsize=16)
plt.xlabel('
',fontsize=16)
#plt.title('Discernment Index') #'Bruce Centenary',
#plt.axis('equal')
plt.show()
#-----Hacke gain---
x1, y1 = 1., 0.
x2, y2 = 0., 1.
#-------
xm1, ym1 = 1., 0.
xm2, ym2 = 0., 0.7
#-------
xb1, yb1 = 1., 0.
xb2, yb2 = 0., 0.3
#-----G vs D
Gf=(discernment[-1]-discernment[0])/(1.-discernment[0])
xoD=discernment[0]

#---
fig, ax = plt.subplots()
ax.set_facecolor('white')
ax.plot([x1, x2], [y1, y2],'--',color='red',label='High')
ax.plot([xm1, xm2], [ym1, ym2],'--', color='green',label='Medium')
ax.plot([xb1, xb2], [yb1, yb2],'--', color='blue',label='Low')
ax.grid(False)  #--False
#ax.quiver(xoD,yoG,xfD,yfG, angles='xy', scale_units='xy', scale=1, color='black', headlength=3.5)
ax.scatter(xoD, Gf,facecolor='black', edgecolor='red', label='
')
ax.legend(loc='upper right')  #---lower center, lower right, upper left
ax.set_ylabel('
',fontsize=16)
ax.set_xlabel('
',fontsize=16)
plt.show()
#-----
N=3    #--N the conceptual attributes
M=3    #--M are of level
po=0.4 #---initial background
n=80 #Number of students
number_of_lessons=100 #---16 questions/weeks, time of instruction apx 5 weeks
s_sys=[]
for k in range(n):
  s_sys.append(student(M,N,po))
SystemS=s_sys #--Fixed student system
def phase_change(p_mt):
  for nl in range(number_of_lessons): #-peer instruction
    couples=combinations_without_repeats(n)
    for n_collisions in couples:
      r,s=n_collisions[0],n_collisions[1] #interaction between students
      Fis.exchange(r,s) #--
      Int_Inst(r)       #--Interaction with the instructor
  return fis1.OutlierS_States()/n,fis1.D()
#-------Graphs
fig, (ax1, ax2) = plt.subplots(2, 1, sharex=True)
fig1, (ax3) = plt.subplots(1, 1)
#---list of instruction level
colori=['red','blue','green','black','cyan','magenta']
kc=0
level_instruction=[0.1,0.2,0.3,0.4,0.5] #,0.4,0.5,0.6 [0.05,0.1,0.15] [0.1,0.2,0.3,0.4,0.5,0.6]
values_pm=np.linspace(0.,1., num=25, endpoint=True) #--50
for l in level_instruction:
  po=l
  sCoRE,dis=[],[]
  for p_mt in values_pm:
    s_sys=[]
    for k in range(n):
      s_sys.append(student(M,N,po))
    SystemS=s_sys #--Fixed student system
    fis1=samplig(SystemS)
    Fis=peer_instruction(SystemS)#--system
    orchestrare,discern=phase_change(p_mt) #score, dicern
    sCoRE.append(orchestrare)
    dis.append(discern)
  ax2.plot(values_pm,sCoRE,'bo--',color=colori[kc],markersize=3, alpha=1.0,label= '
%0.1f' % po)
  ax1.plot(values_pm,dis,'bo--',color=colori[kc],markersize=3, alpha=1.0,label= '%0.1f' % po)
  #plt.plot(values_pm,sCoRE,'bo--',color=colori[kc],markersize=4, alpha=1.0,label= '
%0.2f' % po)
  #plt.plot(values_pm,dis,'bo--',color=colori[kc+1],markersize=3, alpha=1.0,label= '
%0.2f' % po)
  ax3.plot(dis,sCoRE,'bo',color=colori[kc],markersize=3, alpha=1.0,label= '%0.1f' % po)
  kc+=1
#--------------
ax3.grid(False)
#plt.legend(title='
',loc='upper left')  #---lower center, lower right, upper left
ax3.set_ylabel('
',fontsize=16)
#plt.ylabel('
',fontsize=16)
#plt.xlabel('
',fontsize=16)
ax3.set_xlabel('
',fontsize=16)
ax3.legend(title='
',loc='upper left')  #---lower center, lower right, upper left
#plt.title('Phase change') #'Bruce Centenary',
#plt.ylim(-0.05, 1.05)
#plt.show()

#------


# Plot the first graph on the first subplot
#ax1.set_xlabel('
',fontsize=16)
ax2.grid(True , which='both', axis='both', color='black', linewidth=0.2, linestyle='-')
ax2.set_ylabel('
',fontsize=16)
ax2.set_ylim(-0.01, 1.05)
#ax1.set_title('')

# Plot the second graph on the second subplot
ax1.grid(True , which='both', axis='both', color='black', linewidth=0.2, linestyle='-')
ax2.set_xlabel('
',fontsize=16)
ax1.set_ylabel('
',fontsize=16)
ax1.legend(title='
',loc='lower right')  #---lower center, lower right, upper left
ax1.set_ylim(0, 1.05)
#ax1.legend(title='
',  loc='upper left') #bbox_to_anchor=(1.05, 1),
#ax2.set_title('Graph 2')

# Remove the left spine of the original y-axis
#ax2.yaxis.set_visible(False)
# Adjust the layout and spacing
plt.tight_layout()
plt.show()

#---Monte Carlo

df,df_S = pd.DataFrame(),pd.DataFrame()
#--time
lectures2,sCr2=[],[]
N_sampling=50 #--Number of samples in the Montecarlo system
#---
M=4    #--M are of level, rows
N=3    #--N the conceptual attributes, columns
po=0.5 #---initial background
p_mt=0.93
n=16 #Number of students
number_of_lessons=10 #---cycles PI
#---
for i in range(N_sampling):
  discernment2,sCr2=[],[]
  #-------Reboot the system
  s_sys=[]
  for k in range(n):
    s_sys.append(student(M,N,po))
    SystemS=s_sys #--Fixed student system
#---xxxxxxxxxxxxxxxxxx--------------xxxxxxxxxxxxxxxxxxxxx--------------
  fis2=samplig(SystemS)
  Fis2=peer_instruction(SystemS)#--system
  #---
  for nl in range(number_of_lessons+1):
    sCr2.append(fis2.s_c()[0])
    discernment2.append(fis2.D())#--
    lectures2.append(nl)
    #-peer instruction
    for k in range(8):  #---quality of cognitive confrontation amplificator
      couples=combinations_without_repeats(n)
      for n_collisions in couples:
        r,s=n_collisions[0],n_collisions[1]
        Fis2.exchange(r,s) #--Peer Instruction
        Int_Inst(r)
  #---Dataframe
  column_name = f'D{i+1}'
  column_data = [i, i+1]
  # Asignar los valores a las columnas correspondientes
  df.loc[:, column_name] = discernment2
  df_S.loc[:, column_name] = sCr2
#----
# calculate the mean and standard deviation of each row
row_means = df.mean(axis=1)
row_stds = df.std(axis=1)
#--
row_means_S = df_S.mean(axis=1)
row_stds_S = df_S.std(axis=1)
# add the mean and standard deviation as new columns
df['Mean'] = row_means
df['Std'] = row_stds
df.reset_index(level=0, inplace=True)
#--
df_S['Mean_S'] = row_means_S
df_S['Std_S'] = row_stds_S
df_S.reset_index(level=0, inplace=True)

#----------
Delta=1
indices_to_keep = list(range(1,number_of_lessons+Delta, Delta))
# Select the desired rows using boolean indexing
reduced_df = df[df.index.isin(indices_to_keep)]
df=reduced_df
#--
reduced_df_S= df_S[df_S.index.isin(indices_to_keep)]
df_S=reduced_df_S
#-------------Experimental
row_list_average=[0.6064814814814815,0.6388888888888888,0.6481481481481481,0.7407407407407407,0.7731481481481481,0.888888888888889,0.837962962962963,0.8194444444444444,0.8657407407407407,0.9166666666666667]
row_list_std=[0.2607979033861433,0.2510153136763686,0.13112017824544941,0.13791842018537226,0.15647874203168596,0.09752305016361233,0.10199204717196483,0.10808640212647301,0.12162217086647249,0.05324409543913043]

#---
plt.style.use('default')
plt.figure(edgecolor='red') #navy,gray edgecolor facecolor='gray'

plt.errorbar(df['index'],df['Mean'], yerr=df['Std'],fmt='o', capsize=4 ,linestyle='none',label='Std_Model')
plt.plot(df['index'],df['Mean'],'--',color="black",markersize=4, alpha=0.5,label='Average_Model')

plt.errorbar(x=df['index'],y=row_list_average, yerr=row_list_std,fmt='o', capsize=4,linestyle='none',label='Std_Exp')
plt.plot(df['index'],row_list_average,'--',color="green",markersize=4, alpha=0.5,label='Average_Exp')

plt.grid(False)
plt.legend(loc='lower right')  #---lower center, lower right, upper left
plt.ylabel('
',fontsize=16)
plt.xlabel('Cycles PI',fontsize=16)
plt.ylim(0.2, 1.15)
plt.show()
#----
spr=row_list_average
teo=df['Mean'].tolist()
def v_True(spr,teo):
  v=0.0
  for k in range(len(spr)):
    e_relative=np.abs((spr[k]-teo[k])/teo[k])
    v=v+(1-e_relative)
  v=v/len(spr)
  return v
print("Truth value=",round(v_True(spr,teo),2))

\end{lstlisting}

\chapter{Instruments for the 6 PI cycles} \label{app:6test}
During this research, $6$ PI cycles were executed, and a similar instrument was applied in each one. These instruments are shown below.

\section{Cycle 1}
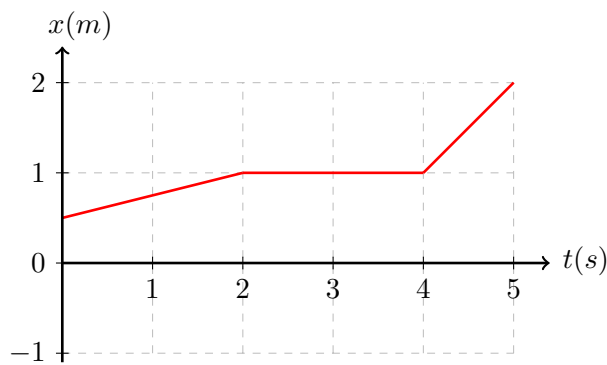
\begin{figure}[!htp]
	\begin{center}
			\begin{tikzpicture}[scale=1.2]
			 \draw[dashed,gray!50,step=1.0] (0,-1) grid (5,2);
			\draw [red,line width=1.0pt] (0cm,0.5cm) -- (2cm,1cm);
			\draw [red,line width=1.0pt] (2cm,1cm) -- (4cm,1cm);
			\draw [red,line width=1.0pt] (4cm,1cm) -- (5cm,2cm);				 
			\draw [->,line width=1.0pt] (0cm,0cm) -- (5.4cm,0cm);
			\draw [->,line width=1.0pt] (0cm,-1.1cm) -- (0cm,2.4cm);
			\coordinate [label=below:\textcolor{black} {$t(s)$}] (x) at  (5.8cm,0.3cm);
			\coordinate [label=below:\textcolor{black} {$x(m)$}] (x) at  (0.2cm,2.9cm);
			 \foreach \x/\xtext in {1/1, 2/2, 3/3, 4/4, 5/5}  
			 \draw[shift={(\x,0)}] (0pt,2pt)--(0pt,-2pt) node[below] {$\xtext$}; 
			 \foreach \y/\ytext in {-1/-1, 0/0, 1/1,2/2}  
			 \draw[shift={(0.0,\y)}] (2pt,0pt)--(-2pt,0pt) node[left] {$\ytext$}; 
			\end{tikzpicture}
	\caption{El sistema de referencia crece de izquierda a derecha y tiene como origen el extremo izquierdo 
	         del marco de la ventana.}
	\label{mur1}
	\end{center}
\end{figure}

\textbf{\textit{Responda todas las preguntas de acuerdo con esta situación}}:\\
Un gato camina por el marco horizontal de una ventana en una casa grande. Su movimiento se representa por 
la gráfica en la figura \ref{mur1}.  

\begin{enumerate}
   
    
 \item ($1pt$) Cuando se comienza a observar el movimiento de gato este se encuentra en la posición 
         
	\begin{multicols}{5}
		\begin{itemize}
			\item [A-)] $0,5m$.
			\item [B-)] $1m$. 
			\item [C-)] $1,5m$.
			\item [D-)] $2m$. 
			\item [E-)] $-0,5m$. 
		\end{itemize}
	\end{multicols}   

\item ($1pt$) Finalizando el  movimiento descrito por la gráfica, el gato se mueve
         
	\begin{multicols}{3}
		\begin{itemize}
			\item [A-)] hacia abajo.
			\item [B-)] horizontalmente.
			\item [C-)] hacia arriba.
			\item [D-)] en círculo.
			\item [E-)] oscilando.
		\end{itemize}
	\end{multicols}  
   
\item ($1pt$) El tiempo que le tarda al gato en alcanzar una posición de $1,5m$ es
         
	\begin{multicols}{5}
		\begin{itemize}
			\item [A-)] $5s$.
			\item [B-)] $4,5s$. 
			\item [C-)] $4s$.
			\item [D-)] $3,5s$. 
			\item [E-)] $3s$. 
		\end{itemize}
	\end{multicols}                                              		
\item ($2pt$) Según la gráfica,  cuando se deja de estudiar el movimiento del gato  su velocidad es

	\begin{multicols}{5}
		\begin{itemize}
			\item [A-)] $1m/s$.
			\item [B-)] $2m/s$. 
			\item [C-)] $1,5m/s$.
			\item [D-)] $0,25m/s$. 
			\item [E-)] $0m/s$. 
		\end{itemize}
	\end{multicols}   
\item ($2pt$) Durante los dos primeros segundos del movimiento, el gato se desplazó 

	\begin{multicols}{5}
		\begin{itemize}
			\item [A-)] $-0,5m$.
			\item [B-)] $1m$. 
			\item [C-)] $1,5m$.
			\item [D-)] $2m$. 
			\item [E-)] $0,5m$.
		\end{itemize}
	\end{multicols}   
\item ($2pt$) El tiempo que el gato permaneció en reposo fue 

	\begin{multicols}{5}
		\begin{itemize}
			\item [A-)] $5s$.
			\item [B-)] $4s$. 
			\item [C-)] $3s$.
			\item [D-)] $2s$. 
			\item [E-)] $0s$.
		\end{itemize}
	\end{multicols}  

\item ($4pt$) Justo cuando el tiempo es $4,5s$ el gato  

	\begin{multicols}{2}
		\begin{itemize}
			\item [A-)] se mueve hacia abajo.
			\item [B-)] se mueve a la derecha.
			\item [C-)] no se mueve.
			\item [D-)] se mueve en círculos.
			\item [E-)] se mueve a la izquierda.
		\end{itemize}
	\end{multicols}     
\item ($4pt$) El desplazamiento total del gato fue   

	\begin{multicols}{5}
		\begin{itemize}
			\item [A-)] $-0,5m$.
			\item [B-)] $1m$. 
			\item [C-)] $1,5m$.
			\item [D-)] $2m$. 
			\item [E-)] $1,25m$. 
		\end{itemize}
	\end{multicols}  
\item ($4pt$) Para responder esta pregunta tenga en cuenta el siguiente convenio:

\begin{center}
    \begin{tabular}{ | c | c | c |c | c |c |}
    \hline
    \textbf{Línea}       & \textbf{En $(+)$} & \textbf{En $(-)$} & \textbf{ De $(+)$ a $(-)$} &\textbf{De $(-)$ a $(+)$} & \textbf{En $0$}  \\ \hline
    \textit{Horizontal}      & $1$     & $2$      & $-------$   &$-------$   & $3$  \\ \hline
    \textit{Inclinada hacia arriba}      & $4$     & $5$     & $6$  &$ 10$              & $-----$  \\ \hline
    \textit{Inclinada hacia abajo}      & $7$      & $8$       & $9$  & $11$               & $-----$  \\ \hline
    \end{tabular}
\end{center}

La gráfica de velocidad contra tiempo que representa el movimiento del gato tiene la secuencia de líneas  

	\begin{multicols}{3}
		\begin{itemize}
			\item [A-)] $8\rightarrow 9\rightarrow 5$
			\item [B-)] $7\rightarrow 11\rightarrow 5$ 
			\item [C-)] $1\rightarrow 3\rightarrow 1$
			\item [D-)] $11\rightarrow 10\rightarrow 3$ 
			\item [E-)] $6\rightarrow 5\rightarrow 6$ 
		\end{itemize}
	\end{multicols}
\end{enumerate}

\section{Cycle 2}

\textbf{\textit{Responda todas las preguntas de acuerdo con esta situación}}:\\

Un atleta corre por una pista recta de izquierda a derecha con una rapidez constante de $6\frac{m}{s}$, 
transcurridos $10$ segundos escucha la orden de su entrenador y se devuelve inmediatamente por 
donde venía con una rapidez de $4\frac{m}{s}$, continúa con esta rapidez durante otros $15$ segundos en 
los cuales alcanza el final de la pista. Para el análisis del problema tome el sistema de 
referencia creciendo hacia la derecha y con origen en el extremo izquierdo de la pista.   


\begin{enumerate}
    
\item ($1pt$) Transcurridos $3s$ la velocidad del atleta es 
         
	\begin{multicols}{5}
		\begin{itemize}
			\item [A-)] $3\frac{m}{s}$.
			\item [B-)] $6\frac{m}{s}$. 
			\item [C-)] $-6\frac{m}{s}$.
			\item [D-)] $18\frac{m}{s}$. 
			\item [E-)] $1,8\frac{m}{s}$. 
		\end{itemize}
	\end{multicols}   

\item ($1pt$) Finalizando el  movimiento, el atleta se mueve
         
	\begin{multicols}{3}
		\begin{itemize}
			\item [A-)] hacia abajo.
			\item [B-)] a la derecha.
			\item [C-)] hacia arriba.
			\item [D-)] a la izquierda.
			\item [E-)] oscila.
		\end{itemize}
	\end{multicols}  
   
\item ($1pt$) Cuando la velocidad del atleta es $-4\frac{m}{s}$ ha transcurrido un tiempo de 
         
	\begin{multicols}{5}
		\begin{itemize}
			\item [A-)] $12s$.
			\item [B-)] $3,75s$. 
			\item [C-)] $-60s$.
			\item [D-)] $60s$. 
			\item [E-)] $2s$. 
		\end{itemize}
	\end{multicols}                                              		
\item ($2pt$) Entre el segundo $6$ y $8$ el atleta se desplazó
	\begin{multicols}{5}
		\begin{itemize}
			\item [A-)] $36m$.
			\item [B-)] $-12m$. 
			\item [C-)] $48m$.
			\item [D-)] $12m$. 
			\item [E-)] $30m$. 
		\end{itemize}
	\end{multicols}   
\item ($2pt$) Durante los dos últimos segundos del movimiento, el atleta se desplazó 

	\begin{multicols}{5}
		\begin{itemize}
			\item [A-)] $12m$.
			\item [B-)] $-12m$. 
			\item [C-)] $8m$.
			\item [D-)] $-8m$. 
			\item [E-)] $0m$.
		\end{itemize}
	\end{multicols}   
\item ($2pt$) El tiempo necesario para que el atleta se desplace $-16m$ es  

	\begin{multicols}{5}
		\begin{itemize}
			\item [A-)] $5s$.
			\item [B-)] $4s$. 
			\item [C-)] $3s$.
			\item [D-)] $2s$. 
			\item [E-)] $1s$.
		\end{itemize}
	\end{multicols}  

\item ($4pt$) La posición del atleta en un tiempo de $5s$ es  

	\begin{multicols}{2}
		\begin{itemize}
			\item [A-)] $36m$.
			\item [B-)] $-12m$. 
			\item [C-)] $48m$.
			\item [D-)] $12m$. 
			\item [E-)] $30m$. 
		\end{itemize}
	\end{multicols}     
\textbf{\textit{Responda las preguntas $8$ y $9$ de acuerdo con esta tabla}}:\\
\begin{center}
    \begin{tabular}{ | c | c | c |c | c |c |}
    \hline
    \textbf{Línea}       & \textbf{En $(+)$} & \textbf{En $(-)$} & \textbf{ De $(+)$ a $(-)$} &\textbf{De $(-)$ a $(+)$} & \textbf{En $0$}  \\ \hline
    \textit{Horizontal}      & $1$     & $2$      & $-------$   &$-------$   & $3$  \\ \hline
    \textit{Inclinada hacia arriba}      & $4$     & $5$     & $6$  &$10$              & $-----$  \\ \hline
    \textit{Inclinada hacia abajo}      & $7$      & $8$       & $9$  & $11$               & $-----$  \\ \hline
    \end{tabular}
\end{center}

\item ($4pt$) La gráfica de velocidad contra tiempo que representa el movimiento del atleta tiene la 
             secuencia de líneas  

	\begin{multicols}{3}
		\begin{itemize}
			\item [A-)] $2\rightarrow 1$
			\item [B-)] $7\rightarrow 5$ 
			\item [C-)] $1\rightarrow 2$
			\item [D-)] $9\rightarrow 3$ 
			\item [E-)] $1\rightarrow 3\rightarrow 2$ 
		\end{itemize}
	\end{multicols}  
\item ($4pt$) La gráfica de posición contra tiempo que representa el movimiento del atleta tiene la secuencia de líneas  

	\begin{multicols}{3}
		\begin{itemize}
			\item [A-)] $4\rightarrow 7$
			\item [B-)] $7\rightarrow 5$ 
			\item [C-)] $1\rightarrow 2$
			\item [D-)] $4\rightarrow 8$ 
			\item [E-)] $1\rightarrow 4$  
		\end{itemize}
	\end{multicols}
\end{enumerate}

\section{Cycle 3}

\textbf{\textit{Responda todas las preguntas de acuerdo con esta situación}}:\\

Un atleta realiza su entrenamiento en un parque a través de un sendero recto. Pasa por un puente horizontal sobre el riachuelo caminando hacia el norte con ritmo constante, luego  se detiene y después continua su recorrido por el mismo camino, como muestra la gráfica de velocidad contra tiempo en la figura \ref{mur3}.  

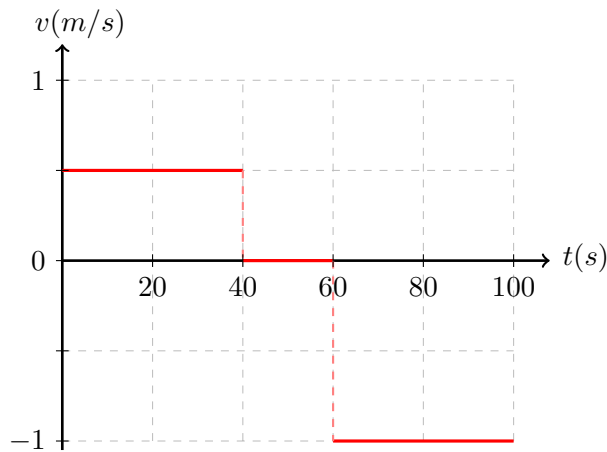
\begin{figure}[!htp]
	\begin{center}
			\begin{tikzpicture}[scale=1.2]
			 \draw[dashed,gray!50,step=1.0] (0,-2) grid (5,2);
			 \draw [->,line width=1.0pt] (0cm,0cm) -- (5.4cm,0cm);
			\draw [->,line width=1.0pt] (0cm,-2.1cm) -- (0cm,2.4cm);
			\draw [red,line width=1.2pt] (0cm,1cm) -- (2cm,1cm);
			\draw [red,line width=1.2pt] (2cm,0cm) -- (3cm,0cm);
			\draw [red,line width=1.2pt] (3cm,-2cm) -- (5cm,-2cm);
			\draw [dashed,red!50,line width=0.8pt] (2cm,1cm) -- (2cm,0cm);
			\draw [dashed,red!50,line width=0.8pt] (3cm,0cm) -- (3cm,-2cm);	
			\coordinate [label=below:\textcolor{black} {$t(s)$}] (x) at  (5.8cm,0.3cm);
			\coordinate [label=below:\textcolor{black} {$v(m/s)$}] (x) at  (0.2cm,2.9cm);
			 \foreach \x/\xtext in {1/20, 2/40, 3/60, 4/80, 5/100}  
			 \draw[shift={(\x,0)}] (0pt,2pt)--(0pt,-2pt) node[below] {$\xtext$}; 
			 \foreach \y/\ytext in {-2/-1,-1/, 0/0, 1/,2/1}  
			 \draw[shift={(0.0,\y)}] (2pt,0pt)--(-2pt,0pt) node[left] {$\ytext$}; 
			\end{tikzpicture}
	\caption{El sistema de referencia crece de sur a norte y tiene como origen el puente sobre el riachuelo.}
	\label{mur3}
	\end{center}
\end{figure}


\begin{enumerate}

    
\item ($1ps$) La velocidad del hombre a los $20s$ es 
         
	\begin{multicols}{5}
		\begin{itemize}
			\item [A-)] $10m/s$.
			\item [B-)] $-10m/s$. 
			\item [C-)] $0,5m/s$.
			\item [D-)] $-0,5m/s$. 
			\item [E-)] $0m/s$. 
		\end{itemize}
	\end{multicols}   

\item ($1ps$) El hombre está en reposo en el tiempo
         
	\begin{multicols}{3}
		\begin{itemize}
			\item [A-)] $50s$.
			\item [B-)] $80s$. 
			\item [C-)] $20s$.
			\item [D-)] $3,5s$. 
			\item [E-)] $90s$. 
		\end{itemize}
	\end{multicols}  
   
\item ($1ps$) El hombre
         
	\begin{multicols}{2}
		\begin{itemize}
			\item [A-)] camina más rápido a la derecha que a la izquierda.
			\item [B-)] camina más lento a la izquierda que a la derecha.
			\item [C-)] camina más de prisa de regreso que de ida. 
			\item [D-)] camina igual de rápido a la derecha que a la izquierda.
			\item [E-)] se mantiene más tiempo en reposo que caminando.
		\end{itemize}
	\end{multicols}                                              		
\item ($2ps$) El desplazamiento del hombre entre $60s$ y $80s$ es

	\begin{multicols}{5}
		\begin{itemize}
			\item [A-)] $-20m$.
			\item [B-)] $80m$. 
			\item [C-)] $1m$.
			\item [D-)] $20m$. 
			\item [E-)] $0m$.
		\end{itemize}
	\end{multicols}   

\item ($2ps$) El tiempo que el hombre duró en reposo fue  

	\begin{multicols}{5}
		\begin{itemize}
			\item [A-)] $50s$.
			\item [B-)] $80s$. 
			\item [C-)] $20s$.
			\item [D-)] $100s$. 
			\item [E-)] $60s$. 
		\end{itemize}
	\end{multicols}
	
\item ($2ps$) El hombre pasa sobre el riachuelo cuando el observador activa el cronómetro para registrar el tiempo. De acuerdo con la gráfica \ref{mur3} al final de su paseo el hombre está  

	\begin{multicols}{2}
		\begin{itemize}
			\item [A-)] al norte del riachuelo.
			\item [B-)] al sur del riachuelo. 
			\item [C-)] justo sobre el riachuelo.
		\end{itemize}
	\end{multicols}  
\item ($4ps$) La posición final del hombre es
 
	\begin{multicols}{5}
		\begin{itemize}
			\item [A-)] $20m$.
			\item [B-)] $-20m$.
			\item [C-)] $60m$.
			\item [D-)] $0,5m$.
			\item [E-)] $-60m$.
		\end{itemize}
	\end{multicols}  
	
\item ($4ps$) El intervalo de tiempo en el cual el hombre tuvo un desplazamiento nulo fue  entre 

	\begin{multicols}{5}
		\begin{itemize}
			\item [A-)] $20s$ y $50s$.
			\item [B-)] $60s$ y $100s$.
			\item [C-)] $20s$ y $100s$.
			\item [D-)] $30s$ y $70s$.
			\item [E-)] $20s$ y $70s$.
		\end{itemize}
	\end{multicols}  
\item ($4ps$) Para responder esta pregunta tenga en cuenta el siguiente convenio:

\begin{center}
    \begin{tabular}{ | c | c | c |c | c |c |}
    \hline
    \textbf{Línea}       & \textbf{En $(+)$} & \textbf{En $(-)$} & \textbf{ De $(+)$ a $(-)$} &\textbf{De $(-)$ a $(+)$} & \textbf{En $0$}  \\ \hline
    \textit{Horizontal}      & $1$     & $2$      & $-------$   &$-------$   & $3$  \\ \hline
    \textit{Inclinada hacia arriba}      & $4$     & $5$     & $6$  &$10$              & $-----$  \\ \hline
    \textit{Inclinada hacia abajo}      & $7$      & $8$       & $9$  & $11$               & $-----$  \\ \hline
    \end{tabular}
\end{center}

La gráfica de posición contra tiempo que representa el movimiento del hombre tiene la secuencia de líneas  

	\begin{multicols}{3}
		\begin{itemize}
			\item [A-)] $4\rightarrow 1\rightarrow 1$
			\item [B-)] $7\rightarrow 11\rightarrow 5$ 
			\item [C-)] $4\rightarrow 1\rightarrow 9$
			\item [D-)] $4\rightarrow 11\rightarrow 3$ 
			\item [E-)] $6\rightarrow 4\rightarrow 6$ 
		\end{itemize}
	\end{multicols}

\end{enumerate}

\section{Cycle 4}

\textit{\textbf{Responda todas las preguntas de acuerdo con esta situación:}}\\
Un motociclista realiza una rodada por una carretera recta. Su movimiento se representa por la gráfica abajo mostrada. 
Responda todas las preguntas de acuerdo con esta situación.  

\begin{center}
			\begin{tikzpicture}[scale=1.2]
			 \draw[dashed,gray!50,step=1.0] (0,-2) grid (8,4);
			\draw [red,line width=1.0pt] (0cm,0cm) -- (2cm,4cm);
			\draw [red,line width=1.0pt] (2cm,4cm) -- (3cm,4cm);
			\draw [red,line width=1.0pt] (3cm,4cm) -- (6cm,-2cm);				 
			\draw [->,line width=1.0pt] (0cm,0cm) -- (8.4cm,0cm);
			\draw [->,line width=1.0pt] (0cm,-2.1cm) -- (0cm,4.4cm);
			\coordinate [label=below:\textcolor{black} {$t(h)$}] (x) at  (8.8cm,0.3cm);
			\coordinate [label=below:\textcolor{black} {$x(km)$}] (x) at  (0.2cm,5.0cm);
			 \foreach \x/\xtext in {1/, 2/1, 3/, 4/2, 5/, 6/3, 7/, 8/4 }  
			 \draw[shift={(\x,0)}] (0pt,2pt)--(0pt,-2pt) node[below] {$\xtext$}; 
			 \foreach \y/\ytext in { -2/-50, -1/-25, 0/0, 1/25, 2/50, 3/75, 4/100}  
			 \draw[shift={(0.0,\y)}] (2pt,0pt)--(-2pt,0pt) node[left] {$\ytext$}; 
			\end{tikzpicture}\\
			Figrua 1: El sistema de referencia crece de izquierda a derecha y tiene como origen un punto sobre la carretera.
\end{center}


\begin{enumerate}
    
    
\item ($1ps$) Cuando ha pasado un tiempo de hora y media la posición del motociclista es 
         
	\begin{multicols}{5}
		\begin{itemize}
			\item [A-)] $0 km$.
			\item [B-)] $100 km$. 
			\item [C-)] $150 km$.
			\item [D-)] $50 km$. 
			\item [E-)] $- 50 km$. 
		\end{itemize}
	\end{multicols}   

\item ($1ps$) Al cabo de una hora y 20 minutos se puede decir que el motociclista  
         
	\begin{multicols}{3}
		\begin{itemize}
			\item [A-)] está detenido.
			\item [B-)] se está regresando.
			\item [C-)] avanza con velocidad constante.
			\item [D-)] se mueve en círculo.
			\item [E-)] se encuentra oscilando.
		\end{itemize}
	\end{multicols}  
   
\item ($1ps$) Cuando transcurrieron justamente dos horas y media de movimiento el motociclista 
         
	\begin{multicols}{2}
		\begin{itemize}
			\item [A-)] se detuvo.
			\item [B-)] empezó a retroceder . 
			\item [C-)] cambió su velocidad.
			\item [D-)] pasó por el punto de partida.
			\item [E-)] hizo un círculo. 
		\end{itemize}
	\end{multicols}                                              		
\item ($1ps$) Según la gráfica, el desplazamiento del motociclista las primeras dos horas de movimiento es

	\begin{multicols}{5}
		\begin{itemize}
			\item [A-)] $100 km$.
			\item [B-)] $250 km$. 
			\item [C-)] $50 km$.
			\item [D-)] $150 km$. 
			\item [E-)] $-50 km$. 
		\end{itemize}
	\end{multicols}   
\item ($2ps$) La velocidad que tiene el motociclista justo a las 2,5 horas de iniciar su movimiento es

	\begin{multicols}{5}
		\begin{itemize}
			\item [A-)] $0 km/h$.
			\item [B-)] $100 km/h$. 
			\item [C-)] $-100km/h$.
			\item [D-)] $50 km/h$. 
			\item [E-)] $-50 km/h$.
		\end{itemize}
	\end{multicols}   
\item ($2ps$) El tiempo que el motociclista permanece en reposo es 

	\begin{multicols}{5}
		\begin{itemize}
			\item [A-)] $1.5h$.
			\item [B-)] $2.5h$. 
			\item [C-)] $0h$.
			\item [D-)] $0.5h$. 
			\item [E-)] $3h$.
		\end{itemize}
	\end{multicols}  

\item ($4ps$)  Según la gráfica, el desplazamiento total del motociclista es 

	\begin{multicols}{5}
		\begin{itemize}
			\item [A-)] $-100 km$.
			\item [B-)] $150 km$. 
			\item [C-)] $50 km$.
			\item [D-)] $250 km$. 
			\item [E-)] $-50 km$. 
		\end{itemize}
	\end{multicols} 
\item ($4ps$)  La velocidad que tiene el motociclista justo a los treinta minutos de iniciar el movimiento es   

	\begin{multicols}{5}
		\begin{itemize}
			\item [A-)] $100 km/h$.
			\item [B-)] $75 km/h$. 
			\item [C-)] $-100km/h$.
			\item [D-)] $50 km/h$. 
			\item [E-)] $-50 km/h$.
		\end{itemize}
	\end{multicols}  
\item ($4ps$)  Para responder esta pregunta tenga en cuenta el siguiente convenio:

\begin{center}
    \begin{tabular}{ | c | c | c |c | c |c |}
    \hline
    \textbf{Línea}       & \textbf{En $(+)$} & \textbf{En $(-)$} & \textbf{ De $(+)$ a $(-)$} &\textbf{De $(-)$ a $(+)$} & \textbf{En $0$}  \\ \hline
    \textit{Horizontal}      & $1$     & $2$      & $-------$   &$-------$   & $3$  \\ \hline
    \textit{Inclinada hacia arriba}      & $4$     & $5$     & $6$  &$10$              & $-----$  \\ \hline
    \textit{Inclinada hacia abajo}      & $7$      & $8$       & $9$  & $11$               & $-----$  \\ \hline
    \end{tabular}
\end{center}

La gráfica de velocidad contra tiempo que representa el movimiento del motociclista tiene la secuencia de líneas  

	\begin{multicols}{3}
		\begin{itemize}
			\item [A-)] $8\rightarrow 9\rightarrow 5$
			\item [B-)] $1\rightarrow 3 \rightarrow 1$ 
			\item [C-)] $1\rightarrow 3\rightarrow 2$
			\item [D-)] $1\rightarrow 2\rightarrow 3$ 
			\item [E-)] $6\rightarrow 5\rightarrow 6$ 
		\end{itemize}
	\end{multicols}
\end{enumerate}

\section{Cycle 5}

\textbf{\textit{Responda todas las preguntas de acuerdo con esta situación}}:\\

En una carrera de $100m$ lisos un atleta al escuchar el disparo se queda quieto
durante 
$2s$, luego parte súbitamente recorriendo $50m$ en $5s$, súbitamente
cambia su rapidez y finaliza al pasar por la meta justo $8s$ después. Para el análisis del problema tome el sistema de 
referencia creciendo hacia la derecha y con origen en el extremo izquierdo de la pista.    


\begin{enumerate}
    
\item $(1ps)$  Lo que se desplaza el atleta los últimos $8s$ de movimiento es
         
	\begin{multicols}{5}
		\begin{itemize}
			\item [A-)] $50m$.
			\item [B-)] $100m$. 
			\item [C-)] $-50m$.
			\item [D-)] $25m$. 
			\item [E-)] $-20m$. 
		\end{itemize}
	\end{multicols}   

\item $(1ps)$ Finalizando el  movimiento, el atleta se mueve
         
	\begin{multicols}{3}
		\begin{itemize}
			\item [A-)] hacia abajo.
			\item [B-)] a la derecha.
			\item [C-)] hacia arriba.
			\item [D-)] a la izquierda.
			\item [E-)] oscila.
		\end{itemize}
	\end{multicols}  
   
\item $(1ps)$ Cuando el atleta se ha desplazado los primeros $50m$ ha transcurrido un tiempo de
         
	\begin{multicols}{5}
		\begin{itemize}
			\item [A-)] $2s$.
			\item [B-)] $5s$. 
			\item [C-)] $7s$.
			\item [D-)] $15s$. 
			\item [E-)] $10s$. 
		\end{itemize}
	\end{multicols}                                              		
\item $(2ps)$ La velocidad que lleva el atleta a los 5s de sonar el disparo de salida es
	\begin{multicols}{5}
		\begin{itemize}
			\item [A-)] $7,1\frac{m}{s}$.
			\item [B-)] $-10\frac{m}{s}$.
			\item [C-)] $6,7\frac{m}{s}$.
			\item [D-)] $10\frac{m}{s}$.
			\item [E-)] $-5\frac{m}{s}$. 
		\end{itemize}
	\end{multicols}   
\item $(2ps)$ La velocidad que lleva el atleta a los 10s de sonar el disparo de salida es

	\begin{multicols}{5}
		\begin{itemize}
			\item [A-)] $6,25\frac{m}{s}$.
			\item [B-)] $-10\frac{m}{s}$.
			\item [C-)] $-6,25\frac{m}{s}$.
			\item [D-)] $10\frac{m}{s}$.
			\item [E-)] $-5\frac{m}{s}$. 
		\end{itemize}
	\end{multicols}   
\item $(2ps)$ El tiempo que marca el reloj cuando el atleta atraviesa la meta es  

	\begin{multicols}{5}
		\begin{itemize}
			\item [A-)] $5s$.
			\item [B-)] $8s$. 
			\item [C-)] $13s$.
			\item [D-)] $15s$. 
			\item [E-)] $10s$.
		\end{itemize}
	\end{multicols}  

\item $(4ps)$ La posición del atleta cuando ha transcurrido un tiempo de $11s$ a partir del disparo de salida es  

	\begin{multicols}{2}
		\begin{itemize}
			\item [A-)] $31,25m$.
			\item [B-)] $100m$. 
			\item [C-)] $25m$.
			\item [D-)] $81,25m$. 
			\item [E-)] $75m$. 
		\end{itemize}
	\end{multicols}     
\textbf{\textit{Responda las preguntas $8$ y $9$ de acuerdo con esta tabla}}:\\
\begin{center}
    \begin{tabular}{ | c | c | c |c | c |c |}
    \hline
    \textbf{Línea}       & \textbf{En $(+)$} & \textbf{En $(-)$} & \textbf{ De $(+)$ a $(-)$} &\textbf{De $(-)$ a $(+)$} & \textbf{En $0$}  \\ \hline
    \textit{Horizontal}      & $1$     & $2$      & $-------$   &$-------$   & $3$  \\ \hline
    \textit{Inclinada hacia arriba}      & $4$     & $5$     & $6$  &$10$              & $-----$  \\ \hline
    \textit{Inclinada hacia abajo}      & $7$      & $8$       & $9$  & $11$               & $-----$  \\ \hline
    \end{tabular}
\end{center}

\item $(4ps)$ La gráfica de velocidad contra tiempo que representa el movimiento del atleta tiene la 
             secuencia de líneas  

	\begin{multicols}{3}
		\begin{itemize}
			\item [A-)] $2\rightarrow 1$
			\item [B-)] $7\rightarrow 5$ 
			\item [C-)] $1\rightarrow 2$
			\item [D-)] $9\rightarrow 3$ 
			\item [E-)] $1\rightarrow 1$ 
		\end{itemize}
	\end{multicols}  
\item $(4ps)$ La gráfica de posición contra tiempo que representa el movimiento del atleta tiene la secuencia de líneas  

	\begin{multicols}{3}
		\begin{itemize}
			\item [A-)] $4\rightarrow 7$
			\item [B-)] $5\rightarrow 5$ 
			\item [C-)] $1\rightarrow 2$
			\item [D-)] $4\rightarrow 4$ 
			\item [E-)] $1\rightarrow 4$  
		\end{itemize}
	\end{multicols}

\end{enumerate}

\section{Cycle 6}

\textbf{\textit{Responda todas las preguntas de acuerdo con esta situación}}:\\

Un ciclista sale de su casa y realiza un recorrido a través de un sendero recto. Llega al parque y allí se detiene para luego continuar por el mismo sendero, como muestra la gráfica de velocidad contra tiempo en la figura \ref{mur3}.  

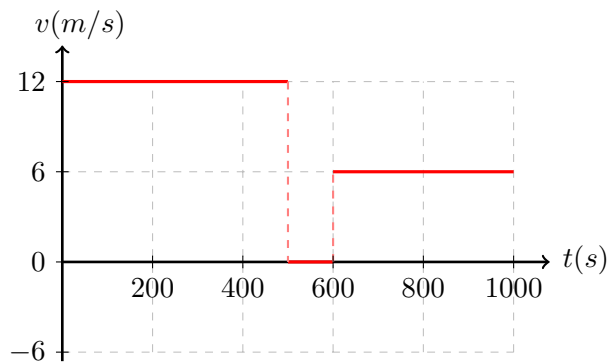
\begin{figure}[!htp]
	\begin{center}
			\begin{tikzpicture}[scale=1.2]
			 \draw[dashed,gray!50,step=1.0] (0,-1) grid (5,2);
			 \draw [->,line width=1.0pt] (0cm,0cm) -- (5.4cm,0cm);
			\draw [->,line width=1.0pt] (0cm,-1.1cm) -- (0cm,2.4cm);
			\draw [red,line width=1.2pt] (0cm,2cm) -- (2.5cm,2cm);
			\draw [red,line width=1.2pt] (2.5cm,0cm) -- (3cm,0cm);
			\draw [red,line width=1.2pt] (3cm,1cm) -- (5cm,1cm);
			\draw [dashed,red!50,line width=0.8pt] (2.5cm,2cm) -- (2.5cm,0cm);
			\draw [dashed,red!50,line width=0.8pt] (3cm,0cm) -- (3cm,1cm);	
			\coordinate [label=below:\textcolor{black} {$t(s)$}] (x) at  (5.8cm,0.3cm);
			\coordinate [label=below:\textcolor{black} {$v(m/s)$}] (x) at  (0.2cm,2.9cm);
			 \foreach \x/\xtext in {1/200, 2/400, 3/600, 4/800, 5/1000}  
			 \draw[shift={(\x,0)}] (0pt,2pt)--(0pt,-2pt) node[below] {$\xtext$}; 
			 \foreach \y/\ytext in {-1/-6, 0/0, 1/6,2/12}  
			 \draw[shift={(0.0,\y)}] (2pt,0pt)--(-2pt,0pt) node[left] {$\ytext$}; 
			\end{tikzpicture}
	\caption{El sistema de referencia crece de sur a norte y tiene como origen la casa del ciclista.}
	\label{mur3}
	\end{center}
\end{figure}


\begin{enumerate}
    
    
\item ($1ps$) La velocidad del ciclista a los $300s$ es 
         
	\begin{multicols}{5}
		\begin{itemize}
			\item [A-)] $12m/s$.
			\item [B-)] $-12m/s$. 
			\item [C-)] $0m/s$.
			\item [D-)] $6m/s$. 
			\item [E-)] $-6m/s$. 
		\end{itemize}
	\end{multicols}   

\item ($1ps$) El ciclista está en reposo en el tiempo
         
	\begin{multicols}{3}
		\begin{itemize}
			\item [A-)] $250s$.
			\item [B-)] $850s$. 
			\item [C-)] $550s$.
			\item [D-)] $0s$. 
			\item [E-)] $1000s$. 
		\end{itemize}
	\end{multicols}  
   
\item ($1ps$) El ciclista dura detenido
         
	\begin{multicols}{2}
		\begin{itemize}
			\item [A-)] $450s$.
			\item [B-)] $50s$. 
			\item [C-)] $600s$.
			\item [D-)] $0s$. 
			\item [E-)] $100s$. 
		\end{itemize}
	\end{multicols}                                              		
\item ($2ps$) El desplazamiento del ciclista desde que inicia hasta que llega al parque es

	\begin{multicols}{5}
		\begin{itemize}
			\item [A-)] $-6000m$.
			\item [B-)] $2400m$. 
			\item [C-)] $6000m$.
			\item [D-)] $4800m$. 
			\item [E-)] $-48000m$.
		\end{itemize}
	\end{multicols}   

\item ($2ps$) El desplazamiento del ciclista una vez retoma su movimiento después del descanso es  

	\begin{multicols}{5}
		\begin{itemize}
			\item [A-)] $-6000m$.
			\item [B-)] $2400m$. 
			\item [C-)] $6000m$.
			\item [D-)] $4800m$. 
			\item [E-)] $-24000m$.
		\end{itemize}
	\end{multicols}
	
\item ($2ps$) La posición del ciclista al cabo de $300s$ es 

	\begin{multicols}{5}
		\begin{itemize}
			\item [A-)] $-8400m$.
			\item [B-)] $8400m$. 
			\item [C-)] $3600m$.
			\item [D-)] $4800m$. 
			\item [E-)] $-3600m$.
		\end{itemize}
	\end{multicols}
\item ($4ps$) El desplazamiento total del ciclista  es  

	\begin{multicols}{5}
		\begin{itemize}
			\item [A-)] $-8400m$.
			\item [B-)] $8400m$. 
			\item [C-)] $3600m$.
			\item [D-)] $4800m$. 
			\item [E-)] $-3600m$.
		\end{itemize}
	\end{multicols}
	
\item ($4ps$) La posición del ciclista a los $800s$ es 

	\begin{multicols}{5}
		\begin{itemize}
			\item [A-)] $-6000m$.
			\item [B-)] $1200m$. 
			\item [C-)] $7200m$.
			\item [D-)] $4800m$. 
			\item [E-)] $-1200m$.
		\end{itemize}
	\end{multicols}  
\item ($4ps$) Para responder esta pregunta tenga en cuenta el siguiente convenio:

\begin{center}
    \begin{tabular}{ | c | c | c |c | c |c |}
    \hline
    \textbf{Línea}       & \textbf{En $(+)$} & \textbf{En $(-)$} & \textbf{ De $(+)$ a $(-)$} &\textbf{De $(-)$ a $(+)$} & \textbf{En $0$}  \\ \hline
    \textit{Horizontal}      & $1$     & $2$      & $-------$   &$-------$   & $3$  \\ \hline
    \textit{Inclinada hacia arriba}      & $4$     & $5$     & $6$  &$10$              & $-----$  \\ \hline
    \textit{Inclinada hacia abajo}      & $7$      & $8$       & $9$  & $11$               & $-----$  \\ \hline
    \end{tabular}
\end{center}

La gráfica de posición contra tiempo que representa el movimiento del hombre tiene la secuencia de líneas  

	\begin{multicols}{3}
		\begin{itemize}
			\item [A-)] $4\rightarrow 1\rightarrow 4$
			\item [B-)] $7\rightarrow 4\rightarrow 4$ 
			\item [C-)] $4\rightarrow 0\rightarrow 4$
			\item [D-)] $4\rightarrow 11\rightarrow 4$ 
			\item [E-)] $6\rightarrow 4\rightarrow 4$ 
		\end{itemize}
	\end{multicols}
 
\end{enumerate}
\chapter{Data from the 6 PI cycles} \label{app:data_PI}

Below are the data taken in the field that correspond to the performance on App.(\ref{app:6test}) instruments in his solitary reflection and then after the discussion with a peer.

\begin{table}
\centering

\begin{tabular}{|c|c|c|c|c|c|c|c|c|c|c|c|c|}
\hline
$N$& Date & id & test & $a_0$ & $a_1$ & $a_2$ & $a_3$ & $a_4$ & $a_5$ & $a_6$ & $a_7$ & $a_8$ \\
\hline
1 & 8/8/2022 9:18:44 & bjimenezb1\_XXXX & 0 & A & C & B & D & B & D & B & D & A \\
2 & 8/8/2022 12:18:38 & jmurillot\_XXXX & 0 & A & C & B & A & E & D & B & C & C \\
3 & 8/8/2022 14:45:24 & jramirezm14\_XXXX & 0 & A & C & B & B & E & D & E & C & C \\
4 & 8/8/2022 14:46:25 & sgiraldoa\_XXXX & 0 & A & C & B & B & E & D & E & C & C \\
5 & 8/8/2022 14:46:31 & mfonsecai\_XXXX & 0 & A & C & B & B & E & D & E & C & C \\
6 & 8/8/2022 16:31:47 & sriverosr1\_XXXX & 0 & A & B & B & B & E & D & B & C & C \\
7 & 8/8/2022 17:12:21 & etrivinor\_XXXX & 0 & A & B & B & A & E & D & B & C & C \\
8 & 8/8/2022 17:44:02 & gmorenov1\_XXXX & 0 & A & B & B & E & E & D & B & C & B \\
9 & 8/8/2022 17:57:18 & cmendezp1\_XXXX & 0 & A & B & B & A & E & D & B & C & C \\
10 & 8/8/2022 18:00:01 & mtorresc7\_XXXX & 0 & A & B & B & A & E & D & B & C & D \\
11 & 8/8/2022 18:45:03 & gcastillov\_XXXX & 0 & A & B & B & A & E & D & B & C & C \\
12 & 8/8/2022 18:59:26 & mgomezv3\_XXXX & 0 & A & C & B & A & E & D & B & C & C \\
13 & 8/8/2022 19:33:09 & scanor\_XXXX & 0 & A & B & B & D & E & D & B & C & C \\
14 & 8/8/2022 19:47:29 & soliverosc\_XXXX & 0 & A & B & B & A & E & C & B & C & C \\
15 & 8/8/2022 21:07:34 & agarciac5\_XXXX & 0 & A & C & B & B & E & D & B & C & C \\
16 & 8/8/2022 21:36:02 & kleond\_XXXX & 0 & A & C & B & D & E & D & B & C & C \\
17 & 8/8/2022 21:39:01 & msuarezo\_XXXX & 0 & A & C & B & D & E & D & B & C & C \\
18 & 8/8/2022 23:32:47 & aflorezr2\_XXXX & 0 & A & C & B & B & E & D & B & D & C \\
\hline
\end{tabular}
\caption{Individual responses in cycle $1$.}
\end{table}


\begin{table}
\centering
\begin{tabular}{|c|c|c|c|c|c|c|c|c|c|c|c|c|}
\hline
$N$& Date & id & test & $a_0$ & $a_1$ & $a_2$ & $a_3$ & $a_4$ & $a_5$ & $a_6$ & $a_7$ & $a_8$ \\
\hline
19 & 9/8/2022 9:55:36 & mfonsecai\_XXXX & 1 & A & D & E & B & D & A & D & C & A \\
20 & 9/8/2022 9:56:23 & sgiraldoa\_XXXX & 1 & A & D & A & D & D & A & D & E & A \\
21 & 9/8/2022 9:56:35 & scanor\_XXXX & 1 & B & D & A & D & D & B & E & C & A \\
22 & 9/8/2022 9:56:41 & etrivinor\_XXXX & 1 & B & D & A & D & D & B & E & C & A \\
23 & 9/8/2022 9:57:14 & jramirezm14\_XXXX & 1 & B & D & A & D & C & B & E & A & A \\
24 & 9/8/2022 9:57:19 & jmurillot\_XXXX & 1 & B & D & A & D & D & B & E & C & A \\
25 & 9/8/2022 9:57:23 & agarciac5\_XXXX & 1 & B & B & E & A & C & D & B & C & A \\
26 & 9/8/2022 9:57:41 & soliverosc\_XXXX & 1 & B & D & A & D & C & C & A & D & B \\
27 & 9/8/2022 9:57:41 & bjimenezb1\_XXXX & 1 & B & D & A & D & D & B & E & C & A \\
28 & 9/8/2022 9:57:56 & gmorenov1\_XXXX & 1 & B & B & A & D & D & B & E & C & B \\
29 & 9/8/2022 9:58:04 & kleond\_XXXX & 1 & B & D & A & D & D & B & E & A & D \\
30 & 9/8/2022 9:58:13 & aflorezr2\_XXXX & 1 & B & D & D & D & B & B & E & E & B \\
31 & 9/8/2022 9:58:17 & cmendezp1\_XXXX & 1 & B & D & A & D & D & B & E & C & A \\
32 & 9/8/2022 9:58:17 & gcastillov\_XXXX & 1 & B & D & A & D & D & B & E & C & A \\
33 & 9/8/2022 9:58:41 & mtorresc7\_XXXX & 1 & B & B & A & D & C & B & E & E & C \\
34 & 9/8/2022 9:59:21 & sriverosr1\_XXXX & 1 & B & D & A & D & C & B & E & E & E \\
35 & 9/8/2022 10:16:31 & scanor\_XXXX & 1 & B & D & A & D & C & B & E & D & A \\
\hline
\end{tabular}
\caption{Responses after discussion with a peer in cycle $1$.}
\end{table}
\begin{table}
\centering

\begin{tabular}{|c|c|c|c|c|c|c|c|c|c|c|c|c|}
\hline
$N$& Date & id & test & $a_0$ & $a_1$ & $a_2$ & $a_3$ & $a_4$ & $a_5$ & $a_6$ & $a_7$ & $a_8$ \\
\hline
36 & 9/8/2022 10:16:35 & gmorenov1\_XXXX & 2 & B & B & A & D & D & B & E & C & A \\
37 & 9/8/2022 10:16:36 & cmendezp1\_XXXX & 2 & B & D & A & D & D & B & E & C & A \\
38 & 9/8/2022 10:16:37 & soliverosc\_XXXX & 2 & B & D & A & D & C & B & E & C & A \\
39 & 9/8/2022 10:17:19 & jmurillot\_XXXX & 2 & B & D & A & D & D & B & E & C & A \\
40 & 9/8/2022 10:18:16 & bjimenezb1\_XXXX & 2 & B & D & A & D & D & B & E & C & A \\
41 & 9/8/2022 10:18:21 & mtorresc7\_XXXX & 2 & B & B & A & D & C & B & E & C & A \\
42 & 9/8/2022 10:18:51 & mgomezv3\_XXXX & 2 & B & D & A & D & D & B & E & C & A \\
43 & 9/8/2022 10:20:38 & etrivinor\_XXXX & 2 & B & D & A & D & D & B & E & C & A \\
44 & 9/8/2022 10:21:11 & gcastillov\_XXXX & 2 & B & D & A & D & D & B & E & C & A \\
45 & 9/8/2022 10:21:26 & kleond\_XXXX & 2 & B & D & A & D & D & B & E & C & D \\
46 & 9/8/2022 10:21:29 & jramirezm14\_XXXX & 2 & B & D & A & D & C & B & D & C & A \\
47 & 9/8/2022 10:21:31 & sriverosr1\_XXXX & 2 & B & D & A & D & D & B & E & C & A \\
48 & 9/8/2022 10:21:34 & sgiraldoa\_XXXX & 2 & B & D & A & D & C & B & D & C & A \\
49 & 9/8/2022 10:22:01 & mfonsecai\_XXXX & 2 & B & D & E & A & C & B & B & C & A \\
50 & 9/8/2022 10:22:25 & msuarezo\_XXXX & 2 & B & D & A & D & D & B & E & C & D \\
51 & 9/8/2022 10:22:37 & agarciac5\_XXXX & 2 & B & D & E & A & C & B & D & C & A \\
52 & 9/8/2022 10:26:25 & mgomezv3\_XXXX & 2 & B & D & A & D & D & B & E & C & A \\
53 & 9/8/2022 10:27:16 & aflorezr2\_XXXX & 2 & B & D & D & D & B & B & E & C & A \\
\hline
\end{tabular}
\caption{Individual responses in cycle $2$.}
\label{tab:my-tablevc}
\end{table}


\begin{table}
\centering
\begin{tabular}{|c|c|c|c|c|c|c|c|c|c|c|c|c|}
\hline
N & Date & id & test & $a_0$ & $a_1$ & $a_2$ & $a_3$ & $a_4$ & $a_5$ & $a_6$ & $a_7$ & $a_8$ \\
\hline
54 & 10/8/2022 7:58:01 & scanor\_XXXX & 3 & C & A & C & A & C & B & B & D & C \\
55 & 10/8/2022 9:20:33 & bjimenezb1\_XXXX & 3 & C & A & C & A & C & C & A & D & C \\
56 & 10/8/2022 12:57:23 & kleond\_XXXX & 3 & C & A & C & A & C & B & B & E & C \\
57 & 10/8/2022 14:20:27 & jmurillot\_XXXX & 3 & C & A & C & A & C & B & B & E & C \\
58 & 10/8/2022 14:39:32 & mgomezv3\_XXXX & 3 & C & A & C & A & C & B & B & E & C \\
59 & 10/8/2022 16:15:38 & gmorenov1\_XXXX & 3 & C & A & C & A & C & B & B & D & C \\
60 & 10/8/2022 18:50:41 & cmendezp1\_XXXX & 3 & C & A & C & A & C & B & B & E & C \\
61 & 10/8/2022 19:48:40 & mtorresc7\_XXXX & 3 & C & C & C & D & C & B & E & D & C \\
62 & 10/8/2022 20:08:15 & etrivinor\_XXXX & 3 & C & A & C & D & C & B & B & B & C \\
63 & 10/8/2022 21:15:44 & msuarezo\_XXXX & 3 & C & A & C & A & C & B & B & E & C \\
64 & 10/8/2022 21:33:29 & mfonsecai\_XXXX & 3 & C & C & C & D & C & C & A & C & C \\
65 & 10/8/2022 21:33:54 & sgiraldoa\_XXXX & 3 & C & C & C & D & C & C & A & C & C \\
66 & 10/8/2022 21:35:13 & agarciac5\_XXXX & 3 & C & C & C & D & C & C & A & C & C \\
67 & 10/8/2022 22:24:09 & jramirezm14\_XXXX & 3 & C & C & C & D & C & C & A & C & C \\
68 & 10/8/2022 22:24:09 & jramirezm14\_XXXX & 3 & C & C & C & D & C & C & A & C & C \\
69 & 10/8/2022 22:55:43 & aflorezr2\_XXXX & 3 & A & A & D & A & C & B & D & B & C \\
70 & 10/8/2022 23:41:38 & gcastillov\_XXXX & 3 & C & A & C & A & C & A & B & E & C \\
71 & 10/8/2022 0:00:00 & sriverosr1\_XXXX & 3 & C & A & C & D & C & C & D & B & C \\
\hline
\end{tabular}
\caption{Responses after discussion with a peer in cycle $2$.}
\end{table}

\begin{table}[htbp]
\centering
\begin{tabular}{|c|c|c|c|c|c|c|c|c|c|c|c|c|}
\hline
N & Date & id & test & $a_0$ & $a_1$ & $a_2$ & $a_3$ & $a_4$ & $a_5$ & $a_6$ & $a_7$ & $a_8$ \\
\hline
72 & 16/8/2022 9:27:48 & jmurillot\_XXXX & 4 & B & A & D & C & C & D & E & A & C \\
73 & 16/8/2022 9:29:35 & cmendezp1\_XXXX & 4 & B & A & D & C & C & D & B & A & C \\
74 & 16/8/2022 9:30:11 & gmorenov1\_XXXX & 4 & B & A & D & C & A & D & E & D & C \\
75 & 16/8/2022 9:30:12 & mfonsecai\_XXXX & 4 & B & C & D & C & C & A & E & D & C \\
76 & 16/8/2022 9:30:20 & soliverosc\_XXXX & 4 & B & A & D & C & A & D & D & B & C \\
77 & 16/8/2022 9:30:21 & sriverosr1\_XXXX & 4 & B & A & D & D & E & D & D & D & E \\
78 & 16/8/2022 9:30:50 & sgiraldoa\_XXXX & 4 & B & C & C & D & D & A & A & D & C \\
79 & 16/8/2022 9:30:57 & etrivinor\_XXXX & 4 & B & A & D & C & C & D & E & A & C \\
80 & 16/8/2022 9:31:12 & bjimenezb1\_XXXX & 4 & B & A & D & C & A & D & E & A & C \\
81 & 16/8/2022 9:31:13 & mtorresc7\_XXXX & 4 & B & A & D & C & B & D & E & A & B \\
82 & 16/8/2022 9:31:23 & scanor\_XXXX & 4 & B & A & D & C & A & D & D & A & C \\
83 & 16/8/2022 9:31:46 & agarciac5\_XXXX & 4 & B & C & D & C & C & D & D & D & C \\
84 & 16/8/2022 9:32:52 & gcastillov\_XXXX & 4 & B & A & D & C & C & D & E & A & C \\
85 & 16/8/2022 9:33:47 & aflorezr2\_XXXX & 4 & A & C & D & E & E & D & C & D & C \\
86 & 16/8/2022 9:34:29 & msuarezo\_XXXX & 4 & B & A & B & C & A & D & E & D & C \\
87 & 16/8/2022 9:35:29 & mgomezv3\_XXXX & 4 & B & A & D & B & C & D & E & A & C \\
\hline
\end{tabular}
\caption{Individual responses in cycle $3$.}
\label{tab:my-666table}
\end{table}

\begin{table}[htbp]
\centering
\begin{tabular}{|c|c|c|c|c|c|c|c|c|c|c|c|c|}
\hline
N & Date & id & test & $a_0$ & $a_1$ & $a_2$ & $a_3$ & $a_4$ & $a_5$ & $a_6$ & $a_7$ & $a_8$ \\
\hline
88 & 16/8/2022 9:38:03 & kleond\_XXXX & 5 & B & A & D & D & D & D & D & B & \\ \hline
89 & 16/8/2022 10:01:32 & soliverosc\_XXXX & 5 & B & A & D & C & C & D & E & A & C \\ \hline
90 & 16/8/2022 10:01:42 & sgiraldoa\_XXXX & 5 & B & A & D & C & C & D & E & A & C \\ \hline
91 & 16/8/2022 10:01:46 & scanor\_XXXX & 5 & B & A & D & C & C & D & E & A & C \\ \hline
92 & 16/8/2022 10:01:57 & mtorresc7\_XXXX & 5 & B & A & D & C & C & D & E & A & C \\ \hline
93 & 16/8/2022 10:02:06 & cmendezp1\_XXXX & 5 & B & A & D & C & C & D & E & A & C \\ \hline
94 & 16/8/2022 10:02:23 & jmurillot\_XXXX & 5 & B & A & D & C & C & D & E & A & C \\ \hline
95 & 16/8/2022 10:02:50 & bjimenezb1\_XXXX & 5 & B & A & D & C & C & D & E & A & C \\ \hline
96 & 16/8/2022 10:02:53 & etrivinor\_XXXX & 5 & B & A & D & C & C & D & E & A & C \\ \hline
97 & 16/8/2022 10:02:55 & mgomezv3\_XXXX & 5 & B & A & D & C & C & D & E & A & C \\ \hline
98 & 16/8/2022 10:03:42 & agarciac5\_XXXX & 5 & B & A & D & C & C & D & E & A & C \\ \hline
99 & 16/8/2022 10:03:45 & gcastillov\_XXXX & 5 & B & A & D & C & C & D & E & A & C \\ \hline
100 & 16/8/2022 10:04:00 & mfonsecai\_XXXX & 5 & B & A & D & C & C & D & E & D & D \\ \hline
101 & 16/8/2022 10:04:03 & sriverosr1\_XXXX & 5 & B & A & D & C & C & D & E & A & C \\ \hline
102 & 16/8/2022 10:04:27 & gmorenov1\_XXXX & 5 & B & A & D & C & A & D & E & D & C \\ \hline
103 & 16/8/2022 10:04:39 & kleond\_XXXX & 5 & B & A & D & D & D & D & D & C & \\ \hline
104 & 16/8/2022 10:04:43 & aflorezr2\_XXXX & 5 & B & A & D & C & E & D & E & D & C \\ \hline
105 & 16/8/2022 10:05:27 & msuarezo\_XXXX & 5 & B & A & D & C & A & D & E & D & C \\ \hline
\end{tabular}%
\caption{Responses after discussion with a peer in cycle $3$.}
\end{table}

\begin{table}[htbp]
\centering
\begin{tabular}{|c|c|c|c|c|c|c|c|c|c|}
\hline
\textbf{N} & \textbf{Date} & \textbf{id} & \textbf{test} & $\mathbf{a_0}$ & $\mathbf{a_1}$ & $\mathbf{a_2}$ & $\mathbf{a_3}$ & $\mathbf{a_4}$ & $\mathbf{a_5}$ \\ \hline
106 & 17/8/2022 9:24:38 & mgomezv3\_XXXX & 6 & A & B & C & D & A & D \\ \hline
107 & 17/8/2022 12:08:09 & bjimenezb1\_XXXX & 6 & B & B & C & D & A & D \\ \hline
108 & 17/8/2022 15:00:24 & jmurillot\_XXXX & 6 & A & B & C & D & A & D \\ \hline
109 & 17/8/2022 18:40:16 & gmorenov1\_XXXX & 6 & A & B & C & A & D & D \\ \hline
110 & 17/8/2022 18:42:07 & gcastillov\_XXXX & 6 & A & B & C & D & A & D \\ \hline
111 & 17/8/2022 19:19:43 & sriverosr1\_XXXX & 6 & A & B & C & D & A & D \\ \hline
112 & 17/8/2022 19:41:21 & mtorresc7\_XXXX & 6 & A & B & C & D & A & D \\ \hline
113 & 17/8/2022 19:45:58 & soliverosc\_XXXX & 6 & A & B & B & D & A & C \\ \hline
114 & 17/8/2022 19:46:45 & scanor\_XXXX & 6 & A & B & B & D & A & C \\ \hline
115 & 17/8/2022 21:19:01 & agarciac5\_XXXX & 6 & A & B & C & D & A & D \\ \hline
116 & 17/8/2022 21:29:21 & sgiraldoa\_XXXX & 6 & A & B & C & D & A & D \\ \hline
117 & 17/8/2022 21:31:52 & cmendezp1\_XXXX & 6 & A & B & C & D & A & D \\ \hline
118 & 17/8/2022 22:07:34 & jramirezm14\_XXXX & 6 & B & A & D & C & C & D \\ \hline
119 & 17/8/2022 22:08:09 & jramirezm14\_XXXX & 6 & A & B & C & D & A & D \\ \hline
120 & 17/8/2022 22:27:00 & msuarezo\_XXXX & 6 & A & B & C & C & A & D \\ \hline
121 & 17/8/2022 22:39:25 & kleond\_XXXX & 6 & A & A & C & C & A & D \\ \hline
122 & 17/8/2022 22:54:36 & etrivinor\_XXXX & 6 & A & D & B & D & A & D \\ \hline
\end{tabular}
\caption{Individual responses in cycle $4$.}
\label{tab:my-t6yable}
\end{table}

\begin{table}[htbp]
\centering
\begin{tabular}{|c|c|c|c|c|c|c|c|c|c|c|c|c|}
\hline
N & Date & id & test & $a_0$ & $a_1$ & $a_2$ & $a_3$ & $a_4$ & $a_5$ & $a_6$ & $a_7$ & $a_8$ \\
\hline
123 & 17/8/2022 0:00:00 & mfonsecai\_XXXX & 7 & A & B & C & D & A & D & E & E & D \\
\hline
124 & 22/8/2022 9:50:00 & mtorresc7\_XXXX & 7 & A & C & E & C & B & C & B & C & A \\
\hline
125 & 22/8/2022 11:36:20 & jmurillot\_XXXX & 7 & A & C & E & C & B & C & B & C & A \\
\hline
126 & 22/8/2022 13:32:34 & kleond\_XXXX & 7 & A & C & E & C & B & C & B & D & A \\
\hline
127 & 22/8/2022 13:46:33 & sriverosr1\_XXXX & 7 & A & C & E & C & B & C & B & C & A \\
\hline
128 & 22/8/2022 14:53:39 & cmendezp1\_XXXX & 7 & A & C & E & C & B & C & B & C & A \\
\hline
129 & 22/8/2022 16:09:20 & bjimenezb1\_XXXX & 7 & A & C & E & C & B & C & B & C & A \\
\hline
130 & 22/8/2022 16:32:50 & etrivinor\_XXXX & 7 & A & C & E & C & B & C & B & C & A \\
\hline
131 & 22/8/2022 16:55:36 & gmorenov1\_XXXX & 7 & A & C & E & C & B & C & B & B & A \\
\hline
132 & 22/8/2022 18:27:12 & mgomezv3\_XXXX & 7 & A & C & E & C & B & C & B & C & A \\
\hline
133 & 22/8/2022 20:48:46 & aflorezr2\_XXXX & 7 & A & C & E & C & B & C & B & B & A \\
\hline
134 & 22/8/2022 21:02:57 & scanor\_XXXX & 7 & A & C & E & C & B & C & B & C & A \\
\hline
135 & 22/8/2022 21:03:39 & sgiraldoa\_XXXX & 7 & A & C & E & D & B & C & B & C & A \\
\hline
136 & 22/8/2022 21:03:49 & mfonsecai\_XXXX & 7 & A & C & E & D & B & C & B & C & A \\
\hline
137 & 22/8/2022 21:06:09 & agarciac5\_XXXX & 7 & A & C & E & D & B & C & B & C & A \\
\hline
138 & 22/8/2022 21:30:34 & jramirezm14\_XXXX & 7 & A & C & E & D & B & C & B & C & A \\
\hline
139 & 22/8/2022 21:55:15 & msuarezo\_XXXX & 7 & A & C & E & C & B & C & B & D & A \\
\hline
140 & 22/8/2022 23:03:07 & soliverosc\_XXXX & 7 & A & C & E & C & B & C & B & C & A \\
\hline
141 & 22/8/2022 23:10:55 & gcastillov\_XXXX & 7 & A & C & E & C & B & C & B & C & A \\
\hline
\end{tabular}
\caption{Responses after discussion with a peer in cycle $4$.}
\label{tab:my-toiiuable}
\end{table}

\begin{table}[htbp]
\centering
\begin{tabular}{|c|c|c|c|c|c|c|c|c|c|}
\hline
N & Date & id & test & $a_0$ & $a_1$ & $a_2$ & $a_3$ & $a_4$ & $a_5$ \\
\hline
142 & 29/8/2022 1:34:00 & gcastillov\_XXXX & 8 & D & A & D & E & B & B \\
\hline
143 & 29/8/2022 8:06:15 & jmurillot\_XXXX & 8 & D & A & D & E & B & B \\
\hline
144 & 29/8/2022 15:24:31 & gmorenov1\_XXXX & 8 & D & B & D & E & B & B \\
\hline
145 & 29/8/2022 17:25:52 & cmendezp1\_XXXX & 8 & D & B & D & E & B & B \\
\hline
146 & 29/8/2022 17:58:49 & mgomezv3\_XXXX & 8 & D & B & D & E & B & B \\
\hline
147 & 29/8/2022 18:09:13 & bjimenezb1\_XXXX & 8 & D & B & D & E & B & B \\
\hline
148 & 29/8/2022 19:10:40 & scanor\_XXXX & 8 & D & B & A & E & B & B \\
\hline
149 & 29/8/2022 20:06:29 & mtorresc7\_XXXX & 8 & B & A & A & C & B & B \\
\hline
150 & 29/8/2022 20:09:50 & aflorezr2\_XXXX & 8 & D & B & D & E & B & B \\
\hline
151 & 29/8/2022 20:10:50 & soliverosc\_XXXX & 8 & D & B & A & E & B & B \\
\hline
152 & 29/8/2022 20:10:50 & soliverosc\_XXXX & 8 & D & B & A & E & B & B \\
\hline
153 & 29/8/2022 20:30:32 & etrivinor\_XXXX & 8 & D & B & D & E & B & B \\
\hline
154 & 29/8/2022 20:33:13 & kleond\_XXXX & 8 & D & E & D & D & A & B \\
\hline
155 & 29/8/2022 21:40:00 & jramirezm14\_XXXX & 8 & D & A & D & D & B & B \\
\hline
156 & 29/8/2022 21:40:04 & sgiraldoa\_XXXX & 8 & D & A & D & D & B & B \\
\hline
157 & 29/8/2022 21:41:18 & mfonsecai\_XXXX & 8 & D & A & D & D & B & B \\
\hline
158 & 29/8/2022 21:41:58 & agarciac5\_XXXX & 8 & D & A & D & D & B & B \\
\hline
159 & 29/8/2022 22:00:43 & msuarezo\_XXXX & 8 & D & E & D & E & B & B \\
\hline
160 & 29/8/2022 22:09:16 & sriverosr1\_XXXX & 8 & B & A & A & B & C & B \\
\hline
161 & 29/8/2022 22:00:43 & msuarezo\_XXXX & 8 & D & E & D & E & B & B \\
\hline
\end{tabular}
\caption{Individual responses in cycle $5$.}
\end{table}
\begin{table}[htbp]
\centering
\begin{tabular}{|c|c|c|c|c|c|c|c|c|c|}
\hline
N & Date & id & test & $a_0$ & $a_1$ & $a_2$ & $a_3$ & $a_4$ & $a_5$ \\
\hline
162 & 30/8/2022 9:34:04 & cmendezp1\_XXXX & 9 & E & D & A & B & B & C \\
\hline
163 & 30/8/2022 9:34:23 & jmurillot\_XXXX & 9 & E & D & A & B & B & C \\
\hline
164 & 30/8/2022 9:36:55 & soliverosc\_XXXX & 9 & E & D & E & E & C & B \\
\hline
165 & 30/8/2022 9:37:28 & scanor\_XXXX & 9 & E & D & A & B & B & C \\
\hline
166 & 30/8/2022 9:38:30 & bjimenezb1\_XXXX & 9 & E & D & A & B & B & C \\
\hline
167 & 30/8/2022 9:38:42 & mtorresc7\_XXXX & 9 & E & D & A & B & B & C \\
\hline
168 & 30/8/2022 9:39:04 & aflorezr2\_XXXX & 9 & E & D & A & E & B & C \\
\hline
169 & 30/8/2022 9:39:05 & kleond\_XXXX & 9 & E & D & A & B & B & C \\
\hline
170 & 30/8/2022 9:39:56 & etrivinor\_XXXX & 9 & E & B & A & B & C & C \\
\hline
171 & 30/8/2022 9:40:04 & gcastillov\_XXXX & 9 & E & D & A & B & B & C \\
\hline
172 & 30/8/2022 9:41:18 & sriverosr1\_XXXX & 9 & E & D & A & B & B & C \\
\hline
173 & 30/8/2022 9:41:48 & jramirezm14\_XXXX & 9 & E & D & C & E & C & B \\
\hline
174 & 30/8/2022 9:42:23 & mgomezv3\_XXXX & 9 & E & D & A & B & B & C \\
\hline
175 & 30/8/2022 9:42:35 & gmorenov1\_XXXX & 9 & E & D & A & E & C & D \\
\hline
176 & 30/8/2022 9:44:09 & sgiraldoa\_XXXX & 9 & E & D & C & E & C & B \\
\hline
177 & 30/8/2022 9:44:49 & agarciac5\_XXXX & 9 & E & D & E & B & C & C \\
\hline
178 & 30/8/2022 9:45:19 & msuarezo\_XXXX & 9 & E & D & D & E & D & B \\
\hline
179 & 30/8/2022 9:45:55 & mfonsecai\_XXXX & 9 & E & D & C & B & C & C \\
\hline
\end{tabular}
\caption{Responses after discussion with a peer in cycle $5$.}
\label{tab:another-table}
\end{table}

\begin{table}[htbp]
\centering
\begin{tabular}{|c|c|c|c|c|c|c|c|c|c|}
\hline
N & Date & id & test & $a_0$ & $a_1$ & $a_2$ & $a_3$ & $a_4$ & $a_5$ \\
\hline
180 & 30/8/2022 10:34:27 & bjimenezb1\_XXXX & 10 & E & D & A & B & B & C \\
\hline
181 & 30/8/2022 10:34:28 & mgomezv3\_XXXX & 10 & E & D & A & B & B & C \\
\hline
182 & 30/8/2022 10:34:37 & scanor\_XXXX & 10 & E & D & A & B & B & C \\
\hline
183 & 30/8/2022 10:34:47 & soliverosc\_XXXX & 10 & E & D & A & B & B & C \\
\hline
184 & 30/8/2022 10:34:49 & jmurillot\_XXXX & 10 & E & D & A & B & B & C \\
\hline
185 & 30/8/2022 10:35:02 & mtorresc7\_XXXX & 10 & E & D & A & B & B & C \\
\hline
186 & 30/8/2022 10:36:13 & gmorenov1\_XXXX & 10 & E & D & A & B & B & C \\
\hline
187 & 30/8/2022 10:37:46 & cmendezp1\_XXXX & 10 & E & D & A & B & B & C \\
\hline
188 & 30/8/2022 10:37:57 & mfonsecai\_XXXX & 10 & E & D & A & B & B & C \\
\hline
189 & 30/8/2022 10:38:37 & aflorezr2\_XXXX & 10 & E & D & A & B & B & C \\
\hline
190 & 30/8/2022 10:39:27 & agarciac5\_XXXX & 10 & E & D & A & B & B & C \\
\hline
191 & 30/8/2022 10:39:35 & sriverosr1\_XXXX & 10 & E & D & A & B & B & C \\
\hline
192 & 30/8/2022 10:39:47 & jramirezm14\_XXXX & 10 & E & D & A & B & B & C \\
\hline
193 & 30/8/2022 10:39:48 & sgiraldoa\_XXXX & 10 & E & D & A & B & B & C \\
\hline
194 & 30/8/2022 10:40:27 & etrivinor\_XXXX & 10 & E & D & A & B & B & C \\
\hline
195 & 30/8/2022 10:40:41 & kleond\_XXXX & 10 & E & D & A & B & B & C \\
\hline
196 & 30/8/2022 10:41:07 & gcastillov\_XXXX & 10 & E & D & A & B & B & C \\
\hline
197 & 30/8/2022 10:42:48 & msuarezo\_XXXX & 10 & E & D & A & E & B & C \\
\hline
\end{tabular}
\caption{Individual responses in cycle $6$.}
\end{table}

\begin{table}[htbp]
\centering
\begin{tabular}{|c|c|c|c|c|c|c|c|c|}
\hline
N & Date & id & test & $a_0$ & $a_1$ & $a_2$ & $a_3$ & $a_4$ \\
\hline
198 & 31/8/2022 9:45:44 & gcastillov\_XXXX & 11 & A & B & A & B & B \\
\hline
199 & 31/8/2022 11:35:53 & bjimenezb1\_XXXX & 11 & A & B & A & B & B \\
\hline
200 & 31/8/2022 14:33:36 & jmurillot\_XXXX & 11 & A & B & A & B & B \\
\hline
201 & 31/8/2022 15:01:47 & kleond\_XXXX & 11 & A & B & A & B & B \\
\hline
202 & 31/8/2022 15:46:42 & cmendezp1\_XXXX & 11 & A & B & A & B & B \\
\hline
203 & 31/8/2022 16:30:44 & gmorenov1\_XXXX & 11 & A & B & A & B & B \\
\hline
204 & 31/8/2022 17:52:28 & mtorresc7\_XXXX & 11 & A & B & A & B & B \\
\hline
205 & 31/8/2022 18:17:28 & mgomezv3\_XXXX & 11 & A & B & A & B & B \\
\hline
206 & 31/8/2022 19:20:22 & etrivinor\_XXXX & 11 & C & B & D & B & B \\
\hline
207 & 31/8/2022 19:21:38 & sriverosr1\_XXXX & 11 & A & B & A & B & B \\
\hline
208 & 31/8/2022 21:03:31 & msuarezo\_XXXX & 11 & A & B & A & B & B \\
\hline
209 & 31/8/2022 21:35:32 & scanor\_XXXX & 11 & A & B & A & B & B \\
\hline
210 & 31/8/2022 21:55:22 & aflorezr2\_XXXX & 11 & A & B & A & B & B \\
\hline
211 & 31/8/2022 21:55:22 & aflorezr2\_XXXX & 11 & A & B & A & B & B \\
\hline
212 & 31/8/2022 22:21:48 & mfonsecai\_XXXX & 11 & A & B & A & B & B \\
\hline
213 & 31/8/2022 22:22:50 & sgiraldoa\_XXXX & 11 & A & B & A & B & B \\
\hline
214 & 31/8/2022 22:23:05 & agarciac5\_XXXX & 11 & A & B & A & B & B \\
\hline
215 & 31/8/2022 22:59:19 & jramirezm14\_XXXX & 11 & A & B & A & B & B \\
\hline
216 & 31/8/2022 23:19:57 & soliverosc\_XXXX & 11 & A & B & A & B & B \\
\hline
\end{tabular}
\caption{Responses after discussion with a peer in cycle $6$.}
\end{table}


\begin{table}[htbp]
\centering
\begin{tabular}{|c|c|c|c|c|c|c|c|c|c|}
\hline
Test & $r_0$ & $r_1$ & $r_2$ & $r_3$ & $r_4$ & $r_5$ & $r_6$ & $r_7$ & $r_8$ \\
\hline
1 & A & B & B & A & E & D & B & C & C \\
\hline
2 & B & D & A & D & D & B & E & C & A \\
\hline
3 & C & A & C & A & C & B & B & E & C \\
\hline
4 & B & A & D & C & C & D & E & A & C \\
\hline
5 & A & B & C & D & A & D & E & E & D \\
\hline
6 & A & C & E & C & B & C & B & C & A \\
\hline
\end{tabular}
\caption{Table of Test Data}
\label{tab:test-data}
\end{table}
\chapter{Research instruments in learning the fall of bodies} \label{app:indagCaida}

The instruments of research was presented to the students as shown below.

Los instrumentos de indagación se etiquetaron como experimentos $\alpha, \beta ,\gamma,\delta$ y $\epsilon$. Cuya presentación fué la siguiente:\\

\emph{En el siguiente texto se describen tres experimentos con sus resultados, de acuerdo con ellos marque verdadero  \textbf{V}  o falso  \textbf{F} para 
cada una de las afirmaciones abajo formuladas.} 

\section*{Experimentos $\alpha$}
\subsection*{Experimento $\alpha-1$}
En el salón de clase un profesor toma una hoja de papel limpia,  la corta en dos pedazos ligeramente desiguales, toma el más grande y forma una bola arrugando el papel. Luego
toma en una mano el trozo de papel arrugado  y en la otra el papel sin arrugar, estira las manos a la misma altura del piso y deja caer simultáneamente los dos trozos de papel.  Llega al piso primero el trozo de papel arrugado en forma de bola. 
\begin{enumerate}
\item El resultado se debe a que siempre los cuerpos  pesados caen más rápido que los cuerpos livianos.  \textbf{V}  \textbf{F} 
\item Los resultados del experimento están de acuerdo con la ley de caída de los cuerpos de Galileo, pues los trozos de papel caen con la misma aceleración independiente de su peso,
      pero como tienen diferente tamaño no caen simultáneamente. \textbf{V}  \textbf{F}
\item Como todos los cuerpos con la misma forma siempre caen al mismo tiempo y en este experimento los trozos de papel no tienen la misma forma, entonces por esta razón no caen al
      mismo tiempo. \textbf{V}  \textbf{F}
\end{enumerate}
\subsection*{Experimento $\alpha-2$}
Ahora el profesor recoge los trozos de papel del piso, desarruga y alisa  el trozo arrugado y arruga el otro formando también una bola. Luego toma en una mano el trozo de papel
arrugado y en la otra el papel sin arrugar, estira las manos a la misma altura del piso y deja caer simultáneamente los dos trozos de papel.  Y de nuevo,  llega al piso primero
el trozo de papel arrugado.
\begin{enumerate}
\item Los resultados del experimento están de acuerdo con la ley de caída de los cuerpos de Galileo, pues los trozos de papel caen con la misma aceleración independiente de su peso,
      pero como tienen diferente tamaño no caen simultáneamente. \textbf{V}  \textbf{F}
\item El experimento demuestra que es posible que un cuerpo liviano llegue antes al piso que uno pesado.\textbf{V}  \textbf{F}
\item Como cuerpos con la misma forma siempre caen al mismo tiempo y en este experimento los trozos de papel no tienen la misma forma, entonces por esta razón no caen al mismo
     tiempo. \textbf{V}  \textbf{F}
\end{enumerate}
\subsection*{Experimento $\alpha-3$}
Finalmente, el profesor recoge los trozos de papel del piso, los arruga formando con cada trozo una pequeña bola y tomando en cada mano un  trozo los deja caer simultáneamente
desde la misma altura.  Los trozos de papel llegan al tiempo al piso.
\begin{enumerate}
\item Lo observado en el experimento está de acuerdo con la ley de caída de los cuerpos  de Galileo, ya que según tal ley  todos los cuerpos caen con la misma aceleración $g=9,8m/s^2$
      y por tanto llegan al mismo tiempo al piso. \textbf{V}  \textbf{F}
\item De acuerdo con el experimento es posible que dos cuerpos de diferente peso lleguen simultáneamente al piso. \textbf{V}  \textbf{F}
\item Como cuerpos con la misma forma siempre caen al mismo tiempo y en este experimento los trozos de papel  tienen la misma forma, entonces por esta razón  caen al mismo
      tiempo. \textbf{V}  \textbf{F}
\end{enumerate}
\section*{Experimentos $\beta$}

\subsection*{Experimento $\beta-1$}
El profesor toma una moneda en una mano y en la otra un trocito de papel  más pequeño que la moneda, los deja caer simultáneamente desde unos $50cm$ de altura. Cae primero la moneda. 
\begin{enumerate}
\item El resultado se debe a que siempre los cuerpos  pesados caen antes que los cuerpos livianos.  \textbf{V}  \textbf{F} 
\item Los resultados del experimento  se explican debido a que cuerpos con diferente forma nunca llegarán simultáneamente al suelo, esto sólo ocurriría si
      los cuerpos caen en el vacío; tal como afirma la ley de caída de los cuerpos propuesta por Galileo.\textbf{V}  \textbf{F}
\item El resultado del experimento es el esperado ya que cuerpos con forma diferente nunca llegarán simultáneamente al suelo. \textbf{V}  \textbf{F}
\end{enumerate}
\subsection*{Experimento $\beta-2$}
Ahora, el profesor toma la moneda y encima de ella pone el trocito de papel, así los deja caer desde unos $50cm$ de altura. Se observa que la moneda y el papel no se separan hasta
después de que la moneda toca el piso.
\begin{enumerate}
\item Lo observado en el experimento está de acuerdo con Galileo, ya que según él  todos los cuerpos  han de  llegar al mismo tiempo al piso independiente de su
      peso. \textbf{V}  \textbf{F}
\item El resultado del experimento es el esperado ya que cuerpos con forma diferente nunca llegarán simultáneamente al suelo, pero como están en contacto se puede considerar
      que forman un solo cuerpo y por esto caen simultáneamente al suelo.   \textbf{V}  \textbf{F}
\item El experimento demuestra que dos cuerpos con pesos bastante diferentes pueden caer al piso en el mismo instante.\textbf{V}  \textbf{F}
\end{enumerate}
\subsection*{Experimento $\beta-3$}
Por último, el profesor toma la moneda y debajo de ella pone el trocito de papel, los deja caer desde aproximadamente $1m$ de altura. Se observa que rápidamente la moneda
y el papel se separan y llega primero al piso la moneda.
\begin{enumerate}
\item Cuerpos con forma diferente nunca llegan simultáneamente al suelo, esto es lo que demuestra este experimento. \textbf{V}  \textbf{F}
\item Esto demuestra que en algunos casos es posible que un cuerpo pesado llegue antes al piso que uno liviano. \textbf{V}  \textbf{F}
\item El experimento no se puede explicar con la ley de la caída de los cuerpos de Galileo, pues esta ley sólo es válida cuando los efectos del aire sobre  los cuerpos no cambian
      de manera significativa su aceleración y la mantiene prácticamente igual a $9,8m/s^2$.  \textbf{V}  \textbf{F}
\end{enumerate}
\section*{Experimentos $\gamma$}

\subsection*{Experimento $\gamma-1$}

El profesor usa dos globos para piñata idénticos, un trozo de hilo muy liviano y una canica como materiales de trabajo. Primero infla los globos uno de mayor tamaño que el otro, toma uno en cada mano y
los deja caer desde la misma altura simultáneamente. Llega al piso primero el de menor tamaño.

\begin{enumerate}
\item Como los globos no tienen el mismo tamaño no caen al tiempo, si tuvieran el mismo tamaño caería al tiempo. \textbf{V}  \textbf{F} 
\item Este experimento demuestra que en algunos casos es posible que un cuerpo liviano llegue antes al piso que uno pesado. \textbf{V}  \textbf{F}
\item La ley de caída de los cuerpos propuesta por Galileo afirma que todos los cuerpos independientemente de su peso caen con la misma velocidad, como
      los globos no llegan en el mismo instante al suelo significa que tales globos tienen diferente peso.   \textbf{V}  \textbf{F}
\end{enumerate}
\subsection*{Experimento $\gamma-2$}
En seguida el profesor infla uno de los globos con una canica dentro y el otro simplemente inflado. Los deja caer de la misma altura, y se observa que cae primero el globo
sin canica. 
\begin{enumerate}
\item Para que este resultado se dé es necesario inflar los globos de manera que tengan la misma forma y tamaño, ya que todos los cuerpos con el mismo tamaño y forma llegan en
      el mismo instante al suelo.  \textbf{V}  \textbf{F}
\item Si el globo que contiene la canica se infla más que el que no tiene la canica es posible que lo reportado en el experimento ocurra, ya que al soltar dos cuerpos es posible
      que el más liviano llegue antes al piso.  \textbf{V}  \textbf{F}
\item De acuerdo con Galileo todos los cuerpos caen de manera simultánea e independientemente de sus pesos cuando no hay presencia de aire,  pero como no es el caso 
      entonces cae primero el cuerpo más pesado. \textbf{V}  \textbf{F}
\end{enumerate}
\subsection*{Experimento $\gamma-3$}
  Ahora, el profesor toma los globos infla uno con gran tamaño y otro con tamaño medio, los amarra con un hilo corto muy liviano y los  deja caer simultáneamente desde la
  misma altura.  Luego pregunta por lo que ocurrirá. 
\begin{enumerate}
\item Debido a la ley de caída de los cuerpos de Galileo, ocurrirá que los globos caerán despacio y el hilo entre ellos no se estirará por completo. \textbf{V}  \textbf{F}
\item Como los globos tienen prácticamente la misma forma y tamaños diferentes, el globo de mayor tamaño arrastrara tensando el hilo al globo de menor
      tamaño.\textbf{V}  \textbf{F}
\item Como los efectos del aire sobre el globo más liviano son menores que sobre el más pesado, el globo de menor  peso arrastrara tensando el hilo al globo
      de mayor peso. \textbf{V}  \textbf{F}
\end{enumerate}

\section*{Experimentos $\delta$}
     
\subsection*{Experimento $\delta-1$}
En esta ocasión el profesor toma una bola esférica de icopor y la corta a la mitad, toma una semiesfera en cada mano y las deja caer desde la misma altura y con el frente curvo
  enfrentando el piso. Las semiesferas llegan al piso simultáneamente.
\begin{enumerate}
\item El resultado está de acuerdo con Galileo, pues todos los cuerpos caen simultáneamente independiente de su peso y el medio en que caen. \textbf{V}  \textbf{F}
\item El resultado del experimento está de acuerdo con la ley de caída de los cuerpos que dice que dos cuerpos de la misma forma, tamaño y peso siempre caerán al
      mismo tiempo. \textbf{V}  \textbf{F}
\item El resultado del experimento era de esperarse ya que teóricamente se sabe que la única forma para que dos cuerpos caigan simultáneamente en
      presencia de  aire es que tengan el mismo peso. \textbf{V}  \textbf{F}
\end{enumerate}
\subsection*{Experimento $\delta-2$}
Ahora el profesor deja caer desde la misma altura y de manera simultánea las semiesferas, pero esta vez una por el frente curvo y la otra por el frente plano. Llega
primero al piso la semiesfera que se deja caer con el frente curvo.
\begin{enumerate}
\item La ley de caída de los cuerpos de Galileo afirma que en el vacío todos los cuerpos caen con la misma aceleración independiente de su peso, como los cuerpos
      no están en el vacío tal ley no es aplicable y por tanto no se garantiza que las semiesferas caigan al suelo en el mismo instante. \textbf{V}  \textbf{F}
\item El experimento demuestra que dos cuerpos con la misma forma, tamaño y peso pueden no caer simultáneamente.\textbf{V}  \textbf{F}
\item El experimento demuestra que dos cuerpos del mismo peso  no necesariamente llegan al tiempo al suelo.\textbf{V}  \textbf{F}
\end{enumerate}
\subsection*{Experimento $\delta-3$}
   Finalmente el profesor incrusta monedas a la semiesfera que dejará  caer por el frente plano y deja igual la semiesfera que dejará caer por el frente curvo,  luego
   las deja caer de la misma altura y simultáneamente.  Cae primero la semiesfera con monedas incrustadas.
\begin{enumerate}
\item  El experimento demuestra que siempre que dos cuerpos se dejan caer simultáneamente  llega al suelo primero el más pesado. \textbf{V}  \textbf{F}
\item  El resultado está de acuerdo con la ley de caída de los cuerpos de Galileo,  pues siempre los cuerpos mas pesados llegan primero al piso ya que este es su estado
       natural  y como la semiesfera con monedas pesa más,  tal ley se cumple.\textbf{V}  \textbf{F}
\item  El experimento demuestra que dos cuerpos con la misma forma y tamaño pero peso diferente  nunca caerán simultáneamente al suelo. \textbf{V}  \textbf{F}
\end{enumerate}

\section*{Experimentos $\epsilon$}

\subsection*{Experimento $\epsilon-1$}
Para este experimento el profesor usa una esfera de icopor y una canica, los suelta de unos $15cm$ de altura simultáneamente y observa que caen al piso en el mismo instante.

\begin{enumerate}
   \item El experimento demuestra que dos cuerpos de pesos diferentes pueden llegar al suelo simultáneamente.  \textbf{V}  \textbf{F}
   \item El resultado del experimento está de acuerdo con la ley de caída de los cuerpos propuesta por Galileo, ya que cuando los cuerpos están cayendo atraídos por la tierra
         adquieren la misma aceleración $g=9.8m/s^2$ sin importar su peso y por tanto llegan al piso en el mismo instante.  \textbf{V}  \textbf{F}
   \item El experimento demuestra que dos cuerpos con la misma forma siempre caerán simultáneamente al suelo.\textbf{V}  \textbf{F}
\end{enumerate}
\subsection*{Experimento $\epsilon-2$}
         Ahora el profesor deja caer simultáneamente la canica y la bola de icopor pero de unos $2,5m$ de altura, aproximadamente desde el techo del salón . Se observa que la canica
         llega primero al suelo.

\begin{enumerate}
\item El resultado de este experimento no puede explicarse con la ley de caída de los cuerpos propuesta por Galileo, ya que el aire afecta de manera
      significativa la aceleración de la bola de icopor.  \textbf{V}  \textbf{F}
\item El experimento demuestra que dependiendo de la altura desde la cual se suelten dos cuerpos  estos pueden llegar o no simultáneamente al suelo, además se
      puede inferir que este hecho físico es independiente de los pesos de cada cuerpo.  \textbf{V}  \textbf{F}
\item El experimento demuestra que es posible que cuerpos de la misma forma no caigan  simultáneamente al suelo. \textbf{V}  \textbf{F}
\end{enumerate}
\subsection*{Experimento $\epsilon-3$}
Por último, el profesor deja caer simultáneamente la canica y la bola de icopor pero cada una de diferente altura. Se observa que llegan al suelo en el mismo instante.
\begin{enumerate}
\item Para que el resultado sea posible se debe soltar la canica de mayor altura ya que debido a la ley de caía de los cuerpos de Galileo, todos los cuerpos independientemente
      del medio en que se muevan caerán simultáneamente. \textbf{V}  \textbf{F}
\item  Para que el resultado sea posible se debe soltar la canica de mayor altura ya que todos los cuerpos más pesados llegan antes que los livianos al suelo. \textbf{V}  \textbf{F}
\item  Para que el resultado sea posible se debe soltar la canica de mayor altura, porque esta adquirirá mayor aceleración que la de icopor. Estos
       sin importar que las dos tengan la misma forma. \textbf{V}  \textbf{F}
\end{enumerate}
\chapter{Disciplinary theoretical reference on the fall of bodies}\label{app:rtd_caida}

The document proposed as a disciplinary theoretical reference addressed to the student and the teacher was as follows:\\

\emph{Esta lectura pretende mostrar de manera cualitativa que cuándo se deja caer un cuerpo en condiciones  cotidianas su tiempo de caída  depende de la interacción a distancia
del cuerpo con el aire y con el planeta tierra. Se describen las tres fuerzas que actúan sobre un cuerpo que cae: el peso, la fuerza de empuje y la fuerza de fricción
con el aire. Además, que dependiendo de estas fuerzas el cuerpo adquirirá determinada aceleración que permitirá estimar de manera cualitativa su tiempo de caída.}\\

Cuando en un ambiente tranquilo libre de turbulencias o ventiscas  se deja caer  un cuerpo ya sea en la sala de una vivienda, en el salón de clases o en el laboratorio de
física, el tiempo que tarda en llegar al suelo  depende  de las fuerzas que actúan sobre dicho cuerpo. Estas fuerzas son: el peso, que es la fuerza de atracción gravitacional
que hace la tierra sobre tal cuerpo; la fuerza de empuje, que es la fuerza que ejerce el aire sobre el cuerpo y está dirigida verticalmente hacia arriba  y por último, la
fuerza de fricción con el aire que actúa hacia arriba sobre el cuerpo mientras este cae.\\

Cuando un cuerpo cae sobre la superficie de la tierra sometido únicamente a la fuerza de gravedad, por ejemplo en el vacío,  su aceleración es constante, es decir  no depende
de la altura a la que se deja caer, ni de sus propiedades, está es la conocida ley de caída de los cuerpos propuesta por Galileo. Tal ley es aplicable sólo si se cumplen
principalmente dos exigencias: que los efectos que el aire ocasiona sobre el cuerpo sean despreciables y  que el cuerpo caiga de alturas  considerablemente pequeñas comparadas
con el radio terrestre.  El peso $w$ de un cuerpo con masa $m$ es $w=mg$ con $g=9,8 m/s^2$. \\

Un cuerpo que cae está rodeado de aire, en estas condiciones experimenta una fuerza vertical hacia arriba que
es igual al peso del aire desalojado por tal cuerpo, esta fuerza se conoce como fuerza de empuje y al fenómeno en general como principio de Arquímedes. Para los casos cotidianos
la fuerza de empuje sobre un cuerpo es pequeña comparada con el peso del cuerpo ya que el peso del aire desalojado es muy pequeño comparado con el peso del cuerpo, por esta
razón suele despreciarse la fuerza de empuje, excepto en los casos de cuerpos extremadamente livianos donde el peso del cuerpo es comparable al peso del aire que desaloja.  La
fuerza de empuje está dada por $F_e=m_fg$, donde $m_f$ es la masa del fluido desalojado y $g=9,8 m/s^2$ la aceleración de la gravedad.\\ 

La fuerza de fricción con el medio tiene una dirección opuesta a la velocidad del cuerpo, así que cuando un cuerpo cae la fuerza de fricción con el medio actúa 
hacia arriba. Esta fuerza es  mayor cuando el área transversal al movimiento del cuerpo es mayor, también a mayor  rapidez 
mayor fuerza de fricción y esta fuerza depende de la forma y el tamaño del cuerpo.  De acuerdo con estudios experimentales la magnitud de la fuerza de fricción con el medio
se puede escribir como $f=\frac{1}{2}c\rho A v^2$. Donde $\rho$ es la densidad del aire, $A$ el área del cuerpo transversal al movimiento, $v$ la rapidez del cuerpo y $c$ es el 
coeficiente adimensional de arrastre que depende de la viscosidad del medio, la rapidez, la forma del cuerpo y el frente de movimiento.\\
\begin{figure}[!htp]
\begin{center}
	\begin{tikzpicture}[scale=1.0]
	\fill[gray!80](0.5cm,3.0cm) circle (0.3cm); 
	\draw[->,black,line width=1.5pt] (0.0cm,6cm)--(0.0cm,0.0cm);
	\draw[black,line width=1.0pt] (-0.2cm,5.5cm)--(0.2cm,5.5cm);
	\draw[->,black,line width=1.5pt] (0.5cm,2.95cm)--(0.5cm,1.0cm);
	\draw[->,black,line width=1.5pt] (0.5cm,3.0cm)--(0.5cm,4.0cm);
	\draw[->,black,line width=1.5pt] (0.5cm,3.0cm)--(0.5cm,3.5cm);
	\coordinate [label=below:\textcolor{black} {$o$}] (x) at  (-0.4cm,5.8cm);
	\coordinate [label=below:\textcolor{black} {$x$}] (x) at  (0.0cm,0.0cm);
	\coordinate [label=below:\textcolor{black} {$w$}] (x) at  (0.5cm,1.0cm);
	\coordinate [label=below:\textcolor{black} {$F_e$}] (x) at  (0.8cm,3.7cm);
	\coordinate [label=below:\textcolor{black} {$f$}] (x) at  (0.5cm,4.6cm);
	\end{tikzpicture}
\caption{Caida de un cuerpo esférico en el aire.}
\label{zen}
\end{center}
\end{figure}
Cuando un cuerpo cae  en general actúan sobre este las tres fuerzas anteriormente descritas, ver figura \ref{zen}. De acuerdo con la segunda ley de Newton y la figura \ref{zen} la 
aceleración del cuerpo está dada por
\begin{equation}\label{ace}
 a=g-\frac{F_e}{m}-\frac{f}{m},
\end{equation}
que al reemplazar las diferentes fuerzas toma la forma
\begin{equation}\label{ace2}
 a=\left( 1- \frac{m_f}{m}\right)g-\frac{c\rho A v^2}{2m}.
\end{equation}
Así, cuando el cuerpo se deja caer (velocidad inicial igual a cero) el tiempo de caída  guarda una relación
inversa con la aceleración: a mayor aceleración menor tiempo le toma al cuerpo llegar  al suelo  y a menor aceleración mayor tiempo le toma llegar al suelo.\\

De las ecuaciones (\ref{ace}) y (\ref{ace2}) se infiere en general lo siguiente: primero, el volumen (tamaño) del  cuerpo  afecta significativamente su aceleración, 
pues a mayor volumen mayor aire desplazado y por tanto mayor  fuerza de empuje; segundo,  la geometría (forma) del cuerpo es también importante  porque   dependiendo de la
manera en que cae el cuerpo su área transversal es diferente y  porque su frente de caída o movimiento afecta significativamente el coeficiente de arrastre, estos dos parámetros 
afectan directamente la fuerza de fricción con el aire y por tanto su aceleración;  tercero, la masa del cuerpo también afecta su aceleración pues a mayor masa mayor peso.

\chapter{Experimental results of the study carried out on learning to fall bodies} \label{app:rEx_caida}

The data derived from the social experiment has been structured in tables \ref{tab19}, \ref{tab13}, and \ref{tabUD}. In these tables, the first column represents the student, while the subsequent columns display the results for each trial $t_k$. Odd-numbered trials ($t_k$) correspond to individual assessments, and even-numbered trials refer to group assessments. The results are encoded in the following manner: each digit in the number signifies the level $q_0, q_1, q_2,$ or $q_3$ of Aristotelian, Galilean, or Newtonian thought. For the first two types of thinking, a high level implies the transcendence of associated preconceptions, whereas for Newtonian thought, a high level indicates a comprehensive understanding. The cumulative sum of the digits reflects the quantitative discernment of each student.

The data derived from the social experiment has been structured in tables \ref{tab19}, \ref{tab13}, and \ref{tabUD}. In these tables, the first column represents the student, while the subsequent columns display the results for each trial $t_k$. Odd-numbered trials ($t_k$) correspond to individual assessments, and even-numbered trials refer to group assessments. The results are encoded in the following manner: each digit in the number signifies the level $q_0, q_1, q_2,$ or $q_3$ of Aristotelian, Galilean, or Newtonian thought. For the first two types of thinking, a high level implies the transcendence of associated preconceptions, whereas for Newtonian thought, a high level indicates a comprehensive understanding. The cumulative sum of the digits reflects the quantitative discernment of each student.
\begin{table}[htp]
\centering
\begin{tabular}{|c||c|c|c|c|c|c|c|c|c|c|}
 \hline  
       & $t_1$& $t_2$& $t_3$& $t_4$& $t_5$& $t_6$& $t_7$& $t_8$& $t_9$& $t_{10}$ \\ \hline  \hline
    $1$  & 111	& 300	& 213	& 212	& 322	& 332	& 231	& 333	& 233	& 333	\\ \hline 
	$2$  & 211	& 210	& 121	& 211	& 211	& 332	& 232	& 332	& 233	& 333	\\ \hline	
	$3$  & 122	& 222	& 222	& 332	& 222	& 332	& 332	& 332	& 331	& 332	\\ \hline
	$4$  & 303	& 200	& 212	& 211	& 332	& 332	& 332	& 332	& 332	& 333	\\ \hline
	$5$  & 200	& 300	& 211	& 212	& 332	& 332	& 232	& 332	& 321	& 332	\\ \hline
	$6$  & 131	& 210	& 201	& 211	& 323	& 332	& 233	& 332	& 332	& 333	\\ \hline
	$7$  & 203	& 200	& 111	& 211	& 222	& 321	& 332	& 123	& 321	& 331	\\ \hline
	$8$  & 211	& 300	& 212	& 221	& 221	& 321	& 321	& 223	& 333	& 333	\\ \hline
	$9$  & 121	& 100	& 121	& 212	& 332	& 332	& 311	& 333	& 331	& 331	\\ \hline
	$10$ & 300	& 300	& 311	& 211	& 332	& 332	& 332	& 223	& 333	& 333	\\ \hline
	$11$ & 020 	& 100	& 220	& 211	& 22	& 112	& 111	& 123	& 233	& 331	\\ \hline
	$12$ & 202	& 222	& 111	& 211	& 12	& 332	& 332	& 332	& 332	& 332	\\ \hline
	$13$ & 213	& 210	& 210	& 211	& 233	& 332	& 332	& 332	& 231	& 331	\\ \hline
	$14$ & 300	& 300	& 123	& 212	& 332	& 332	& 231	& 332	& 231	& 332	\\ \hline
	$15$ & 131	& 102	& 122	& 221	& 222	& 332	& 332	& 333	& 323	& 323	\\ \hline
	$16$ & 322	& 323	& 332	& 332	& 332	& 332	& 333	& 223	& 322	& 332	\\ \hline  
\end{tabular}
\caption{Experimental results group $A$.}\label{tab19} 
\end{table}
\begin{table}[htp]
\centering
\begin{tabular}{|c||c|c|c|c|c|c|c|c|c|c|}
\hline  
       & $t_1$& $t_2$& $t_3$& $t_4$& $t_5$& $t_6$& $t_7$& $t_8$& $t_9$& $t_{10}$ \\ \hline  \hline
    $1$& 0& 	0  & 	200& 	223& 	23& 	332& 	322& 	333& 	322& 	333 \\ \hline 
	$2$& 300& 	300& 	212& 	122& 	333& 	333& 	231& 	231& 	332& 	332 \\ \hline 
	$3$& 223& 	303& 	121& 	233& 	101& 	333& 	231& 	332& 	330& 	331 \\ \hline 
	$4$& 0& 	0  & 	121& 	233& 	212& 	332& 	231& 	231& 	333& 	331 \\ \hline 
	$5$& 300& 	300& 	222& 	233& 	233& 	323& 	323& 	232& 	321& 	332 \\ \hline 
	$6$& 313& 	312& 	320& 	212& 	302& 	321& 	121& 	232& 	333& 	332 \\ \hline 
	$7$& 313& 	323& 	223& 	223& 	332& 	332& 	331& 	331& 	332& 	333 \\ \hline 
	$8$& 323& 	323& 	213& 	223& 	211& 	321& 	322& 	333& 	333& 	333 \\ \hline 
	$9$& 120& 	303& 	221& 	122& 	332& 	332& 	332& 	232& 	232& 	332 \\ \hline 
	$10$& 330& 	300& 	211& 	111& 	323& 	222& 	331& 	232& 	323& 	332 \\ \hline 
	$11$& 10& 	323& 	211& 	212& 	322& 	332& 	232& 	331& 	233& 	333 \\ \hline 
	$12$& 200& 	300& 	121& 	111& 	121& 	222& 	221& 	332&    323& 	322 \\ \hline 
	$13$& 301& 	312& 	20 &    223& 	221& 	332& 	322& 	333& 	332& 	332 \\ \hline 
	$14$& 313& 	300& 	123 &   13& 	212& 	102& 	202& 	232& 	123& 	332 \\ \hline 
	$15$& 103& 	303& 	101	&   212& 	213& 	323& 	221& 	231& 	210& 	331 \\ \hline 
	$16$& 302& 	300& 	112 & 	212& 	233& 	332& 	133& 	332& 	231& 	332 \\ \hline 
\end{tabular}
\caption{Experimental results group $B$.}\label{tab13} 
\end{table}
\begin{table}[htp]
\centering
\begin{tabular}{|c||c|c|c|c|c|c|c|c|c|c|}
\hline  
       & $t_1$& $t_2$& $t_3$& $t_4$& $t_5$& $t_6$& $t_7$& $t_8$& $t_9$& $t_{10}$ \\ \hline  \hline
	$1$   &203	& 300	& 213	& 212	& 0		& 332	& 332	& 332	& 322	& 333	\\ \hline
	$2$   &332	& 130	& 0		& 0		& 221	& 332	& 322	& 333	& 333	& 333	\\ \hline
	$3$   &0	& 0		& 101	& 201	& 332	& 332	& 332	& 332	& 323	& 333	\\ \hline
	$4$   &202	& 203	& 222	& 223	& 323	& 332	& 333	& 333	& 333	& 333	\\ \hline
	$5$   &102	& 203	& 222	& 222	& 332	& 332	& 232	& 333	& 332	& 332	\\ \hline
	$6$   &0	& 0		& 0		& 212	& 221	& 332	& 331	& 333	& 313	& 333	\\ \hline
	$7$   &300	& 300	& 111	& 212	& 120	& 332	& 322	& 333	& 333	& 333	\\ \hline
	$8$   &110	& 200	& 211	& 212	& 321	& 332	& 131	& 333	& 331	& 332	\\ \hline
	$9$   &110	& 200	& 203	& 212	& 333	& 332	& 232	& 333	& 333	& 333	\\ \hline
	$10$   &300	& 100	& 213	& 212	& 322	& 332	& 333	& 333	& 333	& 333	\\ \hline
	$11$   &211	& 303	& 232	& 223	& 232	& 332	& 322	& 333	& 332	& 332	\\ \hline
	$12$   &1	& 303	& 222	& 212	& 332	& 332	& 332	& 333	& 331	& 332	\\ \hline
	$13$   &120	& 100	& 213	& 212	& 333	& 332	& 333	& 333	& 333	& 332	\\ \hline
	$14$   &31	& 130	& 112	& 223	& 122	& 332	& 332	& 333	& 333	& 333	\\ \hline
	$15$   &20	& 200	& 223	& 201	& 312	& 332	& 333	& 333	& 333	& 333	\\ \hline
	$16$   &0	& 0		& 331	& 201	& 213	& 332	& 332	& 332	& 332	& 333	\\ \hline	
\end{tabular}
\caption{Experimental results group $C$.}\label{tabUD} 
\end{table}
\end{document}